\documentclass[preprint2]{aastex}
\usepackage{amsmath}
\usepackage{natbib}
\usepackage{graphicx}
\usepackage{txfonts}
\usepackage{subcaption}
\usepackage[pagebackref,breaklinks,colorlinks,citecolor=blue]{hyperref}
\usepackage[all]{hypcap}

\usepackage[nameinlink]{cleveref}
\usepackage{epsfig,color}

\newcommand{\eqa}[1]{\begin{align}  #1 \end{align}} 
\newcommand{\ba}[1]{\left\langle #1 \right\rangle}   
\newcommand{\msun}{\ensuremath{ {\rm M}_{\odot} }}
\newcommand{\mockalph}[1]{}

\newcommand{\permapj}[1]{\textit{\copyright\ AAS. Reproduced with permission from} \citet{#1}.}  
\newcommand{\permjcap}[1]{\textit{\copyright\ SISSA Medialab Srl.  Reproduced by permission of IOP Publishing from}  \citet{#1}. \textit{All rights reserved.}}  
\newcommand{\permmn}[1]{\textit{Reproduced with permission from} \citet{#1}.}  
\onecolumn

\begin{document}
\title{Galaxy alignments: Theory, modelling and simulations}

\author{Alina            Kiessling\altaffilmark{1},           Marcello
  Cacciato\altaffilmark{2},     Benjamin     Joachimi\altaffilmark{3},
  Donnacha  Kirk\altaffilmark{3}, Thomas  D. Kitching\altaffilmark{4},
  Adrienne Leonard\altaffilmark{3}, Rachel Mandelbaum\altaffilmark{5},
  Bj{\"o}rn    Malte     Sch{\"a}fer\altaffilmark{6},    Crist{\'o}bal
  Sif{\'o}n\altaffilmark{2},  Michael L.  Brown\altaffilmark{7}, Anais
  Rassat\altaffilmark{8}} \email{Alina.A.Kiessling@jpl.nasa.gov}

\altaffiltext{1}{Jet  Propulsion Laboratory,  California Institute  of
  Technology,  4800  Oak  Grove   Drive,  Pasadena,  CA,  91109,  USA}
\altaffiltext{2}{Leiden Observatory,  Leiden University, PO  Box 9513,
  2300 RA, Leiden, the Netherlands}
\altaffiltext{3}{Department  of  Physics   and  Astronomy,  University
  College London, Gower Street, London WC1E 6BT, UK}
\altaffiltext{4}{Mullard Space Science  Laboratory, University College
  London, Holmbury St Mary, Dorking, Surrey RH5 6NT, UK}
\altaffiltext{5}{McWilliams  Center   for  Cosmology,   Department  of
  Physics, Carnegie Mellon University, Pittsburgh, PA 15213, USA}
\altaffiltext{6}{Astronomisches   Recheninstitut,    Zentrum   f{\"u}r
  Astronomie der Universit{\"a}t  Heidelberg, Philosophenweg 12, 69120
  Heidelberg, Germany}
\altaffiltext{7}{Jodrell  Bank  Centre  for  Astrophysics,  School  of
  Physics  and  Astronomy,  University  of  Manchester,  Oxford  Road,
  Manchester M13 9PL, UK}
\altaffiltext{8}{Laboratoire    d'astrophysique     (LASTRO),    Ecole
  Polytechnique F{\'e}d{\'e}rale  de Lausanne (EPFL),  Observatoire de
  Sauverny, CH-1290 Versoix, Switzerland}

\begin{abstract}
The shapes of  galaxies are not randomly oriented on  the sky.  During
the galaxy formation  and evolution process, environment  has a strong
influence, as tidal gravitational  fields in the large-scale structure
tend to  align nearby galaxies.   Additionally, events such  as galaxy
mergers affect the relative alignments  of both the shapes and angular
momenta  of  galaxies  throughout their  history.   These  ``intrinsic
galaxy  alignments''  are  known  to   exist,  but  are  still  poorly
understood.  This review will offer  a pedagogical introduction to the
current theories that describe  intrinsic galaxy alignments, including
the  apparent difference  in  intrinsic alignment  between early-  and
late-type galaxies and the latest  efforts to model them analytically.
It  will  then describe  the  ongoing  efforts to  simulate  intrinsic
alignments using  both $N$-body and hydrodynamic  simulations.  Due to
the relative youth  of this field, there  is still much to  be done to
understand intrinsic galaxy alignments  and this review summarises the
current state of the field, providing a solid basis for future work.
\end{abstract}


\tableofcontents

\section{Introduction}
\label{sec:introduction}

The Universe  is filled  with galaxies,  enormous collections  of gas,
dust  and billions  of stars  all held  together by  gravity. Galaxies
reside in the centre of dark matter  haloes that contain up to 95\% of
the   mass    of   the   galaxy   \citep[see    e.g.][and   references
  therein]{CAW+15}.   The current  picture of  structure formation  is
hierarchical  with galaxies  forming through  anisotropic collapse  of
localised overdensities.  Larger dark matter haloes are then formed by
steady mass accretion  of surrounding dark matter  and through mergers
with other haloes (including the galaxies they contain).  These larger
dark matter haloes  contain many tens or hundreds of  galaxies and are
known    as    \emph{groups}   ($\mathtt{\sim}    10^{13}\msun$)    or
\emph{clusters}   ($\mathtt{\gtrsim}10^{14}\msun$),   which  are   the
largest  gravitationally   bound  structures  in  the   Universe  (see
\citealp{D03} for an introduction to structure formation).

Given that the Universe is  homogeneous and isotropic on large scales,
it was commonly assumed for some time that the observed galaxies would
have random  orientations if  a large  enough sample  were considered.
However, it is now known that local physical processes, in addition to
initial conditions, have a  strong influence on the \textit{alignment}
of galaxies  with respect  to their  surrounding environment  and that
these    alignments    can    be    strongly    anisotropic.     These
\textit{intrinsic}  alignments  are  not   restricted  simply  to  the
orientation of  the galaxy, but  also to the rotational  properties of
galaxies.  The  physical processes acting  on galaxies depend  on many
factors  including   whether  the   galaxy  is  early-   or  late-type
(elliptical or spiral disc), red or blue (ellipticals with old stellar
populations  or  spiral discs  with  active  star formation).   It  is
reasonable to assume that these factors may also have an impact on how
the galaxy is aligned with surrounding structures (for a more complete
introduction to galaxy formation and evolution, see \citealp{MBW10}).

Observations are  largely focused  on investigating the  alignments of
galaxies, as opposed  to dark matter haloes, since  these are directly
observed by telescopes\footnote{However, there  are methods to measure
  dark   matter   halo   ellipticities  through   observations,   e.g.
  galaxy-galaxy                                                lensing
  \citep[e.g.][]{NR00,HYG04,MHB+06,PHH+07,UHS+12,SHH+15}}.    However,
cosmic  structures grow  much  larger than  galactic  scales and  this
growth depends sensitively  on the properties of dark  matter and dark
energy  in an  expanding  Universe.  Dark  matter  particles are  only
subject to gravity and can  be approximated as experiencing no elastic
collisions, which  would give rise  to pressure or viscous  forces, in
strong contrast  to the sub-dominant baryonic  component.  Dark energy
influences  the growth  of  structures through  the  evolution of  the
background density  with time  and the relation  between time  and the
scale factor  of the Universe.   Computationally, it is far  easier to
determine the alignments of dark matter haloes in $N$-body dark matter
simulations  that include  dark  energy, since  simulating baryons  is
still   computationally   expensive  and   it   is   unclear  how   to
self-consistently  model the  dominant physical  processes numerically
over the large  dynamic range involved in the problem  at hand.  There
is a  wealth of literature  looking at  the alignments of  dark matter
haloes with respect to substructures, other haloes and even the cosmic
web.  Even though the mass  resolution of early simulations restricted
the studies to  cluster-sized haloes, more recent  works have extended
this to  include group- and  galaxy-sized haloes in addition  to using
semi-analytic models to add some of the key properties of galaxies and
hydrodynamics to add baryonic processes to the simulations.

Understanding  the  alignments of  galaxies,  groups  and clusters  is
important from a formation and  evolution perspective.  However, it is
essential to galaxy surveys that include weak gravitational lensing as
a key  cosmological probe\footnote{e.g. the Kilo  Degree Survey, KiDS:
  http://kids.strw.leidenuniv.nl;   the  Dark   Energy  Survey,   DES:
  http://www.darkenergysurvey.org; the Hyper  Suprime-Cam Survey, HSC:
  http://www.naoj.org/Projects/HSC;  Euclid:  http://www.euclid-ec.org
  and http://sci.esa.int/euclid; the  Large Synoptic Survey Telescope,
  LSST: http://www.lsst.org/lsst;  and the Wide Field  InfraRed Suvery
  Telescope, WFIRST: http://wfirst.gsfc.nasa.gov} for a very different
reason.  Weak gravitational lensing  exploits the correlations between
the shapes of distant galaxy images,  which have been distorted by the
gravitational   deflection  of   light  by   the  intervening   matter
distribution  (see   \citealp{BS01}  for  more  information   on  weak
gravitational lensing).  This distortion  or \textit{shear} signal can
be boosted  by the  intrinsic alignments of  galaxies which  mimic the
gravitational shear, or diminished by intrinsic alignments of galaxies
that  are anti-correlated  with  the gravitational  shear signal.   If
ignored, these can  bias the inference on cosmology  and represent the
largest  astrophysical  systematic  for  upcoming  weak  gravitational
lensing  surveys  \citep[see][]{paper3}.   In order  to  mitigate  the
signal from  these alignments without  losing a significant  amount of
cosmological  information,  it  is  first  necessary  to  formulate  a
reasonably  accurate model  of  the alignments  \citep[see][]{paper3}.
Only then will  it be possible to accurately quantify  the effects and
mitigate the bias that they induce .

This review  is part of a  topical volume on galaxy  alignments and it
provides an introduction  to the theory, modelling  and simulations of
alignments of structures within the  Universe.  Also in this volume is
a basic  overview of the  galaxy alignments \citep{paper1} and  a more
technical review on observational results, the impact on cosmology and
mitigation techniques \citep{paper3}.

This    review     first    defines    alignment     observables    in
\Cref{sec:Observables},  to  give  context  to  the  alignments  being
modelled   or   simulated   throughout.   The   large-,   small-   and
intermediate-scale  theories  and  models   for  alignments  are  then
reviewed in \Cref{sec:Theory}. \Cref{sec:Nbody} summarises the results
of alignments in $N$-body  simulations, while \Cref{sec:Hydro} reviews
results   from   hydrodynamic  simulations.    \Cref{sec:SemiAnalytic}
introduces semi-analytic modelling  and this is followed  by a roadmap
or  wish   list  for  future  investigations   into  galaxy  intrinsic
alignments,     \Cref{sec:Roadmap},    and     final    remarks     in
\Cref{sec:Conclusions}.

\section{Observables}
\label{sec:Observables}

In this section,  the observable quantities that are  measured in real
data and  predicted theoretically  are defined.  These  include galaxy
shapes (\Cref{subsec:shapes}), the relative angles of interest between
the  astronomical  structures   (\Cref{sec:Alignments}),  and  2-point
correlation functions (\Cref{subsec:2point}).

\subsection{Galaxy shapes}\label{subsec:shapes}

When  observing the  sky,  it  is not  possible  to  observe the  full
three-dimensional  shapes  of  galaxies.    The  shapes  measured  are
inherently two-dimensional projections of the three-dimensional shape.
While  galaxies  and  dark  matter  haloes  do  not  in  general  have
elliptical isophotes or isodensity contours,  it is common to describe
their shapes in terms of the  effective ellipticity at some radius (or
averaged   over  a   range  of   radii,  as   in  weak   lensing;  see
\citealp{paper3}).  The observed ellipticity, $\epsilon$, can be split
into two components
\begin{equation}
  \epsilon \simeq \epsilon^{\rm s} + \gamma,
  \label{eq:epsilon}
\end{equation}
where $\epsilon^{\rm  s}$ is  the \textit{intrinsic  ellipticity} that
corresponds to the true shape  of the light distribution, and $\gamma$
is  the \textit{gravitational  shear}\footnote{The term  gravitational
  shear  could be  misleading  here since  the intrinsic  ellipticity,
  $\epsilon^{\rm  S}$, is  also  influenced  by (tidal)  gravitational
  effects as shown  in this section.  However,  the term gravitational
  shear for the effects of gravitational lensing has been in use for a
  long time  and it would  be unwise  to adopt a  different convention
  here.  Throughout,  the term gravitational shear  is synonymous with
  gravitational lensing.},  a distortion to the  galaxy image produced
by  intervening matter  along  the  line of  sight  to  the galaxy  in
question (see  \citealp{paper1} for a discussion  of the approximation
in  this  equation).   There   is  a  fundamental  difference  between
$\epsilon^{\rm s}$  and $\gamma$ in that  $\epsilon^{\rm s}$ describes
the deviation of the true shape of the galaxy itself from circularity,
while $\gamma$ is a distortion  to the \textit{observed} galaxy shape,
and will depend on the matter distribution through each line of sight.

Ellipticities are tensorial quantities that have, in complex notation,
two  components (note  that \autoref{eq:epsilon}  is also  an equation
between complex numbers).  It is standard to define a fixed coordinate
frame on the sky, with a position angle, $\varphi$, defining the angle
of the semi-major axis of the ellipse from one axis of that coordinate
frame.  If the  total magnitude of the ellipticity  is $|\epsilon|$ as
defined in terms of the semi-minor  to semi-major axis ratio $q$ (see,
e.g., \citealt{BJ02}  for many  common ellipticity  definitions), then
the    two   components    of   ellipticity    in   this    coordinate
system\footnote{If a two-component ellipticity seems unfamiliar, it is
  worth considering  that the standard geometric  representation of an
  ellipse using an  axis ratio and a position angle  also requires two
  numbers.}  can  be defined  as $\epsilon_1=|\epsilon|\cos{2\varphi}$
and    $\epsilon_2=|\epsilon|\sin{2\varphi}$,    with   the    complex
ellipticity  denoted   $\epsilon=\epsilon_1+\mathrm{i}\epsilon_2$,  or
equivalently $\epsilon=\left|\epsilon\right|\exp(2\mathrm{i}\varphi)$.
The factor of 2  by which the phase angle is  multiplied takes care of
the internal spin-2 symmetry of the ellipticity field, which is mapped
onto itself after a rotation of the coordinates by $\pi$.

The  notion of  spatial  ellipticity correlations  and  their link  to
correlation functions  is illustrated  here with the  following model:
The ellipticity  of every galaxy can  be seen as being  drawn randomly
from some  underlying distribution,  with a  non-vanishing correlation
function describing  the dependence of the  random processes assigning
ellipticities  to  two  neighbouring   galaxies.   If  $\epsilon$  and
$\epsilon'$   are  the   ellipticities   of   galaxies  at   positions
$\boldsymbol{\alpha}$ and $\boldsymbol{\alpha}^{\prime}$ respectively,
their   ellipticities  are   drawn   from   a  multivariate   Gaussian
distribution $p(\epsilon^\prime,\epsilon)$,  which can be viewed  as a
conditional distribution $p(\epsilon^\prime|\epsilon)$,  such that the
outcome $\epsilon^\prime$  depends on the  value of $\epsilon$.   As a
multivariate distribution $p(\epsilon,\epsilon^\prime)$,
\begin{equation}
\label{eq:epsgaussian}
p(\epsilon,\epsilon^\prime)                                          =
\frac{1}{\sqrt{(2\pi)^4\mathrm{det}(C_\epsilon)}}           \exp\left[
-\frac{1}{2} \left(
\begin{array}{c}
\epsilon\\
\epsilon^\prime
\end{array}
\right)^+
C_\epsilon^{-1}
\left(
\begin{array}{c}
\epsilon\\
\epsilon^\prime
\end{array}
\right)
\right],
\end{equation}
this property is encoded in the covariance matrix $C_\epsilon$,
\begin{equation}
C_\epsilon =
\left(
\begin{array}{cc}
\langle \epsilon\epsilon^* \rangle & \langle \epsilon\epsilon^{\prime *} \rangle \\
\langle \epsilon^*\epsilon^\prime \rangle & \langle \epsilon^\prime\epsilon^{\prime*} \rangle
\end{array}
\right).
\end{equation}
Note  that,  since $\epsilon$  is  a  complex number,  the  covariance
$C_\epsilon$  is Hermitian,  with  the asterisk  denoting the  complex
conjugate  and the  plus  sign in  \Cref{eq:epsgaussian} denoting  the
Hermitian conjugate.   The two  variances $\langle|\epsilon|^2\rangle$
and  $\langle|{\epsilon^\prime}|^2\rangle$  are equal  in  homogeneous
random  fields,  because  the  fluctuation  properties  are  identical
everywhere.   The off-diagonal  element  is  the correlation  function
$\xi_\epsilon =  \langle\epsilon\epsilon^{\prime *}\rangle$,  which is
invariant  under  translation  in homogeneous  fields.   It  generally
decreases  with   increasing  distance   due  to   the  Cauchy-Schwarz
inequality,   $\xi_\epsilon   \leq  \langle|\epsilon|^2\rangle$,   and
describes how rapidly the fluctuating  field loses memory of its value
$\epsilon$ at  $\boldsymbol{\alpha}$ when  increasing the  distance to
$\boldsymbol{\alpha}^\prime$. If  the random  field is  isotropic, the
random field's  correlation properties do not  change under rotations,
and    consequently   $\xi$    only    depends    on   the    distance
$\vartheta\equiv\left|\boldsymbol{\alpha}-\boldsymbol{\alpha}^\prime\right|$.
The averaging brackets $\langle\ldots\rangle$ denote ensemble averages
over  statistically  equivalent  realisations   of  the  random  field
$\epsilon(\boldsymbol{\alpha})$.

\subsection{Alignment angles and types}
\label{sec:Alignments}

\begin{figure}[t]
\begin{center}
\begin{tabular}{cc}
\resizebox{0.3\hsize}{!}{\includegraphics{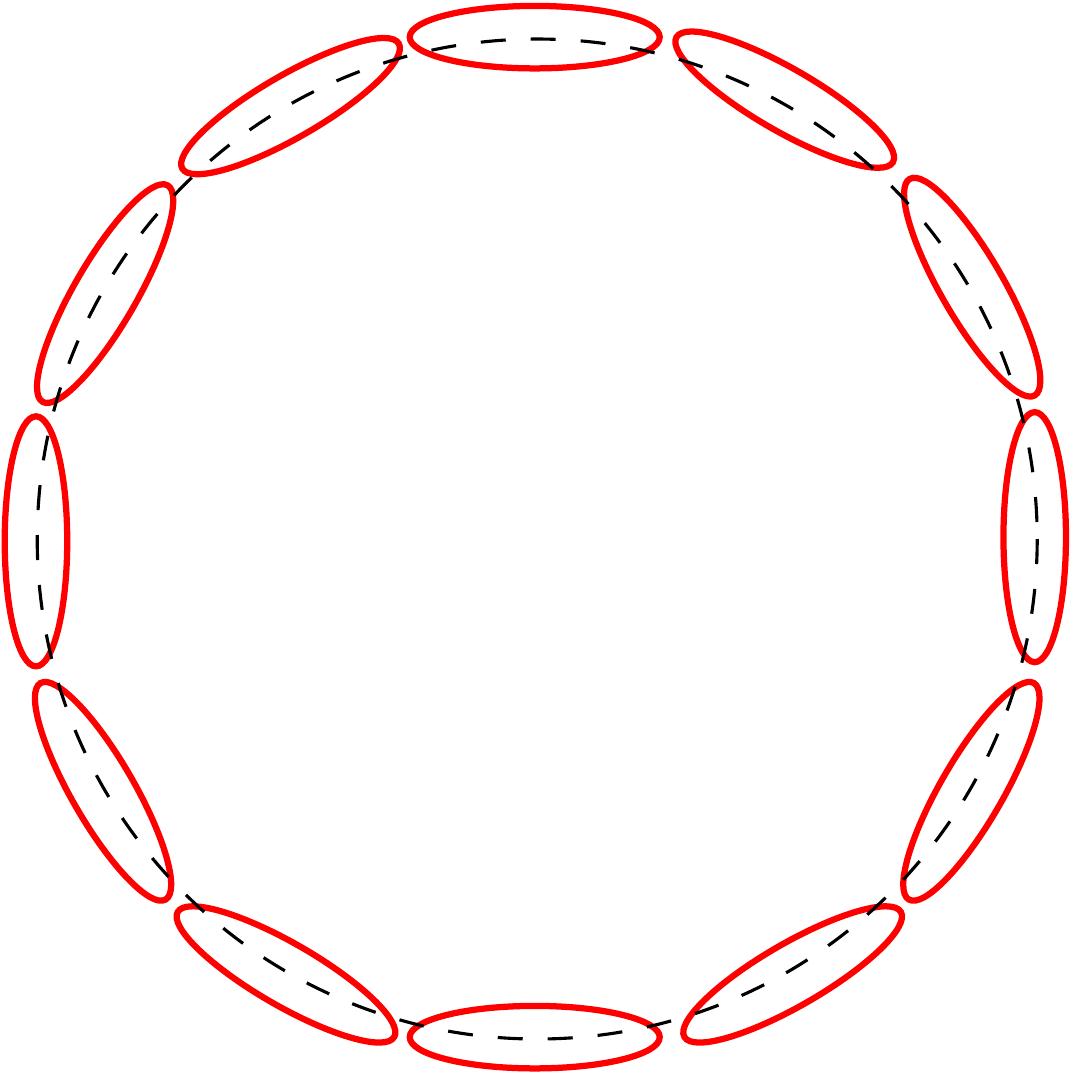}} & \resizebox{0.3\hsize}{!}{\includegraphics{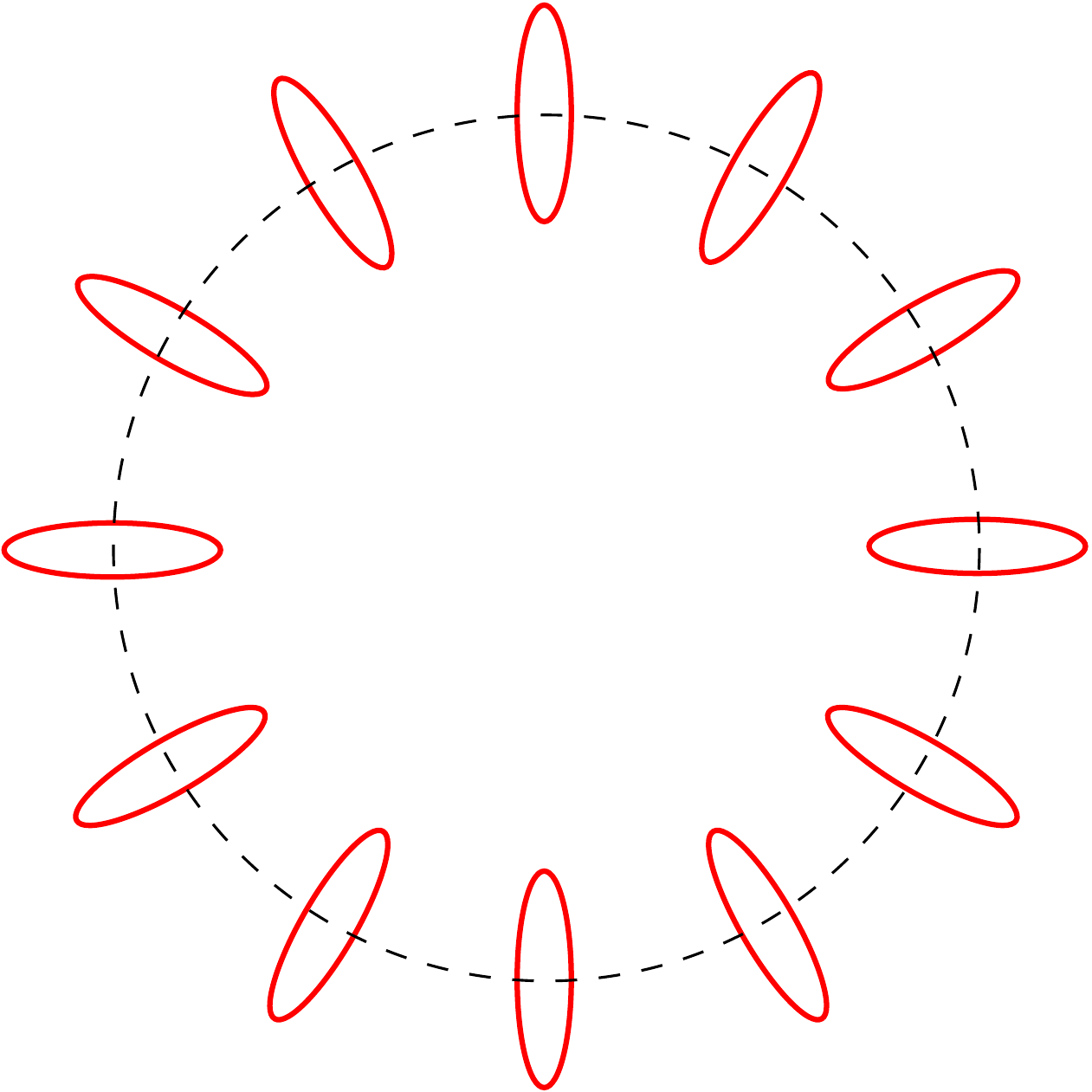}} \\
\resizebox{0.3\hsize}{!}{\includegraphics{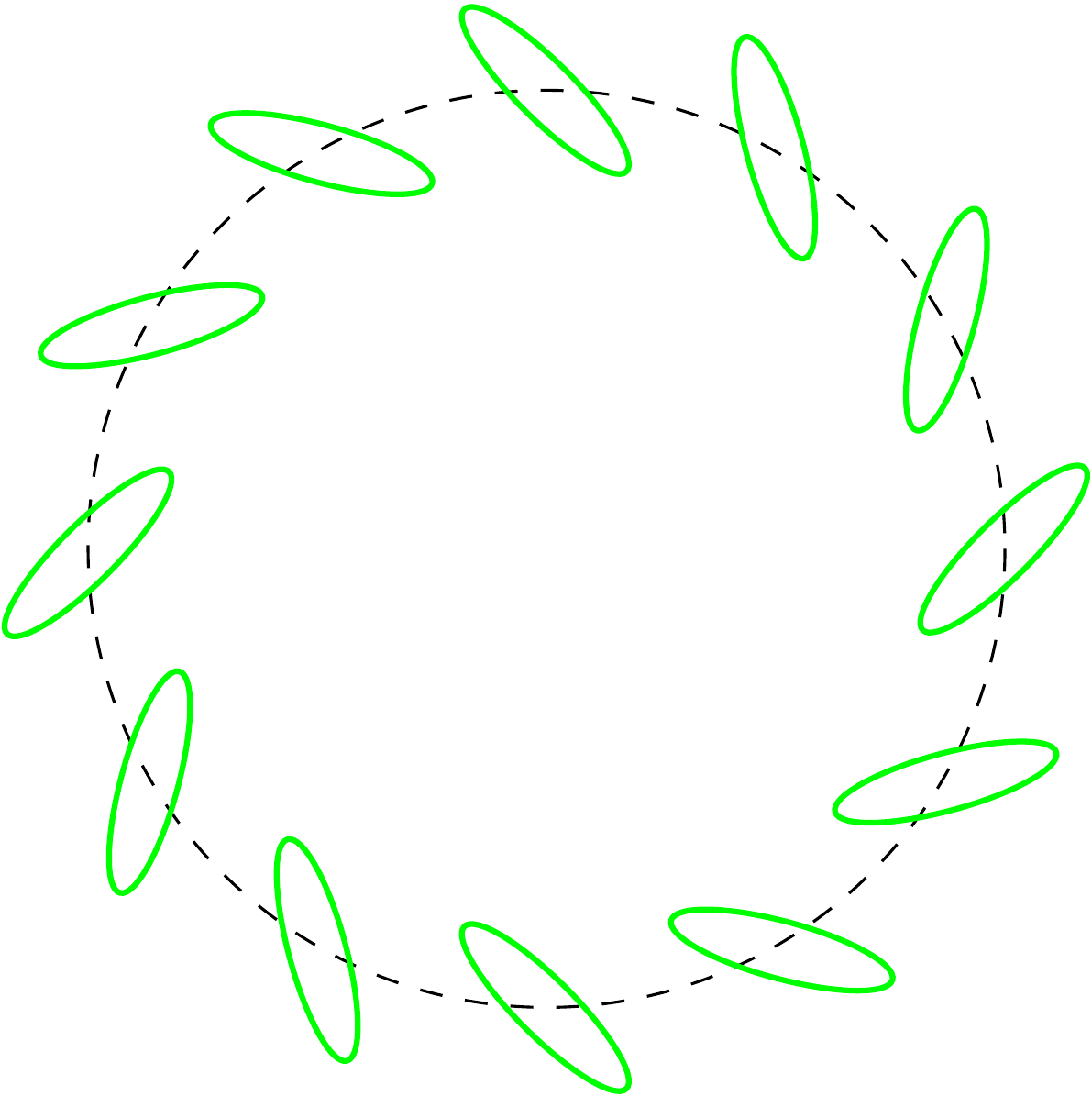}} & \resizebox{0.3\hsize}{!}{\includegraphics{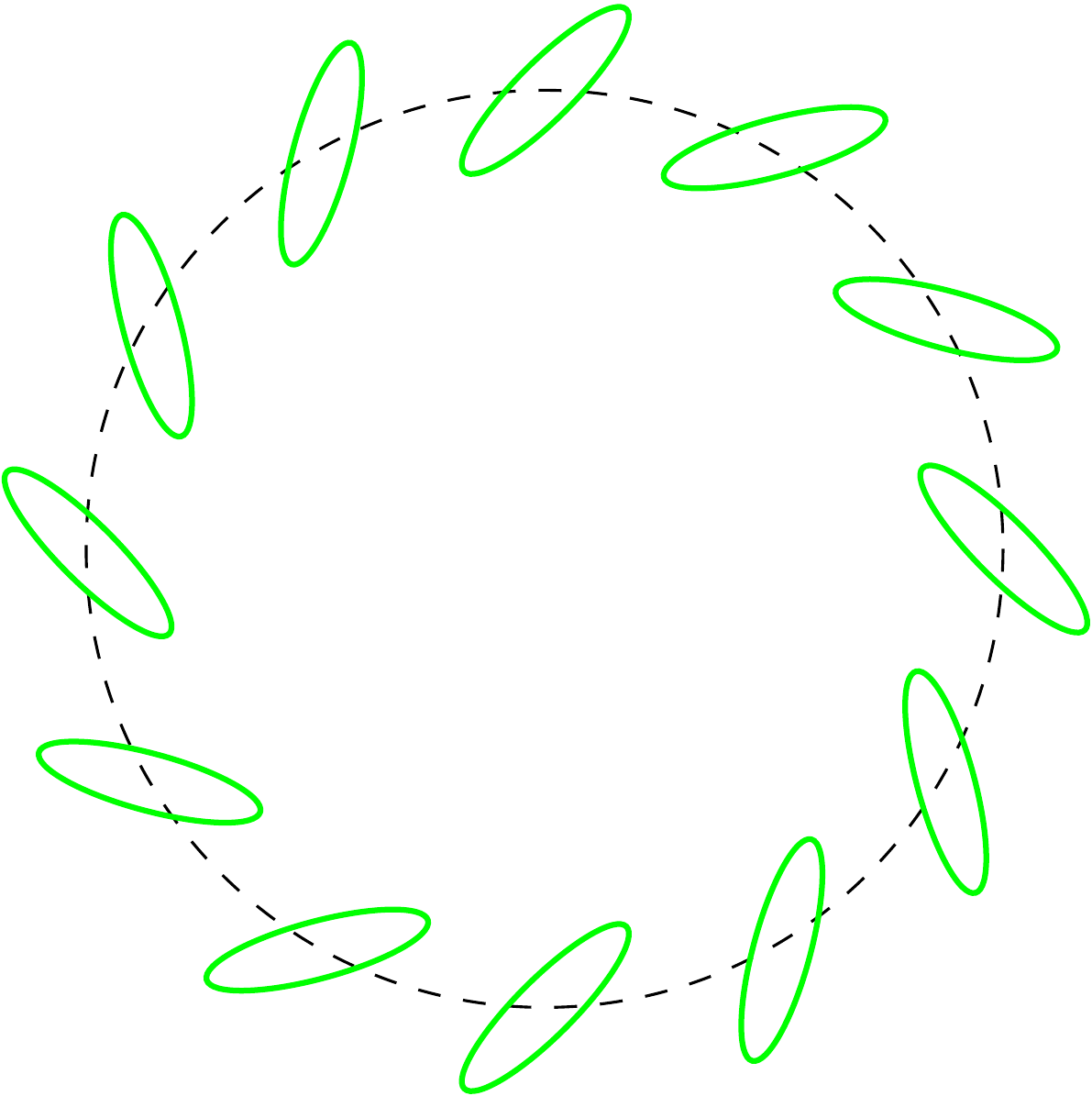}}
\end{tabular}
\caption{Parity-even  $E$-modes  (top  row) and  the  $B$-modes
  (bottom row)  of the ellipticity  field. Measured with respect  to a
  radial  line   from  the  centre   of  the  pattern,   one  observes
  $\epsilon_+>0;   \epsilon_\times=0$   (top   left),   $\epsilon_+<0;
  \epsilon_\times=0$  (top  right), $\epsilon_+=0;  \epsilon_\times<0$
  (bottom left), and $\epsilon_+=0; \epsilon_\times>0$ (bottom right),
  in the convention given by \Cref{eq:epsilon_shape_shape}.}
\label{f:ellipticity_parity_eigenmodes}
\end{center}
\end{figure}

There are two  common alignments of ellipticities  and position angles
that are considered theoretically or observationally: the alignment of
the shapes of two objects, or the  alignment of the shape of an object
with the  position of another object.   In both cases it  is common to
relate any ellipticity to a reference axis such as the line connecting
a pair  of objects,  which makes this  measurement independent  of the
coordinate system  used. The  two ellipticity  components can  then be
defined as
\begin{equation}
  \label{eq:epsilon_shape_shape}
\begin{aligned}
 \epsilon_+            &=            -            \lvert\epsilon\rvert
 \cos\left[2\left(\varphi-\varphi_0\right)\right]\;;
 \\       \epsilon_\times        &=       -       \lvert\epsilon\rvert
 \sin\left[2\left(\varphi-\varphi_0\right)\right]\;,
\end{aligned}
\end{equation}
where $|\epsilon|$ is  the absolute value of  the complex ellipticity,
$\varphi$  the polar  angle of  the ellipticity,  and $\varphi_0$  the
polar angle of  the reference axis (both polar angles  are measured in
the  same coordinate  system). $\epsilon_+$  measures the  ellipticity
component at 0$^\circ$  and 90$^\circ$ from the  reference axis, while
$\epsilon_\times$ measures the ellipticity component at 45$^\circ$ and
$135^{\circ}$ from it.  In  this particular convention, $\epsilon_+>0$
represents tangential alignment,  and $\epsilon_+<0$ radial alignment,
with respect  to a line  connecting the  galaxy to a  reference point,
e.g.   the  centre  of  a  galaxy cluster.   This  is  illustrated  in
\Cref{f:ellipticity_parity_eigenmodes}.  There  are a variety  of sign
conventions in  the literature.  Since gravitational  lensing tends to
cause  tangential alignment  and  intrinsic  alignments are  primarily
radial,  some  studies  will  define the  signs  such  that  intrinsic
alignments are negative and others  such that they are positive. Works
that focus on intrinsic galaxy alignments  tend to use the latter sign
convention  to  yield  a  positive  signal  for  the  expected  radial
alignments, e.g.   in \Cref{f:ModelComparison} this  latter convention
was used.

\begin{figure}[t]
  \begin{center}
    \includegraphics[width=0.5\hsize]{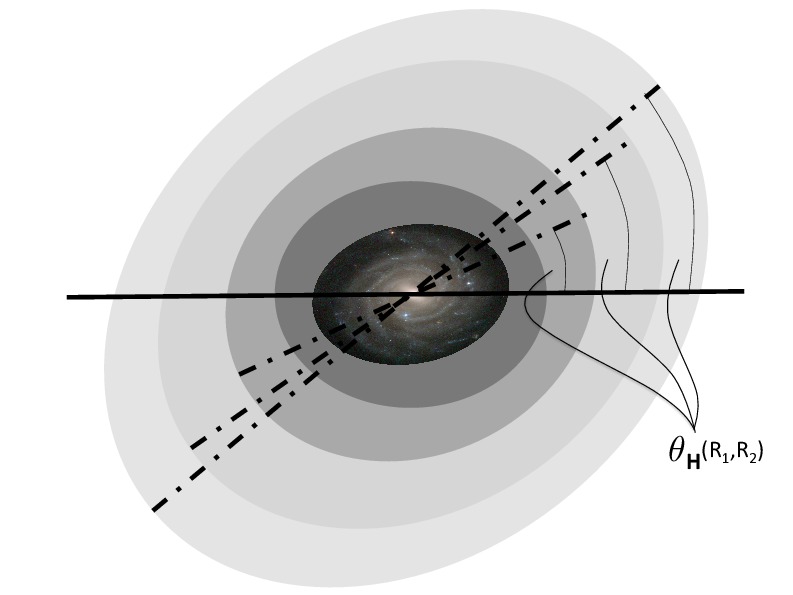}
    \caption{ (Mis)alignment of the dark  matter halo as a function of
      radius ($\rm  R_1,R_2$).  The long, horizontal  solid line shows
      the major  axis of the  innermost radial bin (dark  grey); other
      dashed  lines show  major axes  measured at  larger radii.   See
      \Cref{t:IAObservables}     for     a    complete     list     of
      alignments. \textit{Galaxy image credit: ESA/Hubble \& NASA}.}
    \label{fig:sketch_intra}
  \end{center}
\end{figure}

\begin{figure*}[t]
  \begin{subfigure}{0.48\textwidth}
    \includegraphics[width=3.2in]{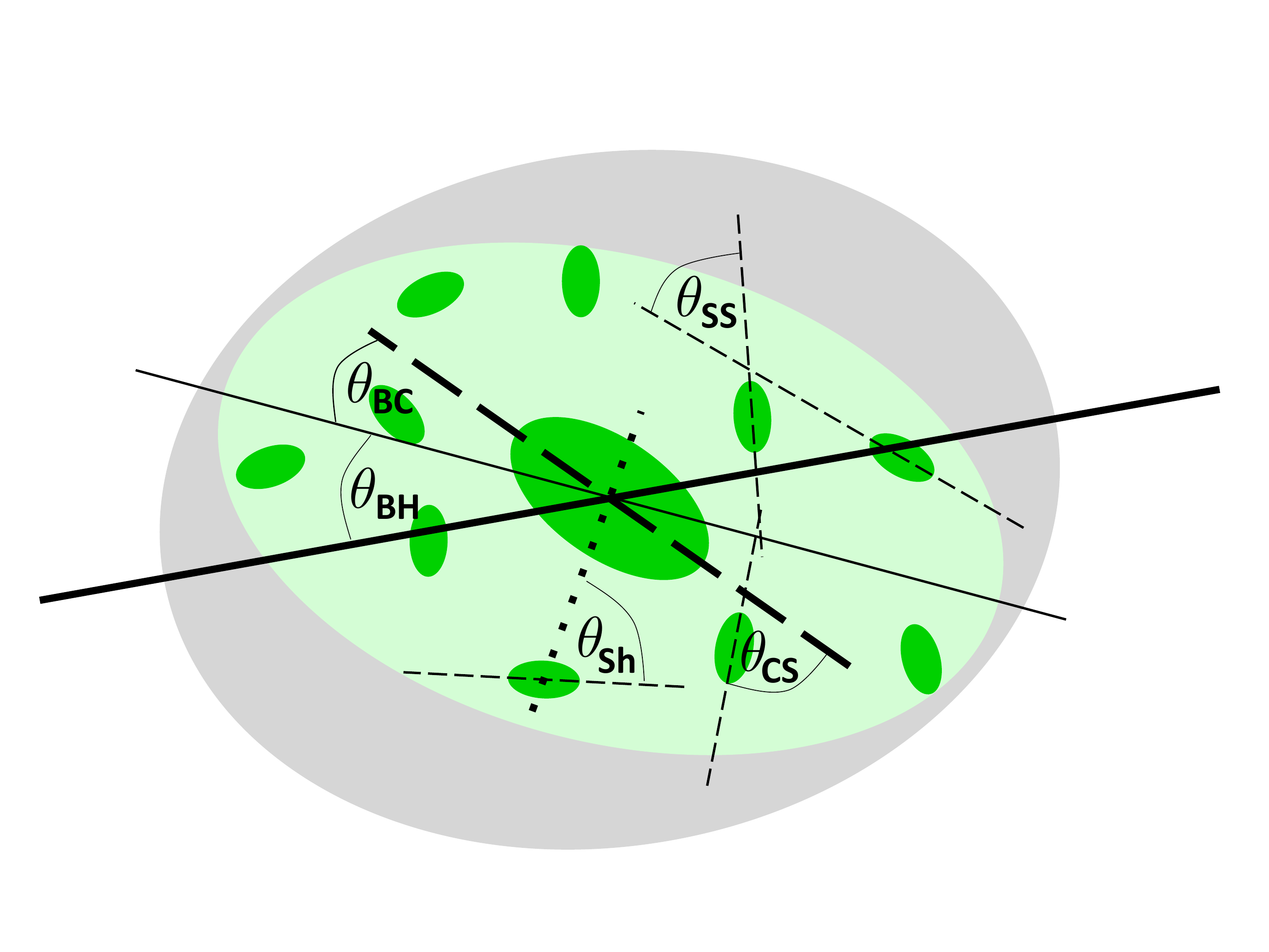}
    \caption{one-halo alignments.}
    \label{fig:sketch_satellites}
  \end{subfigure}
  \begin{subfigure}{0.48\textwidth}
    \includegraphics[width=3.2in]{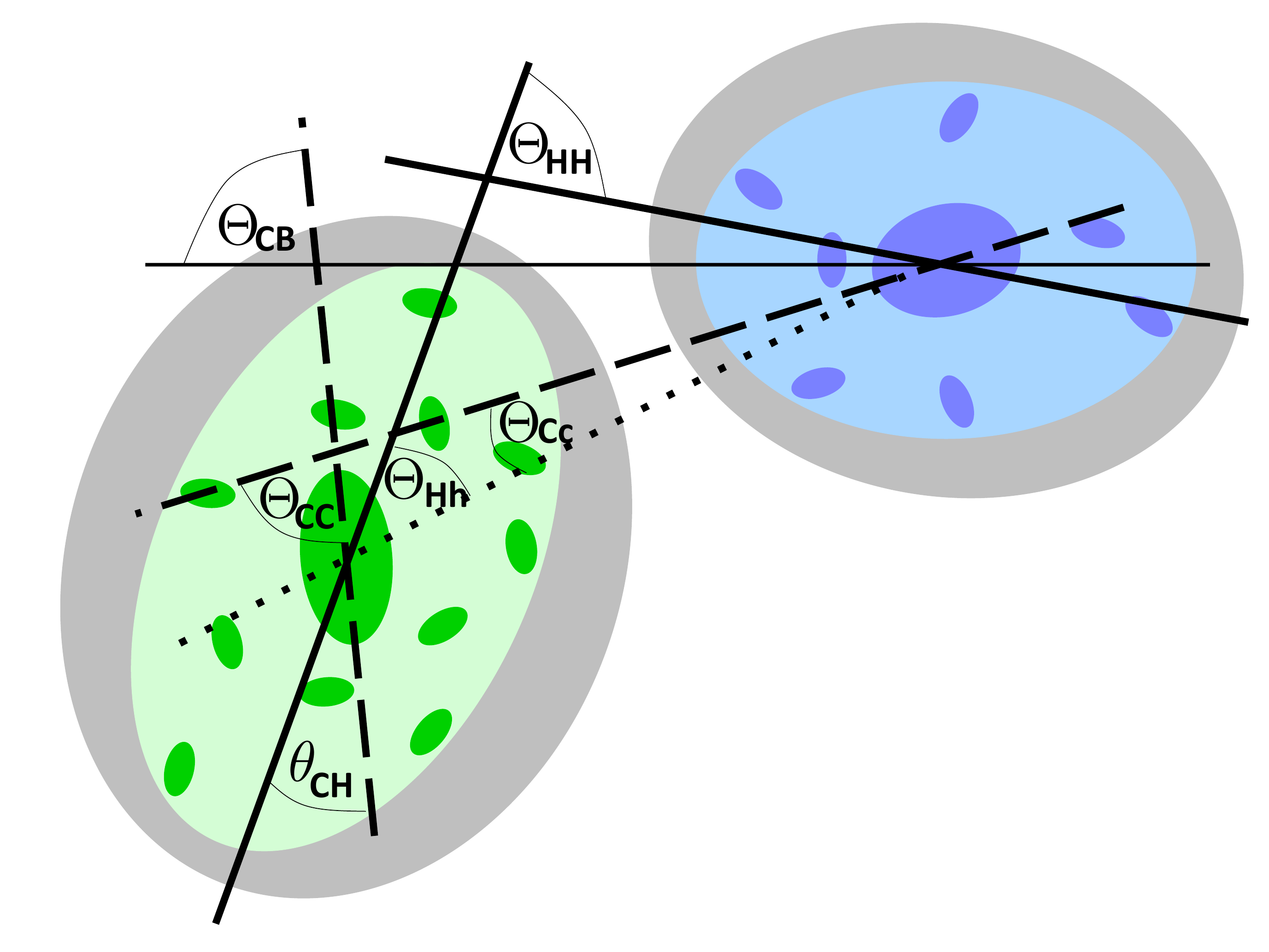}
    \caption{two-halo alignments.}
    \label{fig:sketch_2halo}
  \end{subfigure}
  \caption{ (Mis)alignments,  $\theta$, in one- and  $\Theta$, in two-
    haloes.   \textit{Both:} Thick  solid  and dashed  lines show  the
    semi-major axes of  the dark matter haloes (grey  ellipses; H) and
    central galaxies  (larger dark green  or blue ellipses;  C).  Thin
    solid lines show the semi-major axes of the satellite distribution
    (light green or blue ellipses;  B).  \textit{Left:} One-halo - The
    thin  dashed lines  show  the semi-major  axes  of the  satellites
    (small dark green  ellipses; S).  The thick dotted  line shows the
    vector connecting the satellite and  the centre of the dark matter
    halo.  \textit{Right:} Two-halo - The dotted line shows the vector
    connecting two central galaxies.  See \Cref{t:IAObservables} for a
    complete list of alignments.  }
  \label{fig:sketch_halo}
\end{figure*}

\begin{figure*}[t]
  \begin{subfigure}{0.46\textwidth}
 \includegraphics[width=\hsize]{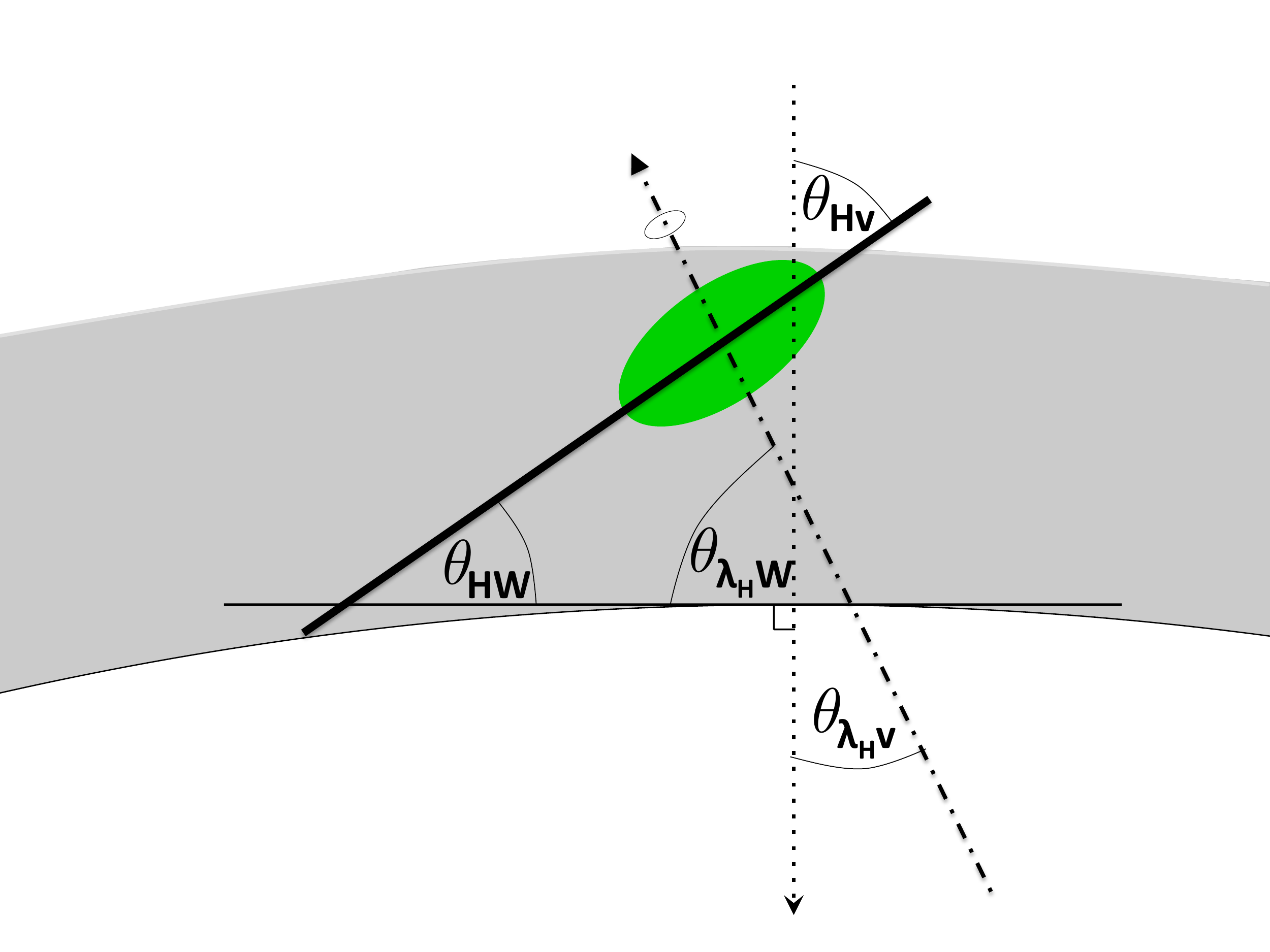}
 \caption{Sheet/Wall \& Void alignments.}
 \label{fig:sketch_void}
\end{subfigure}
  \begin{subfigure}{0.44\textwidth}
 \includegraphics[width=\hsize]{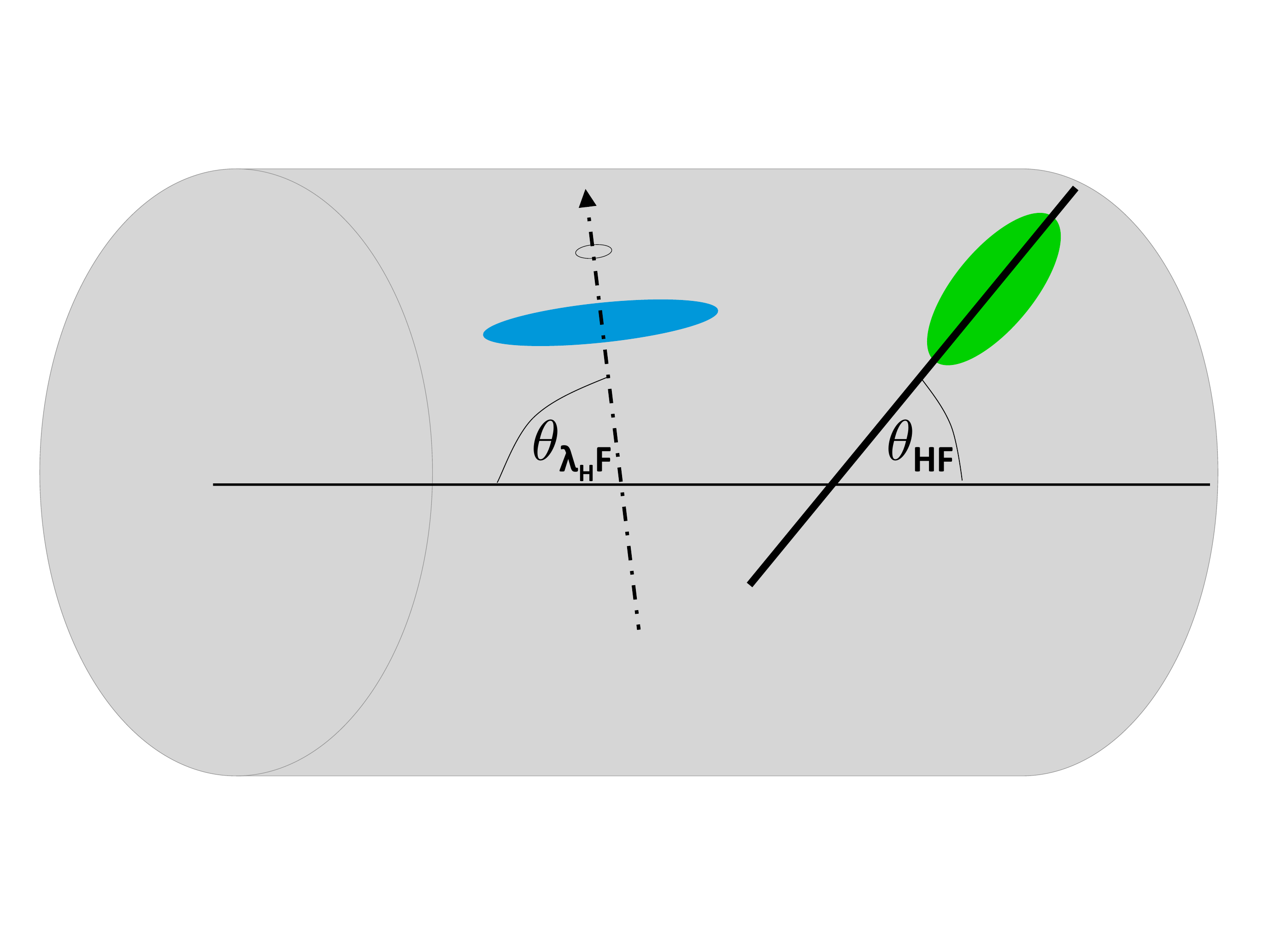}
 \caption{Filament alignments.}
 \label{fig:sketch_filament}
\end{subfigure}
\caption{(Mis)alignments, $\theta$, in a  sheet/wall (W), void (v) and
  filament (F).  The thick lines show  the semi-major axes of the dark
  matter  haloes (green  and blue  ellipses; H),  thin lines  show the
  plane of  the wall or  the shape of the  filament (grey; W,  F), the
  dot-dash line shows the spin  vector of the halo, $\lambda_{\rm H}$,
  and the  dotted line shows the  direction to the centre  of the void
  (v).  \textit{Left:} The sheet/wall is  located at the boundary of a
  spherical void in this figure. When  considering a void that is very
  large, locally the sheet/wall can be  treated as flat.  The plane of
  the sheet/wall is oriented left to right on the page (represented by
  the  thin, horizontal  line) and  normal to  the page  surface.  The
  vector normal  to the  plane of  the sheet/wall  is oriented  top to
  bottom on the page and is  represented by the dashed line (that also
  shows   the   direction  to   the   centre   of  the   void).    See
  \Cref{t:IAObservables} for a complete list of alignments.}
\label{fig:sketch_LSS}
\end{figure*}

In \Cref{fig:sketch_intra,fig:sketch_halo,fig:sketch_LSS}  a selection
of  the   possible  alignments   are  represented,   while  additional
alignments  not  explicitly  shown  in   the  figures  are  listed  in
\Cref{t:IAObservables} (found in \Cref{subsec:measurements}).  Not all
combinations of  alignments are  useful to measure,  so if  a possible
alignment  is absent,  it is  likely  not measured  in simulations  or
observations.   Alignments  of   the  spin  axis  are   not  shown  in
\Cref{fig:sketch_intra,fig:sketch_halo} or  in \Cref{t:IAObservables},
but they can be inferred easily from the figures.

Alignments  can be  considered  on  a wide  range  of  scales. In  the
following                                                  discussion,
\Cref{fig:sketch_intra,fig:sketch_halo,fig:sketch_LSS}             and
\Cref{t:IAObservables},  (mis)alignment angles  within a  single halo,
$\theta$, or between  two haloes, $\Theta$, are  given two sub-indices
for either shape (upper case  letter) or position (lower case letter),
for   a  central   galaxy   (C),  satellite   galaxy  (S),   satellite
distribution\footnote{Note  that  the  terms satellite  and  satellite
  distribution   are   interchangeable   with  subhalo   and   subhalo
  distribution  when measuring  dark  matter  alignments.}  (B),  dark
matter  halo (H)  or  a full  galaxy sample  (G;  which includes  both
satellite  and  central  galaxies),  wall (W),  filament  (F)  or  the
position of the centre  of a void (v).  The first  and second of these
indices  denote  the  object  in  the  first  and  second  columns  of
\Cref{t:IAObservables}, respectively.

The large-scale structure of the  Universe can be classified into four
separate  components called  cosmic  web elements  -  typically it  is
divided  in to  clusters,  filaments, sheets/walls\footnote{The  terms
  ``sheet''  and  ``wall''  are used  interchangeably  throughout  the
  literature  and  this  review.}   and  voids.   In  the  most  basic
classification  of the  large-scale  structure, sheets  are planes  of
structure that delineate  the edge of voids, which  are underdense and
often modelled as spherical or ellipsoidal.  Filaments are cylindrical
structures  (to first  approximation)  and clusters,  the nodes  where
filaments  meet,  are  modelled  as ellipsoidal.   A  more  quantative
description of web classification is given in \Cref{sec:web}.

\Cref{fig:sketch_intra} shows the internal alignments of a dark matter
halo at different radii,  $\theta_{\rm H}({\rm R_1,R_2})$, where ${\rm
  R}_i$  is the  radius  of the  halo shells.   In  this example,  the
orientation of  the inner  dark matter halo  is being  determined with
respect to the shells at larger  radii.  This 3D alignment is measured
in simulations  and the radius may  be an absolute radius  relative to
the  halo centre  or  an isodensity  radius; see  \Cref{sec:Internal}.
Dark matter  haloes exist  over a  huge range of  scales, but  for the
purposes  of this  work, the  scales that  are interesting  range from
subhaloes   ($\mathtt{\sim}  10^{10}\:\msun$)   through  to   clusters
($\mathtt{\gtrsim} 10^{14}\:\msun$). The dominant, most massive galaxy
near the  centre of a  galaxy sized dark matter  halo is known  as the
central galaxy.

It is common for a dark matter halo with a central galaxy to contain a
number (varying between  zero and tens) of  smaller satellite galaxies
(each   surrounded   by  a   dark   matter   subhalo)  as   shown   in
\Cref{fig:sketch_satellites}. These are less  massive than the central
galaxy and reside  in positions throughout the host  halo, rather than
at the centre.  In this figure  the smaller ellipses can be considered
either  a  satellite galaxy  \emph{or}  a  subhalo, depending  on  the
measurement  of interest.   On larger  scales, dark  matter group  and
cluster haloes contain many tens and hundreds of galaxy sized systems.
In  this  case,   the  objects  inside  the   larger  halo  (centrals,
satellites,  subhaloes  and  galaxy-sized   dark  matter  haloes)  are
all considered as substructure within the larger halo.

The shape of the satellite  distribution is defined by identifying the
positions  of the  satellites  within  the halo  and  fitting a  shape
(typically  an ellipsoid)  to their  distribution. In  this case,  the
shape of the individual satellites play  no role in defining the shape
of   the   satellite  distribution.    \Cref{fig:sketch_2halo}   shows
alignments within and between two  haloes.  The shape of the satellite
distribution is  sometimes used as a  proxy for the shape  of the dark
matter  halo  observationally.   Consequently, $\Theta_{\rm  Bb}$  and
$\Theta_{\rm BB}$  are considered equivalent to  $\Theta_{\rm Hh}$ and
$\Theta_{\rm  HH}$  (although  the  simulation  literature  does  find
differences in the shape and orientation of the satellite distribution
compared        with        the         dark        matter,        see
\Cref{sec:Satellite,sec:SHsatellite,sec:LHshape}).

\Cref{fig:sketch_LSS}  shows alignments  between  dark matter  haloes,
walls,  voids and  filaments  (the clusters  are  simply massive  dark
matter  haloes,  see  \Cref{fig:sketch_halo}).  Also  shown  in  these
figures is  the orientation of  the dark matter halo  angular momentum
vector, $\lambda_{\rm H}$,  with the cosmic web  elements. The angular
momentum vector denotes the axis around which the halo is rotating and
is discussed further in \Cref{sec:quadratic,sec:momentum}.

\subsection{Ellipticity 2-point correlations}\label{subsec:2point}

When  considering sets  of objects  for which  ellipticity correlation
functions  are   to  be  measured,  components   of  the  ellipticity,
$\epsilon_+$  and  $\epsilon_\times$,  are   typically  defined  in  a
coordinate    frame    aligned    with    the    separation    vector,
$\boldsymbol{\vartheta}$,  between the  two  galaxies.  This  involves
rotating the two  original components of ellipticity in  the sky frame
($\epsilon_1$ and $\epsilon_2$; see \autoref{eq:epsilon_shape_shape}).
The  correlation  functions  $\xi_+$  and $\xi_-$  of  these  two  new
ellipticity components are defined as
\begin{equation}\label{eq:epsilon_correlation}
\xi_\pm (\vartheta) = \langle \epsilon_+ \epsilon_+^\prime\rangle(\vartheta) \pm
\langle \epsilon_\times\epsilon_\times^\prime\rangle(\vartheta),
\end{equation}
while        the         third        correlation        $\left\langle
\epsilon_+\epsilon_\times^\prime    \right\rangle     (\vartheta)    =
\left\langle             \epsilon_\times\epsilon_+^\prime\right\rangle
(\vartheta)$ is  parity-odd and expected  to vanish due to  the parity
symmetry of the Universe.

Considering the variance of Fourier modes of the random field, defined
in  \Cref{subsec:shapes},   is  a  particularly  useful   concept  for
homogeneous random fields,  because in these cases,  the Fourier modes
with different  wave vectors are  uncorrelated, while the  variance of
equal wave  vectors is  related to  the power  spectrum.  Constructing
power  spectra  (originally  used   in  the  context  of  polarisation
correlations    in    the    cosmic    microwave    background;    see
\citealp{KKS97,S97})  of   the  projected  galaxy   ellipticity  field
$\epsilon(\boldsymbol{\alpha})$ yields two  parity eigen-modes for the
angular power spectrum as a function of multipole $\ell$,
\begin{equation}\label{eq:powerspectra}
C^\epsilon_{E,                        B}(\ell)                       =
\pi\int\vartheta\,\mathrm{d}\vartheta\:\left[\xi_+(\vartheta)J_0(\ell\vartheta)
  \pm \xi_-(\vartheta)J_4(\ell\vartheta)\right],
\end{equation}
with the  Bessel functions of  the first  kind, $J_0$ and  $J_4$.  The
physical interpretation of these  parity eigen-modes is illustrated in
\Cref{f:ellipticity_parity_eigenmodes}. 

Parity-odd $B$-modes are  a typical feature of  some alignment models,
in particular of  the quadratic alignment model  applicable for spiral
galaxies (see  \Cref{sec:quadratic} for  information on  the quadratic
alignment model).  $B$-modes are to lowest order not present in linear
alignment models, but can be generated by introducing weighting to the
ellipticity field, for instance related  to galaxy biasing or peculiar
motion.  $B$-mode generation  through weighting is also  well known in
higher-order  corrections  to  weak gravitational  lensing,  which  is
$B$-mode free  to lowest  order, but  corrections related  to geodesic
corrections or  to clustering can  evoke $B$-mode patterns,  which are
typically small,  amounting to a  signal of $\sim10^{-4}$  relative to
the $E$-modes on  small angular scales, where the  effect is strongest
\citep{CH02,BBV10,KH10}.

While  the formalism  has been  outlined for  angular correlations  of
fields  at two  different  positions  on the  sky,  it generalises  to
correlations of shapes  in three dimensions in  a straightforward way.
Commonly, the  correlations of the  aligning fields are  formulated in
three dimensions (using  physical separations $\boldsymbol{r}$ derived
using known redshifts),  though still using the  projected (2D) shape.
For  example,   $\xi_{g+}(\boldsymbol{r})$  defines   the  correlation
function  of  projected galaxy  ellipticities  with  the positions  of
galaxy overdensities as a  function of 3D separation $\boldsymbol{r}$.
However, note that  the assumption of isotropy is not  a very good one
in  this case,  since  it  is the  unobserved  3D  shapes that  should
correlate   with   an  equal   strength   to   galaxies  at   separate
$\boldsymbol{r}$, not the projected 2D shapes.  Moreover, at the stage
of   expressing  the   relation  between   redshifts  and   distances,
complications  such  as   redshift-space  distortions  enter,  causing
misestimates of radial distances.  Observationally, it is difficult to
model the impact of redshift  space distortions on 2-point correlation
functions,   particularly   on   scales   where   non-linear   density
perturbations are important (see, e.g.,  \citealt{K87} for a review of
redshift  space  distortions  in  general, or  \citealt{SMM14}  for  a
derivation of their lowest-order impact on intrinsic alignment 2-point
correlations).    Without  a   good  model,   the  three   dimensional
ellipticity  correlation functions  are  difficult  to interpret.   To
avoid  both  of  these complications  (anisotropy  and  redshift-space
distortions), a  two-dimensional ellipticity correlation  function for
galaxies as a function of their transverse separation on the sky $r_p$
is commonly calculated,
\begin{equation}
  w_{g+}(r_p) = \int_{-\Pi_{max}}^{+\Pi_{max}} \xi_{g+} (r_p,\Pi) {\rm
    d}\Pi,
\label{eq:wg+}
\end{equation}
determined by projecting  the equivalent three-dimensional correlation
function $\xi_{g+}=\langle\epsilon_+ g\rangle$ between ellipticity and
the galaxy density along the line-of-sight ($\Pi$ is the separation in
the  redshift direction).   This  projected  correlation function  can
similarly be found for other combinations of observables including the
ellipticity   components   $+,\times,\epsilon$,    $g$,   the   galaxy
overdensity with  respect to the mean,  $g=\rho_g/\bar{\rho}_g-1$, and
$\delta$,   the  matter   overdensity  with   respect  to   the  mean,
$\delta=\rho/\bar{\rho}-1$. The densities $\rho_g$  and $\rho$ are the
galaxy density and the matter density respectively.

Identifying intrinsic galaxy alignments  is particularly important for
upcoming weak lensing  surveys, because they mimic  the coherent shape
distortions resulting  from gravitational shear.  A  correlator of two
galaxy ellipticities  can be  taken directly  from the  correlation in
\Cref{eq:epsilon},
\eqa{
\label{eq:iadef}
\underbrace{\ba{{\epsilon} {\epsilon^{\prime}}}} &=
\underbrace{\ba{{\gamma} {\gamma^{\prime}}}} + \underbrace{\bigl\langle
  {\epsilon^{\rm s}} {\epsilon^{\prime\rm s}}} \bigr\rangle +
\underbrace{\bigl\langle {\gamma} {\epsilon^{\prime \rm s}}
  \bigr\rangle+\ba{{\epsilon^{\rm s}} {\gamma^{\prime}}}}{}\;.\\
\nonumber \mbox{observed} &
\hspace*{0.52cm} \mbox{GG} \hspace*{0.82cm} \mbox{II} \hspace*{1.62cm}
\mbox{GI} }  The left side  of the equation  is the correlator  of the
observed  ellipticities.   GG  is   the  gravitational  lensing  shear
correlation  (the  signal  that  is most  important  in  weak  lensing
analyses; see  \citealp{paper1} for a more  comprehensive introduction
to weak  gravitational lensing).   II is  the correlation  between the
intrinsic shapes  of two galaxies,  and GI is the  correlation between
the gravitational shear  of one galaxy and the intrinsic  shape of the
other  galaxy.  Only  one  of the  GI terms  is  non-zero because  any
gravitational shear  associated with a  galaxy closer to  the observer
can not  be correlated with  the intrinsic  shape of a  galaxy further
away from the  observer\footnote{This may not be true  in the presence
  of  photometric redshift  errors,  where the  relative positions  of
  galaxies along the line of sight  may be confused.}.  In addition to
shape correlations, intrinsic alignments  can cause cross correlations
between the shape and the local  density, which we refer to as $\delta
{\rm  I}$-correlations  in this  review  and  these  give rise  to  GI
correlations.

\section{Theory and modelling}
\label{sec:Theory}

The alignments of galaxies with cosmic structures is a phenomenon that
occurs on a variety of scales. In general these alignments are thought
to  be  sourced  through  tidal  interactions  of  galaxies  with  the
gravitational field  of larger structures.   Fundamentally, alignments
fall   into    three   regimes:    On   very   large    scales   above
$\mathtt{\sim}10\:\mathrm{Mpc}$, galaxies are tidally aligned with the
linearly       evolving       cosmic       large-scale       structure
(\Cref{sec:LargeScaleTheory}).          On        scales         below
$\mathtt{\sim}1\:\mathrm{Mpc}$,  galaxies  may align  themselves  with
their host  halo and this  particular alignment can be  described with
the halo model  (\Cref{sec:SmallScaleTheory}).  Intermediate scales of
a  few  Mpc  are  difficult  to grasp  due  to  non-linearly  evolving
structures and effects due to  clustering and a strong peculiar motion
contribution to  the galaxy redshifts;  how to model the  influence of
these  effects on  galaxy alignments  is a  topic of  current research
(\Cref{sec:IntermediateScales}).

There  are  two  theories  commonly  employed  to  explain  how  tidal
interactions determine the alignment and  hence the shape of a galaxy:
stellar  ellipsoids of  elliptical galaxies  may be  tidally distorted
(\Cref{sec:linear}), and  the orientation  of stellar discs  in spiral
galaxies may be determined by the angular momentum direction, which in
turn  follows from  tidal interactions  (\Cref{sec:quadratic}).  While
tidal interaction  processes on  large scales may  be be  described by
perturbation  theory  (\Cref{sec:tidal_alignments}),   the  shape  and
orientation  of the  stellar  distribution  inside a  halo  is a  more
complicated problem that requires  numerical simulations to solve (see
\Cref{sec:Nbody,sec:Hydro,sec:SemiAnalytic}).  It should be emphasised
that alignment  models, which mediate between  an aligning large-scale
field and  an observable galaxy  shape, may  be based on  fields other
than the tidal gravitational fields, for example vorticity.

\subsection{Large-scale alignments}
\label{sec:LargeScaleTheory}

\subsubsection{Tidal interactions of haloes}
\label{sec:tidal_alignments}

\begin{figure*}[t]
  \begin{subfigure}{0.48\textwidth}
 \begin{center}
 \includegraphics[width=0.7\hsize]{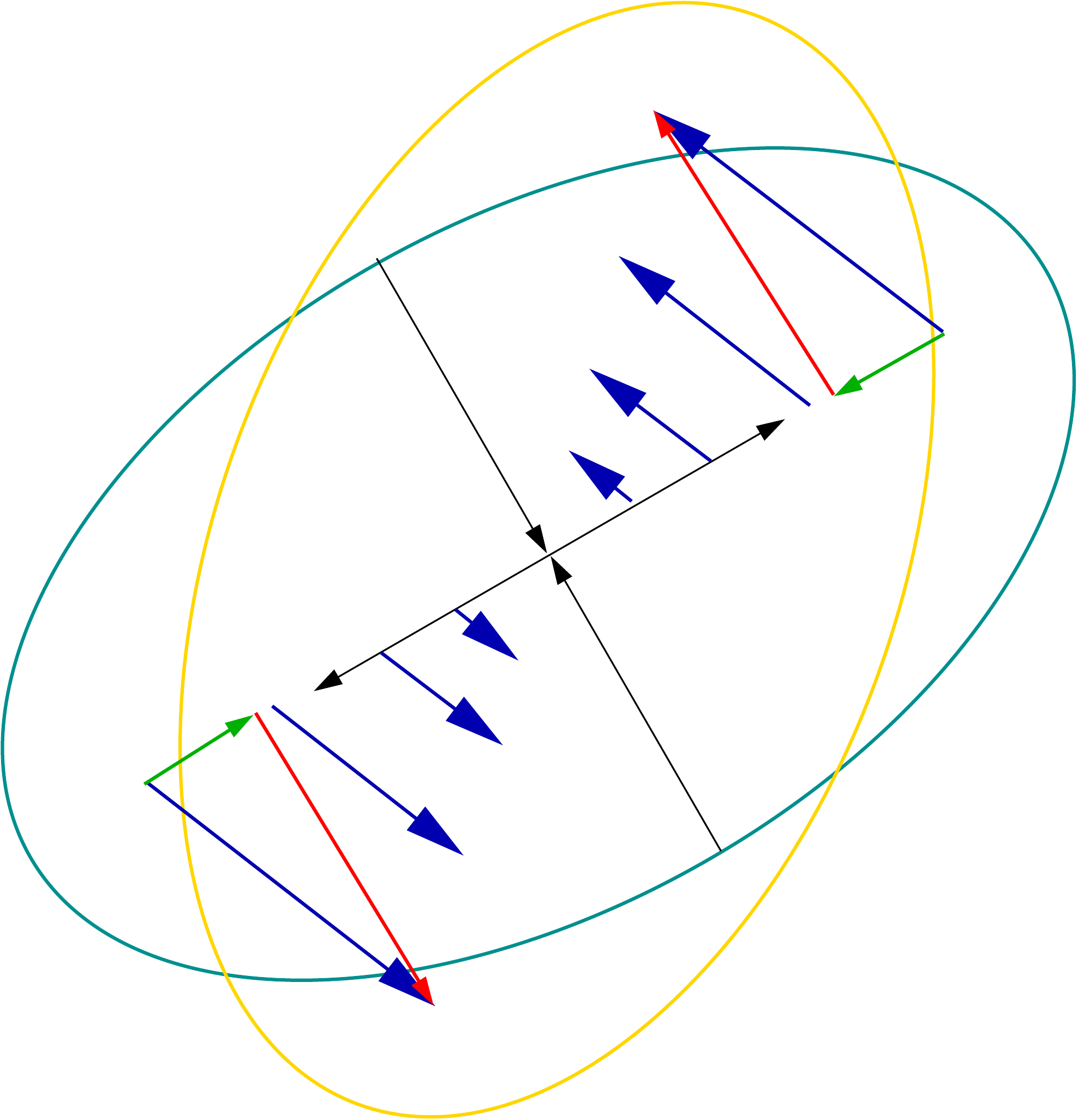}
 \caption{Tidal Torquing}
 \label{f:sketch_tidal_torquing}
 \end{center}
\end{subfigure}
\begin{subfigure}{0.48\textwidth}
\begin{center}
 \includegraphics[width=0.9\hsize]{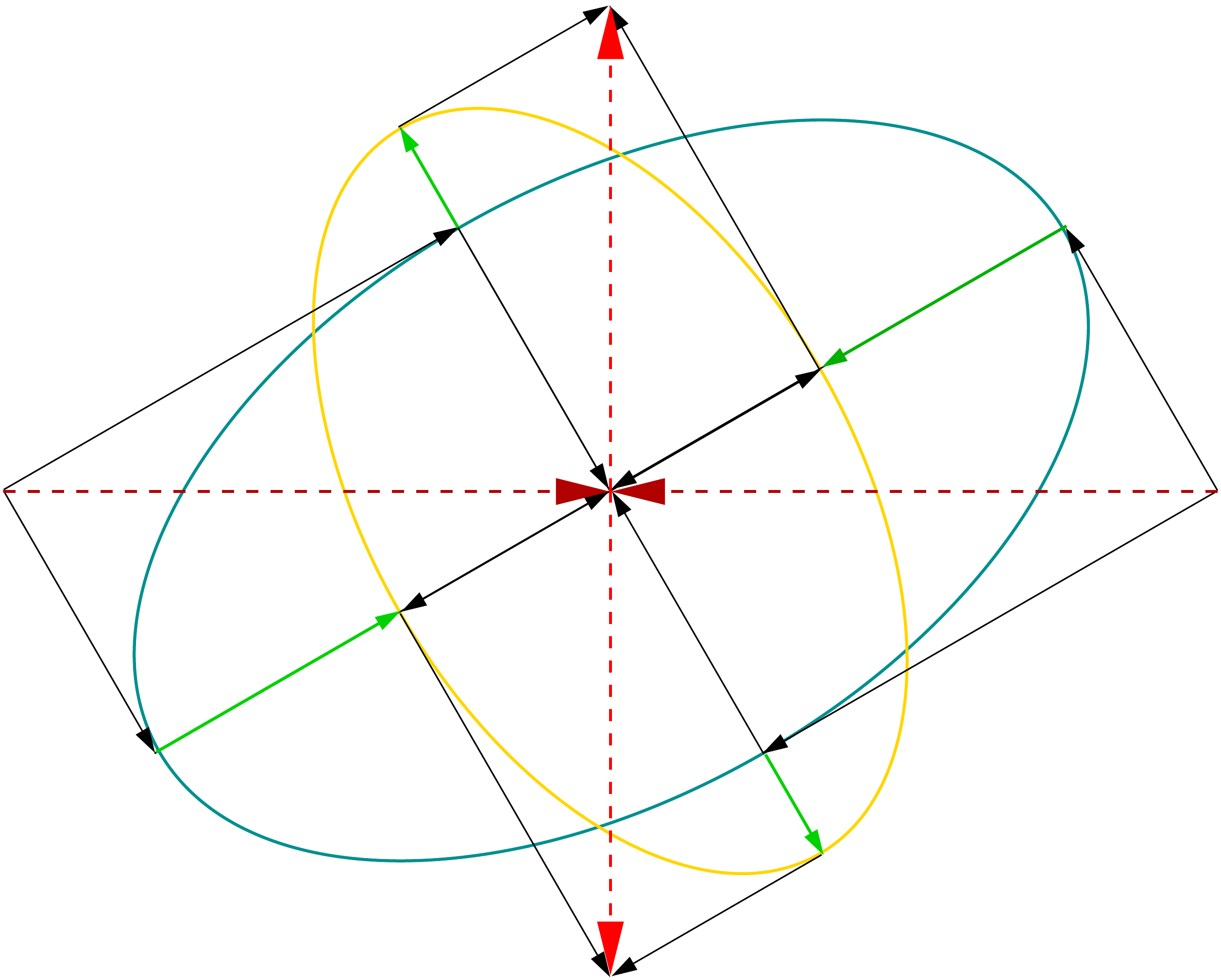}
 \caption{Tidal Stretching}
 \label{f:sketch_tidal_stretching} 
 \end{center}
\end{subfigure}
  \caption{Illustration    of   tidal    stretching   and    torquing.
    \textit{Left:}  Embedding the  halo  (blue ellipse)  into a  tidal
    field which varies across the halo (blue arrows) gives rise to two
    effects: First, there  is a change in shape of  the halo caused by
    the components of the tidal shear inside the principal axis system
    of  the halo,  leading to  an anisotropic  change in  shape (green
    arrows).  Second, there is a shearing deformation of the halo (red
    arrows), which  generates angular momentum.  \textit{Right:}  If a
    halo (blue ellipse) is embedded  into a tidal gravitational field,
    it is tidally stretched to a new shape (yellow ellipse): The tidal
    shear displacement  (red arrows) is projected  into the coordinate
    frame defined by the halo principal axes (black arrows), such that
    the resulting tidal fields cause a  contraction of the halo in one
    direction and a expansion of the halo in the other (green arrows).
  }
\end{figure*}

Tidal interactions  of galaxies with the  cosmic large-scale structure
can be  modelled as  a perturbative  process.  Consider  the position,
$x_i(\boldsymbol{q},a)$, of a particle as  a function of scale factor,
$a$,  and initial  position, $\boldsymbol{q}$.   To lowest  order, the
positions  follow   straight  lines  over  time,   along  a  direction
determined  by the  gradient,  $\partial_i\Psi$,  of the  displacement
potential, $\Psi$, \citep{Z70},
\begin{equation}
x_i (\boldsymbol{q},a) = q_i - D_+(a)\; \partial_i \Psi(\boldsymbol{q})\;,
\end{equation}
where $D_+(a)$  is the growth  function. The interaction of  an entire
protohalo is obtained by  Taylor-expanding the trajectories around the
centre of gravity, $\bar{\boldsymbol{q}}$,
\begin{equation}
\label{eq:zeldovich:taylor}
x_i   (  \boldsymbol{q},a)   \approx   q_i  -   D_+(a)  \left(   \partial_i
\Psi(\bar{\boldsymbol{q}})   +   \sum_j   (q_j   -\bar{q}_j)\;   \partial_i
\partial_j \Psi(\bar{\boldsymbol{q}}) \right)\;,
\end{equation}
which  reveals  the   bulk  motion  of  the   halo  along  $\partial_i
\Psi(\bar{\boldsymbol{q}})$  and   the  differential  motion   of  the
particles  around the  centre, which  is  encoded in  the tidal  field
tensor,   $\partial_i    \partial_j   \Psi(\bar{\boldsymbol{q}})\equiv
T_{ij}(\bar{\boldsymbol{q}})$.  The displacement  potential is related
to the Newtonian gravitational potential, $\Phi$, by a factor of $4\pi
G$, where $G$ is Newton's  gravitational constant.  It should be noted
that  in this  picture  the  protohalo is  treated  as  a test  object
embedded  into a  tidal field,  whereas  in reality  the tidal  fields
themselves determine which particles will ultimately compose the halo.

The haloes principal axis frame  defines a coordinate frame into which
the tidal field, $\partial_i\partial_j\Phi$, can be decomposed. If the
tidal field tensor does not coincide  with the principle axis frame of
the halo, a shearing motion is exerted onto the halo, which ultimately
leads    to    angular    momentum    generation,    as    shown    by
\Cref{f:sketch_tidal_torquing}. As long as merging or accretion do not
play an  important role,  the galactic  disc can  be expected  to form
perpendicular  to the  halo angular  momentum direction:  This is  the
picture behind the quadratic alignment of spiral galaxies.

The component of the tidal  field coinciding with the haloes principal
axis frame gives rise to  an anisotropic deformation as illustrated by
\Cref{f:sketch_tidal_stretching}. This  effect is used to  explain the
alignment of elliptical galaxies with the large-scale structure and is
the basis of  the linear alignment model for  elliptical galaxies, and
even though the haloes and  elliptical galaxies are collapsed objects,
it is still reasonable to expect  a deformation of the structures with
the tidal shear field.

\subsubsection{Linear alignment model for elliptical galaxies}
\label{sec:linear}

\begin{figure}[t]
\begin{center}
\includegraphics[width=9cm]{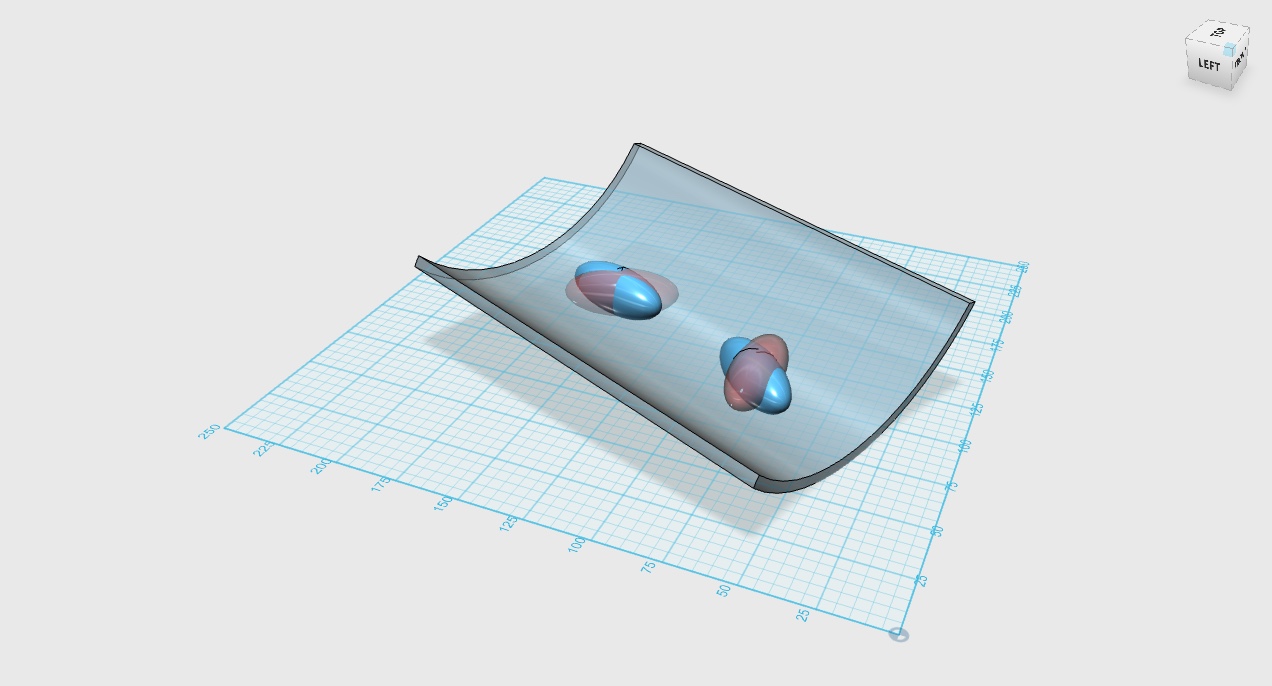}
\caption{Two elliptical galaxies embedded in a gravitational potential
  (grey sheet). They  experience correlated tidal shears  and react to
  the  tidal fields  by changing  their initially  uncorrelated shapes
  (red), to correlated shapes (blue).}
\label{f:elliptical_alignment}
\end{center}
\end{figure}

Elliptical galaxies  are supported by  the velocity dispersion  of the
dark matter particles and of the stars,  which can be assumed to be in
virial equilibrium.   If an elliptical  galaxy is embedded in  a tidal
gravitational field,  it will  distort the  potential of  the galaxy's
dark matter halo  and force the distribution of stars  to assume a new
equilibrium configuration, resulting in a  new shape as illustrated by
\Cref{f:elliptical_alignment}. The reaction of an elliptical galaxy to
an external  tidal field can  be considered instantaneous  because the
dynamical  time scale  of  stars  inside the  galaxy  should be  short
compared  to the  time  scale  on which  tidal  fields  change due  to
structure  formation,   such  that  the  distortion   of  the  stellar
quadrupole grows with the growth  function $D_+$. This picture assumes
that the  shape of an  elliptical galaxy  is perturbed by  an external
tidal shear field while it is not influenced by e.g. merging.

The intrinsic  complex shear, $\gamma^I$ (see  \citealp{paper1} for an
introduction  to $\gamma^I$),  in this  model is  proportional to  the
tidal shear field projected onto the sky,
\begin{equation}
\gamma^I  =  -\frac{C_1}{4\pi   G}\left(  \partial_x^2-  \partial_y^2+
2\mathrm{i} \partial_{xy}^2 \right) \mathcal{S}(\Phi),
\label{eq:linear_shear_relation}
\end{equation}
where  $G$ is  Newton's gravitational  constant  and $x$  and $y$  are
Cartesian  coordinates in  the plane  of  the sky.   Smoothing of  the
gravitational potential,  as indicated by $\mathcal{S}(\Phi)$,  on the
scale of  the halo ensures  that the halo as  a whole reacts  to tidal
fields  and   avoids  substructures   that  would  be   introduced  by
small-scale fluctuations of $\Phi$.   The derivatives $(\partial_x^2 -
\partial_y^2)\mathcal{S}(\Phi)$               and              $(2{\rm
  i}\partial_{xy}^2)\mathcal{S}\Phi$  (with  the prefactor)  give  the
tangential and cross  components of the shear with respect  to the $x$
axis.  $C_1$  is a constant of  proportionality containing information
on  the strength  of  the reaction  that a  galaxy  experiences as  it
adjusts its shape due to  an external tidal shear field.  \citet{BK07}
measured  this through  ellipticity correlations  in SuperCOSMOS  data
\citep{BTH+02}   and    found   $C_1   \simeq   5    \times   10^{-14}
(h^{2}\msun\mathrm{Mpc}^{-3})^{-1}$.  The relation between ellipticity
and tidal shear is,  in the limit of weak fields, a  linear one and it
is  from  this linearity  that  linear  alignments derive  their  name
\citep{CKB01}.   Physically, the  dependence of  the distorted  galaxy
shape  on  the  second  derivatives of  the  gravitational  potential,
corresponds   exactly  to   the   tidal   stretching  illustrated   in
\Cref{f:sketch_tidal_stretching}.   In  addition,  assuming  a  linear
relationship between  the quadrupole  of the  brightness distribution,
which is a symmetric tensor of  rank 2, and the tidal shear transverse
to  the  line of  sight,  which  is a  tensor  of  the same  type,  is
reasonable    motivation     for    \autoref{eq:linear_shear_relation}
\citep{HS04,HS10}\footnote{Note  that   \citet{HS10}  is   an  updated
  version of the original \citet{HS04} paper  that fixes an error in a
  conversion factor that propagated through several equations.}.

\citet{HS04} related the  gravitational potential at an  early time in
the galaxy's  evolution to the  matter density contrast,  $\delta$, on
linear scales via the Poisson equation,
\begin{equation}
\Phi(\boldsymbol{k}) = -4\pi G \frac{\bar{\rho}(z)}{\bar{D}(z)} a^{2}
  k^{-2} \delta(\boldsymbol{k}),
\end{equation}
where $\bar{D}(z)  \propto (1+z) D_+(z)$  is the scaled  growth factor
that is normalized to unity during matter domination, $a$ is the scale
factor, $\boldsymbol{k}$ is the wave  vector and $k$ is the wavenumber
of    the    wave    vector..    The    mean    background    density,
$\bar{\rho}(z)=\Omega_m(z)\rho_{\rm   crit}(z)$,    where   $\rho_{\rm
  crit}(z)$ is the critical density of the Universe, sets the strength
of the  gravitational potential.  The  critical density of  a smoothed
background that produces a spatially flat Universe is,
\begin{equation}
  \rho_{\rm crit}(z) = \frac{3H(z)^2}{8 \pi G},
  \label{eq:crit}
\end{equation}
where $H(z)$ is the Hubble  constant at redshift $z$.

\citet{HS04} assumed  that the \lq primordial\rq\  tidal gravitational
field acting during the formation  of the galaxy determines the galaxy
alignment,  which  would  then  be frozen  in  during  the  subsequent
evolution. As this model is  mostly applicable to elliptical galaxies,
which are  believed to  undergo dramatic  changes in  their morphology
during mergers  at relatively recent  times, it  may be more  valid to
instead assume a quasi-instantaneous response to the local tidal field
at any given time, which would lead to a different redshift scaling of
the  alignment  signal  amplitude. However,  the  redshift  dependence
predicted  by  \citet{HS04},  which  we adopt  in  the  following,  is
consistent with current observations \citep{JMA+11}.

Irrespective of whether the density field is in a linear or non-linear
stage of evolution, the link  between the observable $\gamma^I$ to the
matter  density power  spectrum $P_{\delta  \delta}^{\rm lin}(k)$,  is
best formulated in  Fourier space.  As information on  the tidal shear
field is  only available  at positions where  there are  galaxies, the
density-weighted intrinsic  shear, $\tilde{\gamma}^I$, is  the natural
observable. In the complex notation this is given by
\begin{equation}
  \tilde{\gamma}^I(\boldsymbol{k})                                   =
  \frac{C_{1}\bar{\rho}}{\bar{D}}a^{2}                            \int
  \frac{(k^{2}_{2x}-k^{2}_{2y} + 2\mathrm{i}k_{2x}k_{2y})}             {k^{2}_{2}}
  \delta(\boldsymbol{k}_2) \left[ \delta^{(3)}_D(\boldsymbol{k}_{1}) +
    \frac{b_{g}}{(2\pi)^{3}}   \delta   (\boldsymbol{k}_{1})   \right]
  \mathrm{d}^{3}\boldsymbol{k}_{1},
  \label{eq:weightedshear}
\end{equation}
where   $\boldsymbol{k_2}   =  \boldsymbol{k}   -   \boldsymbol{k}_1$,
$\delta^{(3)}_D$  is the  3D Dirac  delta  function and  $b_g$ is  the
linear biasing factor,  which relates the relative  fluctuation in the
number density  of galaxies  to the local  dark matter  density.  Note
that  the multiplicative  density weighting  becomes a  convolution in
Fourier  space with  $\boldsymbol{k}_1$ as  the integration  variable,
such    that   the    observable    $\tilde{\gamma}^I$   depends    on
$\boldsymbol{k}$.   The  derivatives  of the  gravitational  potential
translate in Fourier space into multiplications with the corresponding
components of  the wave vector,  in this case  $\boldsymbol{k}_2$.  In
this way, the resulting $E$-mode part of the II power spectrum $P_{\rm
  II}(k)$ of $\tilde{\gamma}^I$ is
\begin{equation}
P_{\rm         II}(k)        =         \frac{C^{2}_{1}\bar{\rho}^{2}}
{\bar{D}^{2}}a^{4}\Biggl\{     P_{\delta\delta}^{\rm     lin}(k)     +
        b_{g}^{2}\int         [f_{E}(\boldsymbol{k}_{2})         +
  f_{E}(\boldsymbol{k}_{1})]      \times     f_{E}(\boldsymbol{k}_{2})
\frac{P_{\delta\delta}^{\rm      lin}(k_{1})     P_{\delta\delta}^{\rm
    lin}(k_{2})}     {(2\pi)^{3}}\mathrm{d}^{3}     \boldsymbol{k}_{1}
\Biggr\},
\label{eq:P^EE}
\end{equation}
where $f_E(\boldsymbol{k})$  is a geometric function  that singles out
correlations   between  the   $E$-modes  of   the  ellipticity   field
\citep{HS04}. Note that the $B$-mode correlations are zero, due to the
symmetry of  the tidal shear  tensor.  $B$-modes can be  introduced in
the context of a linear alignment  model by e.g. clustering in analogy
to second-order effects in gravitational lensing.

The second term in brackets in \Cref{eq:P^EE} is caused by the density
weighting and is proportional to the square of the linear matter power
spectrum,    $P_{\delta\delta}^{\rm   lin}(k_{1})P_{\delta\delta}^{\rm
  lin}(k_{2})$.  It  is therefore  sub-dominant compared to  the first
term on large scales and is usually ignored in the literature when the
linear alignment model  is applied.  Being linear in  the tidal field,
there is a cross-correlation between alignments and weak gravitational
lensing  \citep{HS04}, even  for  Gaussian  initial conditions,  which
persists  for  nonlinearly  evolving fields.   While  the  correlation
between shapes in the linear  alignment model is necessarily positive,
the cross-correlation between the overdensity and the density-weighted
intrinsic  shear is  in  general negative  because mass  overdensities
tangentially  align  the lensed  images  of  background objects  while
radially  aligning local  galaxies  if the  overdensities are  massive
enough to define the principal axis of the local tidal quadrupole. The
corresponding power spectrum reads
\begin{equation}\label{eq:pdeltagamma}
P_{\delta     {\rm      I}}(k)     =      -     \frac{C_{1}\bar{\rho}}
{\bar{D}}a^{2}P_{\delta\delta}^{\rm lin}(k),
\end{equation}
which is important  to both weak cosmic  shear \citep{HS04,HMS+04} and
galaxy-galaxy lensing \citep{BMS+12}.  The  parameter $C_{1}$ will, in
general, depend  on galaxy properties including  luminosity, mass, and
formation time.

Correlations of the linear type  can be quite long-ranged: Ellipticity
auto-correlations (II)  have been  measured in  the Sloan  Digital Sky
Survey      \citep[SDSS;][]{YAA+00}      to     reach      out      to
$30\:h^{-1}\mathrm{Mpc}$ \citep{OJL09}, and cross-correlations between
shape   and   density   to  almost   $100\:h^{-1}\mathrm{Mpc}$   (e.g.
\citealp{MHI+06}, and  see also  \citealp{paper3} for  a comprehensive
list of  observations).  The  shape-density correlations give  rise to
intrinsic  shape-lensing (GI)  correlations (see  \autoref{eq:iadef}),
and these  have been  marginally detected in  the Canada-France-Hawaii
Telescope Lensing Survey \citep[CFHTLenS;][]{HGH+13,MZB+15}.

The  linear alignment  model is  characterized by  a single  parameter
(which may  depend on  properties like  luminosity, mass,  etc.)  that
sets  the strength  of the  external tidal  field in  relation to  the
ellipticity of  the distorted galaxy. Recent  analytical studies found
that tidal stretching on a stellar structure in equilibrium may not be
strong  enough  to  explain  the  observed  alignments  of  elliptical
galaxies  \citep{CL15}.  It  should be  noted that  even for  the case
where the particular  model for tidal stretching is  incorrect, it may
still  effectively  describe other  alignment  mechanisms  due to  its
generality.

\subsubsection{Quadratic alignment model for spiral galaxies}
\label{sec:quadratic}

Commonly, spiral galaxy  alignments are explained by  the alignment of
their  angular  momentum  with  the tidal  field  of  the  large-scale
structure, which occurs  due to tidal torquing.  If  the symmetry axis
of the  galactic disc  follows the angular  momentum direction  of the
host halo, the observer will measure ellipticities which depend on the
angle of  inclination of  the galactic  disc.  In  the case  where the
angular momentum points toward the observer, the galaxy is viewed face
on and  will have a small  ellipticity, in contrast to  the case where
the angular momentum  is perpendicular to the line  of sight, implying
that  the  galactic disc  is  viewed  edge on  and  will  have a  high
ellipticity. In  alternative models, the alignment  of spiral galaxies
with the large-scale structure is traced  back to the vorticity of the
surrounding  flow field,  to  the accretion  pattern  of matter  flows
converging on the galaxy.

The  main uncertainties  of this  model are  the extent  to which  the
angular momentum can be predicted perturbatively by tidal torquing and
the orientation of  the stellar disc relative to  the angular momentum
direction of the  host halo.  In fact, there may  be strong deviations
in  the latter  related to  merging,  which can  reorient the  angular
momentum,  and  dissipative  processes,   which  can  destroy  angular
momentum.   Tidal torquing,  however, has  been shown  to predict  the
angular momentum direction of haloes reasonably well, which matters in
this  context,  but  would  fail at  predicting  the  correct  angular
momentum magnitude \citep{CT96}.

The tidal  torquing mechanism combines halo  inertia and gravitational
tidal shear  to generate  angular momentum,  meaning that  the angular
momentum  resulting  from  the  same tidal  fields  can  be  different
depending on the  halo shape.  In order to  capture this, \citet{LP00}
introduced   a    model   that    gives   a    Gaussian   distribution
$p(\boldsymbol{J}|T)$ of angular momenta $\boldsymbol{J}$,
\begin{equation}
p(\boldsymbol{J}|T)     =     \frac{1}{\sqrt{(2\pi)^3\mathrm{det}(C)}}
\exp\left(-\frac{1}{2}    \sum_{\alpha, \alpha^\prime} J_\alpha     (C^{-1})_{\alpha\alpha^\prime}
J_{\alpha^\prime}\right),
\end{equation}
conditional on the tidal shear $T$. The conditionality of the Gaussian
distribution    is    incorporated    in   the    covariance    matrix
$C_{\alpha\alpha^\prime} = \langle J_\alpha J_{\alpha^\prime}\rangle$,
which also depends on the tidal shear,
\begin{equation}
\langle J_\alpha J_{\alpha^\prime}\rangle =
\frac{\langle\boldsymbol{J}^2\rangle}{3}
\left(\frac{1+a_{\rm T}}{3}\delta_{\alpha\alpha^\prime} - a_{\rm T}\sum_\sigma
\hat{T}_{\alpha\sigma}\hat{T}_{\sigma\alpha^\prime}\right),
\label{eqn_ll_crittenden}
\end{equation}
where $\delta_{\alpha \alpha^{\prime}}$ is the Kronecker $\delta$. The
model  is characterised  by the  misalignment parameter,  $a_{\rm T}$,
which allows the variation between  random angular momenta for $a_{\rm
  T}=0$   to   maximally   aligned   angular   momenta   for   $a_{\rm
  T}=3/5$.  Larger values  for $a_{\rm  T}$ would  be in  violation of
keeping $C_{\alpha\alpha^\prime}$ positive definite.

$\hat{T}$              is             the              unit-normalised
($\hat{T}_{\alpha\sigma}\hat{T}_{\sigma\alpha}=1$),          traceless
($\mathrm{tr}\hat{T} =  0$) tidal  shear tensor,  which can  easily be
derived using
\begin{equation}
\tilde{T}_{\alpha\alpha^{\prime}}    =    T_{\alpha\alpha^{\prime}}    -
\frac{\mathrm{Tr}(T)}{3}\delta_{\alpha\alpha^{\prime}},
\end{equation}
and  rescaling   $\hat{T}  =   \tilde{T}/\left|\tilde{T}\right|$  with
$\tilde{T} = \sqrt{\bar{T}_{\alpha\sigma} \bar{T}_{\sigma\alpha}}$. It
should  be emphasised  that  the traceless  unit-normalised tensor  is
derived  from the  initial conditions  for structure  formation, which
determine  the  orientation  of   the  eigensystem  of  $\hat{T}$  and
ultimately the angular momentum direction.

\begin{figure}[t]
\begin{center}
\includegraphics[width=9cm]{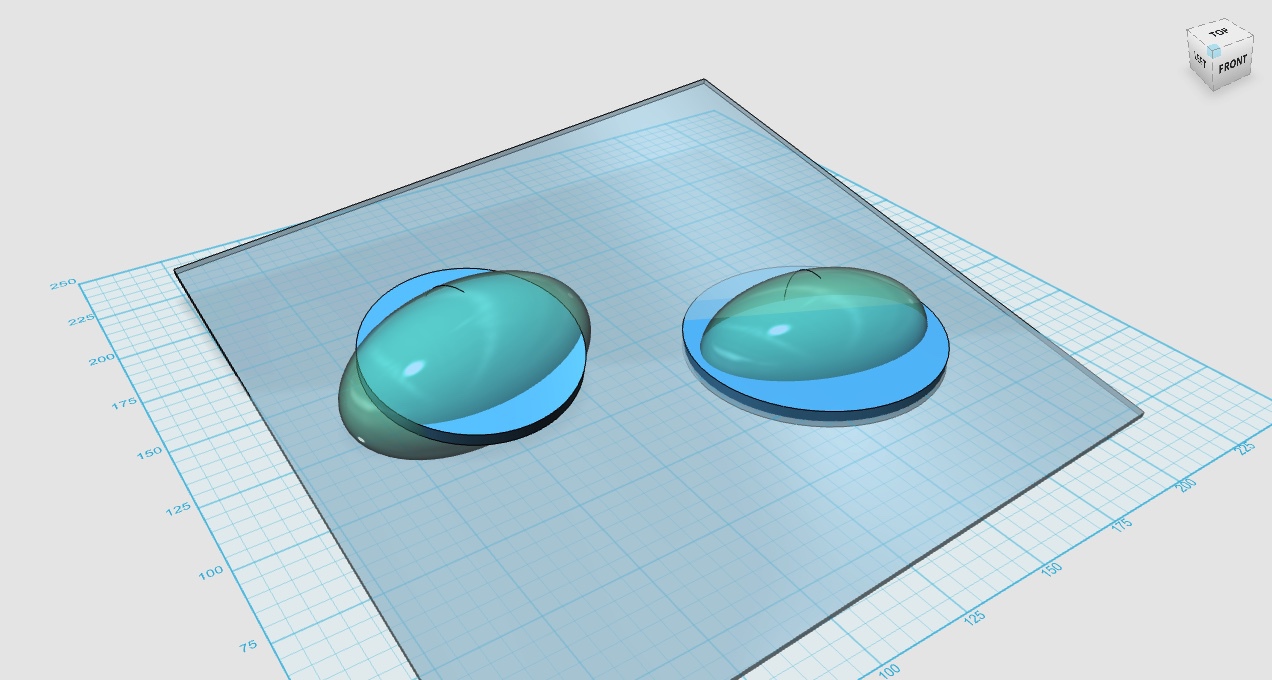}
\caption{Two haloes (green) are  embedded in a gravitational potential
  (grey sheet).  They experience tidal  torquing due to differences in
  the slope  of the gravitational  potential across their  volume, and
  build up  angular momentum by  tidal torquing.  Correlations  in the
  tidal  shear  field  leads  then  to  correlations  in  the  angular
  momentum,  and  ultimately  in  the orientation  of  galactic  discs
  (blue), leading to correlated shapes in the observation.}
\label{f:spiral_alignment}
\end{center}
\end{figure}

Squaring the  tidal shear gives  rise to short-ranged  correlations in
the angular momentum directions, and  ultimately in the derived galaxy
shapes    \citep{SM11}.     This    picture    is    illustrated    in
\Cref{f:spiral_alignment}, where  two neighbouring haloes  are subject
to correlated tidal torquing of their motion along the gradient of the
gravitational potential.   They build  up correlated  angular momenta,
which determines the  orientation of their discs  and therefore, their
shapes. There can  be significant differences in  the torquing process
in filamentary  structures relative  to the  average locations  in the
cosmic structure,  as exemplified by \citep{CPP15},  who introduced an
anisotropic tidal torquing model for describing these situations.

Under  the assumption  that the  symmetry  axis of  the galactic  disc
follows  the  angular momentum  direction  of  the  host halo,  it  is
possible to write down relationships of the form \citep{CKB01}:
\begin{eqnarray}
\gamma^{I}_+ & = & f(J,J_z) (J_x^2-J_y^2) \\
\gamma^{I}_\times & = & 2f(J,J_z) (J_xJ_y),
\end{eqnarray}
where the function $f(J,J_z)$ describes  the scaling of the observable
intrinsic  ellipticity $\gamma^{I}$,  with angular  momentum magnitude
and  direction. In  general, the  observed ellipticity  should show  a
scaling behaviour \citep{CNP+01}
\begin{equation}
|\gamma^{I}| \propto \frac{1-\hat{J}_z^2}{1+\hat{J}_z^2}.
\end{equation}

The angular momentum is related to the tidal shear through
\begin{equation}
  J_\alpha \propto \epsilon_{\alpha\beta\gamma}\sum_\sigma I_{\beta\sigma}
  T_{\sigma\gamma}, 
\end{equation}
where $I_{\beta\sigma}$  is the tensor  of second moments of  the mass
distribution (the  inertia) and $\epsilon_{\alpha\beta\gamma}$  is the
antisymmetric  symbol   (or  Levi-Civita   symbol)  in   3  dimensions
\citep{SM11}.  A direct connection of  the galaxy shape, in analogy to
the linear alignment model, would use the quadratic dependence between
shape and tidal shear according to
\begin{equation}
  \gamma^{I}    =   C_2(\tilde{T}_{x\mu}^{2}-\tilde{T}_{y    \mu}^{2},
  2\tilde{T}_{x\mu}\tilde{T}_{y \mu}),
\label{eqn:QA_gammaI}
\end{equation}
where  the  constant  of  proportionality,  $C_2$,  captures  all  the
processes related to  the angular momentum variance for  a given tidal
shear  and of  the inclination  of the  galactic disc.   The effective
model    in   \Cref{eqn:QA_gammaI}    uses    the   traceless    shear
$\tilde{T}_{\alpha\alpha^{\prime}}$  rather   than  the   tidal  shear
$T_{\alpha\alpha^{\prime}}$ because alignment of spiral galaxies is an
orientation effect where the absolute value of the tidal shears is not
relevant, only  the orientation of  the tidal shear  eigensystem.  The
subscripts  $x$  and $y$  refer  to  the  coordinate system  in  which
$\gamma^I_+$             assumes            the             components
$C_2(\tilde{T}^2_{x\mu}-\tilde{T}^2_{y\mu})$                       and
$2C_2\tilde{T}_{x\mu}$.

Using this model as an effective  model for spiral galaxies in analogy
to  the  alignments of  ellipticals,  the  resulting density  weighted
intrinsic shear $\tilde{\gamma}^I$ is \citep{HS04},
\begin{equation}
\tilde{\gamma}^{I}(\boldsymbol{k})     =    \frac{C_{2}\bar{\rho}^{2}}
      {(2\pi)^{3}\bar{D}^{2}}a^{4}                                \int
      h_{E}(\hat{\boldsymbol{k}}^{\prime}_{1},\hat{\boldsymbol{k}}^{\prime}_{2})\delta(\boldsymbol{k}^{\prime}_{1})\delta(\boldsymbol{k}^{\prime}_{2})
            \left[             \delta_{\rm
          D}^{(3)}(\boldsymbol{k}^{\prime}_{3})                      +
        \frac{b_{g}}{(2\pi)^{3}}   \delta(\boldsymbol{k}^{\prime}_{3})
        \right]       \mathrm{d}^{3}       \boldsymbol{k}^{\prime}_{1}
      \mathrm{d}^{3} \boldsymbol{k}^{\prime}_{2},
\end{equation}
where    $\hat{\boldsymbol{k}}_a    =   \boldsymbol{k}_a/|k_a|$    and
$\boldsymbol{k}_3^{\prime}         =          \boldsymbol{k}         -
\boldsymbol{k}_1^{\prime}-\boldsymbol{k}_2^{\prime}$.      The    term
proportional  to  the  galaxy  bias  $b_{g}$  introduces  a  weighting
proportional to the galaxy density, therefore, two of the three powers
of $\delta$  are due to  the quadratic alignment  and one is  from the
applied density weighting.  Similar to the case for linear alignments,
there is a  geometric factor $h_E$ that depends, in  this case, on the
directions  $\hat{\boldsymbol{k}}$ of  the  wave vector  \citep{HS04}.
The  intrinsic alignment  power spectrum  $P_{\rm II}(k)$  for the  II
correlation takes the form
\begin{equation}
\begin{aligned}
P_{\rm II}(k)                                      =&
\frac{C^{2}_{2}\bar{\rho}^{4}}{\bar{D}^{4}}a^{8}  \Biggl\{2
\int     [h_{E}(\hat{\boldsymbol{k}}_{1}\hat{\boldsymbol{k}}_{2})]^{2}
\frac{P_{\delta\delta}^{\rm           lin}(k_{1})P_{\delta\delta}^{\rm
    lin}(k_{2})}{(2\pi)^{3}}d^{3}\boldsymbol{k}_{1}
\\             &\,\,\,\,\,\,\,\,+             \frac{2}{3}b^{2}_{g}\int
   [h_{E}(\hat{\boldsymbol{k}}^{'}_{1},\hat{\boldsymbol{k}}^{'}_{2}) +
     h_{E}(\hat{k}^{'}_{2},\hat{k}^{'}_{3})   +
     h_{E}(\hat{k}^{'}_{3},\hat{k}^{'}_{1})]^{2}    \;
   \frac{P_{\delta\delta}^{\rm    lin}(k^{'}_{1})P_{\delta\delta}^{\rm
       lin}(k^{'}_{2})P_{\delta\delta}^{\rm
       lin}(k^{'}_{3})}{(2\pi)^{6}}
   d^{3}\boldsymbol{k}^{'}_{1}d^{3}\boldsymbol{k}^{'}_{2} \Biggr\}.
\end{aligned}
\end{equation}

In  this way,  the  ellipticity is  linked to  the  tidal shear  field
through the  angular momentum  direction and it  is possible  to trace
correlations in the ellipticity field back to those in the tidal shear
field,  which in  turn  are  related to  density  correlations by  the
Poisson equation.   This has been demonstrated  by \citet{CNP+01}, who
computed  $E$-  and  $B$-mode ellipticity  correlation  functions  and
showed that they  dominate over the lensing signal  at redshifts below
$z  = 0.3$,  and  are a  small but  significant  contribution to  weak
lensing spectra at redshifts of unity, under the assumption of perfect
alignment  between the  galaxy  and halo  angular  momenta.  The  main
uncertainties  are the  misalignment parameter  $a_{\rm T}$,  which is
measured in numerical simulations by  comparing the tidal shear acting
on a halo with the resulting  angular momentum, and the orientation of
the galactic  disc relative to  the angular momentum  direction.  This
relation is  poorly understood, in  fact many analytical  works assume
perfect alignment  but allow  for a finite  thickness of  the galactic
disc,  which  dilutes the  orientation  effect  and  leads to  a  less
pronounced  ellipticity variation  when tilting  the angular  momentum
direction  \citep[e.g.][]{CNP+01,CNP+2002,CMS12,MS13}.  Following  the
work  of \citet{CNP+01},  \citet{CMS12}  computed angular  ellipticity
power  spectra  and  compared  these with  weak  lensing  shear  power
spectra.   The  intrinsic $E$-mode  (II)  spectra  resulting from  the
quadratic alignments was  found to be an order of  magnitude less than
the  weak lensing  power  spectra for  lensing  surveys reaching  unit
redshift, while  the $B$-mode  spectra would be  an important  test of
systematics.     Generalising   on    this,   \citet{MS13}    computed
three-dimensional  ellipticity (II)  power spectra  and found  them to
differ  from  weak lensing  power  spectra  in  terms of  their  scale
dependence   and  their   correlation  properties   between  different
Fourier-modes, which  arise in  this formalism because  the projection
onto angular correlations does not assume the Limber-approximation.

The density-intrinsic  alignment cross-power  spectrum is zero  in the
simple quadratic alignment model due to the assumed Gaussianity of the
density field.  The correlation would  be proportional to three powers
of the  density field, two from  the alignment model and  one from the
density  field,  and  is  zero,  because odd  moments  of  a  Gaussian
distribution  vanish identically.   However, there  will certainly  be
some non-zero  contribution to  the cross-term  even in  the quadratic
alignment  model  once  the  third- and  higher-order  corrections  to
\Cref{eqn:QA_gammaI}  are   taken  into  account,  as   well  as  from
non-Gaussianity due to non-linear growth of structure.


\subsection{Alignments on scales smaller than host haloes}
\label{sec:SmallScaleTheory}
The  current  understanding  of  galaxy alignments  within  haloes  is
fundamentally  hampered by  the  limited knowledge  of the  non-linear
evolution and baryonic  physics that shapes the  haloes themselves and
the  galaxies  they  host.   Without a  physical  grounding  to  study
intrinsic galaxy  alignments within  a halo, the  best approach  is to
describe alignments through the halo model.

The   halo  model   \citep[e.g.][]{Seljak00,CS02}  is   an  analytical
description of the clustering of dark matter in the Universe, based on
the ansatz that all dark matter particles are contained within haloes.
It further  assumes that the \textit{mass}  of a halo is  the physical
quantity that  drives virtually all  properties of the haloes  (and of
the  galaxies that  inhabit  them).  This  assumption has  motivations
heavily rooted in  the results of $N$-body  simulations.  For example,
the     abundance     \citep[see     e.g.][]{TKK+08},     the     bias
\citep[e.g.][]{TRK+10},  and  the  matter density  profile  \citep[see
  e.g.][]{NFW97,DM14} are all functions of  the halo mass (albeit with
some  scatter).   Perhaps  even   more  interestingly,  properties  of
galaxies such  as stellar mass, luminosity,  star formation efficiency
and size all depend  to first order on the mass  of their host haloes.
Using this  model, predictions for  the clustering of dark  matter, as
well    as   galaxies,    can   be    as   accurate    as   5-10    \%
\citep[e.g.][]{GBS+10,BMC+13}.   The model  has been  further improved
through  incorporation of  the potential  effects of  galaxy formation
(e.g.   baryonic feedback)  into  the clustering  of  dark matter  and
galaxies  \citep[see][]{Fedeli14,FSV+14}.   As the  observational  and
simulated data increase in precision,  it has become apparent that the
halo formation  time plays a  similarly important role in  shaping the
properties of  galaxies such  as halo mass.   This effect  is commonly
known as  \textit{assembly bias}: at  fixed halo mass, the  scatter in
galaxy properties mainly  owes to the different  assembly histories of
haloes \citep[see e.g.][]{GW07,WWDLY13}.

With halo mass  being the central quantity,  other relevant quantities
of the halo model are the number density of haloes of a given mass and
the relation between  mass and observable. With these,  it is possible
to describe the  statistical properties of the observable  (taken at a
single point), while the correlation  functions of the observable need
in addition  the density profile  inside an individual halo  (which is
taken to  scale with halo  mass) and  the correlation function  of the
haloes, while  assuming that the  relation between the  observable and
the halo mass  is a local one.  In this  approximation, the halo model
has been  used for  a number of  applications, including  weak lensing
statistics, and  relevant for this  review, the alignment  of galaxies
inside dark matter structures on scales  from tens of Mpc down to tens
of kpc.   This is possible  due to  the natural halo  model separation
into   intra-  and   inter-halo   properties,   commonly  called   the
\emph{`one-halo'} (1h) and \emph{`two-halo'} (2h) terms, respectively.
\citet{SB10}  pioneered  the modelling  of  galaxy  alignments in  the
language of the halo model. As was commonly done for clustering, their
formalism explicitly accounted for the distinction between central and
satellite  galaxies, which  is  peculiar to  intrinsic alignments  and
effectively  doubles the  number  of contributing  terms  in the  halo
model.

Furthermore,  as  is  evident  from the  equations  that  follow,  the
large-scale terms are a rescaling (based on galaxy bias) of the matter
clustering,  whereas the  small scale  terms are  contributions coming
from haloes  weighted by  their relative abundance.   Accordingly, the
auto- and cross-power spectra of the galaxy alignments read:
\begin{align}
P_{\rm II}(k) = &
\, P^{1h}_{\rm II,cs}(k) +
P^{1h}_{\rm II,ss}(k) +
P^{2h}_{\rm II,cc}(k) +
P^{2h}_{\rm II,cs}(k) +
P^{2h}_{\rm II,ss}(k)\,;  \nonumber \\
P_{\delta{\rm I}}(k) = &
\, P^{1h}_{\delta{\rm I,c}}(k) +
P^{1h}_{\delta{\rm I,s}}(k) +
P^{2h}_{\delta{\rm I,c}}(k) +
P^{2h}_{\delta{\rm I,s}}(k) \, ,
\end{align}
where  the subscripts  `c' and  `s'  stand for  central and  satellite
galaxies,  respectively.  \citet{SB10}  assumed that  satellite galaxy
ellipticities were  uncorrelated with the central  galaxy ellipticity.
This naturally  led to  $P^{1h}_{\rm II,cs}(k)  =0$. The  remaining II
one-halo term is:
\begin{align}
P^{1h}_{\rm II,ss}(k)  = &
\frac{1}{{\bar n}^2_g} \int \langle N_g(N_g-1)|M\rangle \,
{\bar\gamma}^2(M) |w(k|M)|^2 \,  n(M) \, {\rm d}M \, ,
\end{align}
where ${\bar n}_g$ is the  comoving number density of galaxies, $n(M)$
is  the halo  mass function  \citep[e.g.][]{PS74,EPS91,TKK+08,TRK+10},
and $\langle N_g(N_g-1)|M\rangle$  is the second moment  of the galaxy
number distribution at a given halo  mass, $M$.  Here, $w(k|M)$ is the
normalized Fourier transform of the radial 3D-profile of the projected
satellite  galaxy  ellipticities,  whereas  ${\bar\gamma}(M)$  is  the
magnitude of  the intrinsic  shear, $\tilde{\gamma}^I$,  in a  halo of
mass $M$.

The model presented by \citet{SB10} also assumed that central galaxies
acquired  their alignment  in  the  same way  their  host haloes  did.
Correspondingly, the $P^{2h}_{\rm II,cc}(k)$ equals the power spectrum
in  \Cref{eq:P^EE}.  This  is  a  manifestation of  the  fact that  on
sufficiently large scales, the dominant  term for galaxy alignments is
the term that  describes the alignment of their host  haloes. The term
$P^{2h}_{\rm II,ss}(k)$  is formulated  by integrating over  the joint
probability distribution for two haloes  of mass $M_1$ and $M_2$, with
$\langle  N_g|M_1 \rangle$  and $\langle  N_g|M_2\rangle$ the  average
number of galaxies in haloes of mass $M_1$ and $M_2$,
\begin{align}
P^{2h}_{\rm II,ss}(k)  = &
\frac{1}{{\bar n}^2_g} \int \langle N_g|M_1\rangle \,
{\bar\gamma}(M_1) |w(k|M_1)| \,  n(M_1) \, {\rm d}M_1
 \int \langle N_g|M_2\rangle \,
{\bar\gamma}(M_2) |w(k|M_2)| \,  n(M_2) \, {\rm d}M_2
 P_{\rm hh}(k|M_1,M_2) \, ,
\end{align}
where    $P_{\rm     hh}(k|M_1,M_2)=b_{\rm    h}(M_1)b_{\rm    h}(M_2)
P_{\delta\delta}^{\rm lin}(k)$  is the  (dark matter)  halo-halo power
spectrum   and   $b_{\rm   h}(M)$    is   the   halo   bias   function
\citep[][]{TRK+10}.  The two-halo central-satellite term is
\begin{align}
P^{2h}_{\rm II,{cs}}(k)    =    &    \frac{C_1    {\bar
    \rho}a^2}{D_+}\frac{P_{\delta\delta}^{\rm   lin}(k)}{{\bar   n}_g}
\int  \langle N_g|M\rangle  \, {\bar\gamma}(M)  |w(k|M)| \,  b_h(M) \,
n(M) \, {\rm d}M \, .
\end{align}

The terms that define the $P_{\delta {\rm I}}(k)$ power spectrum are:
\begin{align}
P^{1h}_{\delta{\rm I,c}}(k) & = 0 \, , \nonumber \\
P^{1h}_{\delta{\rm I,s}}(k) & =
\frac{1}{{\bar n}_g{\bar \rho}} \int M\, \langle N_g|M\rangle \,
{\bar\gamma}(M) |w(k|M)| \, u(k|M) \,  n(M) \, {\rm d}M \, , \nonumber \\
P^{2h}_{\delta{\rm I,c}}(k) & = - \frac{C_1 {\bar \rho} a^2}{D_+} P_{\delta\delta}^{\rm lin}(k) \, , \nonumber \\
P^{2h}_{\delta{\rm I,s}}(k) & = \frac{1}{{\bar n}_g {\bar \rho}}
\int \langle N_g|M_1\rangle \,
{\bar\gamma}(M_1) |w(k|M_1)| \,  n(M_1) \, {\rm d}M_1  \int \,
M |u(k|M_2)| \,  n(M_2) \, {\rm d}M_2 \; P_{hh}(k|M_1,M_2) \, ,
\end{align}
where  $u(k|M)$ is  the  normalized Fourier  transform  of the  radial
profile  of   the  matter   distribution  in  a   halo  of   mass  $M$
\citep[e.g.][]{NFW97}.   Note  that   $P^{2h}_{\delta{\rm  I,c}}(k)  =
P_{\delta{\rm I}}(k)$ from \Cref{eq:pdeltagamma} since on sufficiently
large scales above around $10~h^{-1}\mathrm{Mpc}$, the central term is
the dominant one.

\citet{SB10} found that  the total halo model  alignment power spectra
are accurately modelled by restricting oneself to the two-halo central
correlation on large scales and  the one-halo satellite correlation on
small scales, i.e.
\begin{align}
P_{\rm  II}(k)  &  \approx  P^{1h}_{\rm  II,ss}(k)\,\,  +  P^{2h}_{\rm
  II,cc}(k)\;;   \nonumber   \\   P_{\delta{\rm   I}}(k)   &   \approx
P^{1h}_{\delta{\rm I,s}}(k)\,\, + P^{2h}_{\delta{\rm I,c}}(k) \;.
\end{align}
The  light blue  dashed  curve in  \Cref{f:ModelComparison} shows  the
satellite  contribution  derived from  $P^{1h}_{\delta{\rm  I,s}}(k)$,
with  the free  amplitude contained  in $\bar{\gamma}$  fitted to  the
data, along with the other halo model parameters.

\subsection{Transition to intermediate scales}
\label{sec:IntermediateScales}

Alignments on intermediate scales of a few Mpc are difficult to treat,
in contrast to alignment models  on large scales and alignments inside
haloes on small scales. Unifying both regimes would yield a consistent
model of alignments over all scales of interest.

Both  the  linear and  quadratic  alignment  models are  derived  from
(Lagrangian) perturbation theory, as a theory of the tidal interaction
of a halo  with the gravitational fields generated  by the large-scale
structure into which the halo is embedded. This naturally implies that
the models can only be applied  on linear scales above several tens of
Mpc. In order  to bridge the gap to fully  non-linear scales below 1-2
Mpc,  where  a  description  according  to the  halo  model  would  be
applicable,  it is  necessary to  modify tidal  interactions and  take
account  of  the baryonic  physics  inside  the  haloes.  It  is  also
necessary  to   have  a   prescription  of   clustering  and   to  add
observational complications such as  redshift-space distortions due to
peculiar motions, where necessary.   These intermediate scales, around
$2-10$Mpc, are referred to as {\it mildly non-linear}, and a number of
methods   have    been   proposed   as    phenomenological   solutions
\citep{TWZ+05,BMC+13}.  Research using  simulations \citep{HBH+04} and
observations \citep{MHI+06} suggest that on these intermediate scales,
intrinsic alignments are stronger than predicted by linear theory.

An  immediate solution  would be  to replace  the linear  matter power
spectrum by  a non-linear one,  which provides stronger  alignments on
small spatial scales \citep{BK07}. In effect this Non-Linear Alignment
(NLA) model  asserts that haloes  experience stronger tidal  fields in
non-linearly evolving  structures while the linear  interaction itself
is not changed, leading to increased alignment at non-linear scales.

However, it  is possible to  imagine that haloes at  close separations
started  interacting directly  with each  other  in a  way that  would
weaken  their  alignment  with  the large-scale  structure,  with  the
interesting  consequence  that  tidal  alignments  would  be  weakened
instead of being strengthened in non-linear structures.  Nevertheless,
the  non-linear alignment  model is  very  easy to  implement and  has
proved itself to  be more consistent with  small-scale observations in
comparison to the linear alignment  model alone, and thus has remained
popular                in                the                literature
\citep{MBB+11,KLB+13,BMS+12,HGH+13,CMS+14,TI14,HT14}.

\begin{figure}[t]
  \begin{center}
    \includegraphics[width=0.7\hsize]{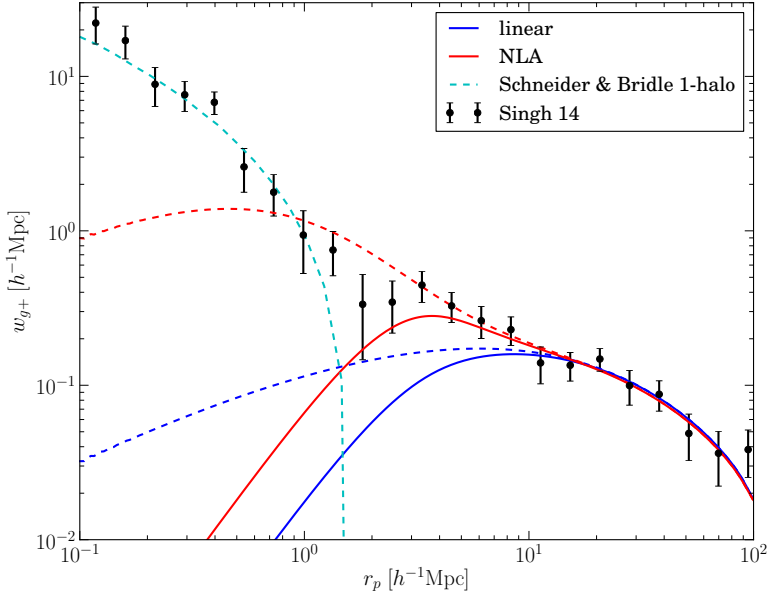}
    \caption{Galaxy  position ellipticity  correlations found  in SDSS
      luminous red galaxies from \citet{SMM14} in comparison to models
      based  on  linear  alignments  (dark  blue  line),  enhanced  by
      non-linear structures (NLA; red  line), and including non-linear
      clustering effects, biasing and  galaxy density weighting (black
      line),  with Gaussian  smoothing  at $k  = 1\:h^{-1}$Mpc  (solid
      lines)  and without  smoothing (dashed  lines).  The  light blue
      dashed  line is  the  fit  to the  one-halo  (1-halo) term  from
      \citet{SB10}  and  the  green   line  shows  a  non-perturbative
      NFW-based  one-halo model  with linear  theory for  the two-halo
      term.   The   black  and   green  models  are   reproduced  from
      \cite{BVS15}. \textit{Figure credit: Jonathan Blazek \& Sukhdeep
        Singh}.}
    \label{f:ModelComparison}
  \end{center}
\end{figure}

The non-linear alignment  model fails to explain  alignments linear in
the  tidal   field  on  a  number   of  counts,  as  pointed   out  by
\citet{BMS11,BVS15}.  While the replacement  of the linear dark matter
spectrum  with  a  nonlinear  one  is  perfectly  suited  to  describe
nonlinearities on  small scales, the  clustering of galaxies  on small
scales introduces a  non-linear weighting of the  ellipticity field by
giving rise to many galaxy pairs with small separations.  Third, there
are non-linear biases  both in the physical  distribution of galaxies,
and  to a  lesser  degree,  distortions of  the  galaxy  field due  to
redshift-space distortions.  These effects have an important influence
on  the shape  of the  resulting ellipticity  correlation function  on
small        scales,        most       notably        on        scales
$\mathtt{\lesssim}10\:\mathrm{Mpc}$,  whereas  on  larger  scales  the
standard linear alignment  model is sufficient.  It is  also common to
smooth  the  tidal  field,  since  fluctuations  on  scales  that  are
sufficiently  small  should  not  have   an  impact  on  the  observed
alignments.

\Cref{f:ModelComparison}  shows  the ellipticity-density  correlation,
$w_{g+}$, for the linear and non-linear alignment (NLA) models and the
fit to the one-halo term from \citet{SB10} with observational data and
best-fit parameters  taken from  the \citet{SMM14} SDSS  analysis.  It
also  shows  two  additional  models  taken  from  \citet{BVS15},  who
developed  analytic models  for galaxy  intrinsic alignments  based on
tidal alignment  theory.  The black  line shows the  best-fit standard
perturbation  theory  model  that  contains  non-linear  contributions
including  non-linear   clustering,  halo  bias  and   galaxy  density
weighting.   The   green  line   shows  a  physically   motivated  and
non-perturbative  one-halo   model  based  on  the   halo  profile  of
Navarro-Frenk-White    \citep[NFW\footnote{\citet{NFW96}   fitted    a
    universal  density  profile  to  dark matter  haloes  in  $N$-body
    simulations.   This profile  is now  known as  the NFW  profile.};
  see][]{NFW96}  combined with  linear theory  for the  two-halo term,
including a  contribution that captures  the enhancement of  the tidal
field in locations where galaxies form. The small wiggles in the green
line at small scales are due to the use of observed galaxy clustering,
since  the model  makes a  prediction  for the  alignment per  galaxy,
rather  than the  total $w_{g+}$  signal.  The  figure also  shows the
effects of smoothing  of the tidal field on the  linear and non-linear
alignment  models, where  the dashed  lines  show the  models with  no
smoothing, while the solid lines show the models with smoothing with a
Gaussian  of  width   $k=1\:h^{-1}$Mpc.   Motivated  by  observational
results,  the NFW-based  model  includes saturation  of the  alignment
(i.e. the alignment  signal does not increase any  further) per galaxy
within the halo virial radius.  The linear alignment model matches the
observations  well on  scales $\mathtt{\gtrsim}10\:h^{-1}$Mpc  and the
one-halo     fit     matches     the    observations     on     scales
$\mathtt{\lesssim}1\:h^{-1}$Mpc, while the non-linear alignment model,
both with  and without smoothing, provides  a reasonable approximation
to the  intermediate scales.  However, the  models from \citet{BVS15},
which include additional non-linear effects,  show a much closer match
to the  data, with the  NFW-based one-halo plus linear  two-halo model
effectively unifying the small, intermediate and large scales.

The alignments of spiral galaxies  in non-linear environments are also
affected  on  small scales  by  galaxy  clustering and  redshift-space
distortions  due  to  peculiar  motion, which  are  partially  already
included in  current analytical  models.  Because the  primary angular
momentum  build-up of  haloes takes  place at  very high  redshifts it
might be  assumed that  the angular  momentum direction  is conserved,
since  secondary   effects,  like  the  shaping   of  the  ellipticity
correlation  function  due to  peculiar  motion  of galaxies,  or  the
lensing mapping  of intrinsically shape-correlated galaxies,  has been
shown to be small \citep{GS13, GS14}.

However, results from numerical simulations are not conclusive on this
issue.    Some   dark   matter  simulations   confirm   this   picture
\citep[e.g.][]{PDH02a}, but recent simulations including baryons found
spin     changes    due     to     the     merging    of     subhaloes
\citep[e.g.][]{DPW+14,Cen14}.  It seems that ultimately the properties
of  the galactic  disc and  its orientation  due to  an initial  tidal
interaction and subsequent merging and  accretion is very difficult to
predict \citep{D+09, HTC10, DDH+14}, which could potentially be solved
by               numerical              simulations               (see
\Cref{sec:Internal,sec:SHinternal,sec:LHspin}).

\subsection{Theory and Modelling Roundup}
\label{sec:roundup}

In this section the established alignment models for galaxies on large
scales  have been  reviewed, starting  from the  idealising theory  of
tidal  interactions,  which give  rise  to  the linear  and  quadratic
alignment models, that  are thought to govern the  alignment of spiral
galaxies  through  the  angular   momentum  generation,  and  that  of
elliptical galaxies through tidal stretching on large scales. The halo
model was also reviewed, for  an effective description of alignment on
small  scales.    The  transition  between  the   two  regimes  nicely
illustrates complications  in constructing alignment models,  which to
some extent are not present on larger scales or are encapsulated in an
effective  description  on  smaller   scales.   These  include  mildly
nonlinear  structure formation,  which is  in principle  accessible by
perturbation theory,  and effects  which shape  intrinsic correlations
such as clustering  and peculiar motion effects, up  to alignment with
nonlinear structures on small scales.

These  issues, as  well as  the idealising  assumptions for  the tidal
interactions themselves,  suggest that improvement can  be expected by
numerical  simulations  of  structure  formation, which  are  able  to
address nonlinear structure formation on  small scales as well as halo
formation, and depending on the simulation, other effects that have an
influence on  the shape and  amplitude of shape correlations  on small
scales.

Specifically,   simulations  fall   into   two  categories:   $N$-body
simulations (see \Cref{sec:Nbody}), where  only the collisionless dark
matter  component is  simulated, aim  at the  statistics of  the tidal
fields  and the  orientation of  dark matter  haloes in  those fields,
ultimately down to scales where  direct interaction between haloes and
the  dynamics   of  tidal  interaction  plays   a  role.   Contrarily,
hydrodynamic  simulations  (see   \Cref{sec:Hydro})  answer  questions
related to the shape of the luminous component of galaxies in relation
to the host  halo properties and the influence  of baryonic components
onto the shape of a galaxy.


\section{$N$-Body simulations}
\label{sec:Nbody}

On  the  largest  scales,  the   evolution  of  the  Universe  can  be
investigated analytically  and with Gaussian random  fields.  However,
on  smaller   scales,  the  Universe  can   be  distinctly  non-linear
(structures would  not collapse without these  non-linear effects) and
it is not possible to probe  the formation and evolution of non-linear
structures analytically, which  is why an alternative  approach had to
be found.   $N$-body simulations  mimic the statistical  properties of
the Universe by  sampling the dark matter density  field with discrete
particles.  These  particles are placed in  a box and since  it is not
computationally possible to represent every  atom in the Universe with
its own particle, each particle represents a sampling of the volume at
that  particular point  in space.   The particles  are generally  very
massive, adding together to make the total density of the volume equal
to  the  average  density  of  the  Universe.   For  example,  a  high
resolution,  cosmological  volume ($\mathtt{\gtrsim}  100\:h^{-1}$Mpc)
simulation    today     might    have     a    particle     mass    of
$\mathtt{\sim}10^6\:h^{-1}\msun$.  This  distribution of  particles is
imprinted with a theoretical power  spectrum of fluctuations from when
the Universe  was still in  a linear  state.  As the  simulation moves
forward through time, the dark  matter particles act under gravity and
evolve over cosmological time.  With  an appropriate choice of initial
condition  power  spectrum  and   subsequent  dark  matter  evolution,
cosmological  $N$-body  simulations  are   able  to  produce  particle
distributions that resemble the  statistical distribution of matter in
our Universe today.

$N$-body  simulations  are used  for  alignment  studies because  they
provide  a data  set with  known parameters,  which is  essential when
trying to understand  the alignment signals.  They can  also provide a
statistical sample --  large numbers of individual  dark matter haloes
can be identified  in a single realization.  There  are many different
measurements  made  in order  to  quantify  the alignment  signal  and
sometimes  many different  ways of  measuring the  same property,  and
these   allow    for   robustness   tests   of    the   results   (see
\Cref{subsec:measurements}).   Most importantly,  $N$-body simulations
are able to provide data on  how the shapes, spins and angular momenta
of  dark matter  haloes  align over  a  wide range  of  scales and  in
different environments (see \Cref{sec:DMResults}).

Another possibility is to investigate a smaller volume simulation with
a much higher resolution. This is  known as a `zoom' simulation, where
a region of interest is  selected within a larger-scale simulation and
then  resimulated  with higher  resolution.   The  advantages of  this
technique are that  the large-scale tidal forces are  still present to
influence the region  of interest, but the higher  resolution allows a
much more detailed study to be performed.

$N$-body simulations alone  can not provide a complete  picture of the
formation  and evolution  of  intrinsic galaxy  alignments.  The  most
important limitation is  that they do not contain any  galaxies. It is
also  now   known  that  baryons   have  an  impact  on   the  shapes,
orientations,  spins  and  angular  momenta  of  dark  matter  haloes.
Investigations  of these  effects must  be studied  using hydrodynamic
simulations  (see   \Cref{sec:Hydro}).   Despite   their  limitations,
$N$-body   simulations  will   remain   an  important   tool  in   the
understanding of  intrinsic alignments.  They  are much faster  to run
than their  hydrodynamic counterparts and  the dark matter  physics is
well  understood  (unlike  baryon physics).   Most  importantly,  they
provide  the  backbone  for  the  semi-analytic  modelling  of  galaxy
properties (see \Cref{sec:SemiAnalytic}), which will likely become the
standard data  sets for the galaxy  intrinsic alignment investigations
of the future.

\subsection{What do people measure?}\label{subsec:measurements} 

\begin{table*}[t]
  \resizebox{\textwidth}{!}{
      \begin{tabular}{llcccl}

        \hline\\

        \textbf{Shape of}  & \textbf{Aligned with} & \textbf{Symbol} & \textbf{Figure(s)} &
        \textbf{Equation(s)} & \textbf{Section(s) or Paper}\\

        \hline \hline

        Halo  & Shape  of itself  at different  radius &  $\theta_{\rm
          H}({\rm                     R_1,R_2})$                     &
        \ref{fig:sketch_intra},\ref{fig:haloalignments},\ref{fig:spinalign},\ref{fig:misalign}
        &             \labelcref{eq:epsilon_shape_shape}             &
        \Cref{sec:Internal,sec:SHinternal} \\

        \hline

        Halo & Halo Shape & $\Theta_{\rm HH}$ & \ref{fig:sketch_2halo}
        & \labelcref{eq:epsilon_shape_shape},\labelcref{eq:haloalign}
        & \Cref{sec:HaloHalo}\\

        \hline

        Halo    &    Halo    Position   &    $\Theta_{\rm    Hh}$    &
        \ref{fig:sketch_2halo}
        & \labelcref{eq:epsilon_shape_shape},\labelcref{eq:axisalign}
        & \Cref{sec:LSS,sec:HaloHalo}
        \\

        \hline

        Halo   &  Sheet/Wall  Shape   &   $\theta_{\rm  HW}$   &
        \ref{fig:sketch_void}  & \labelcref{eq:epsilon_shape_shape}  &
        \Cref{sec:LSS} \\

        \hline

        Halo  &  Void  Position  of   Centre  &  $\theta_{\rm  Hv}$  &
        \ref{fig:sketch_void}  & \labelcref{eq:epsilon_shape_shape}
        & \Cref{sec:LSS} \\

        \hline

        Halo    &   Filament    Shape    &    $\theta_{\rm   HF}$    &
        \ref{fig:sketch_filament} & \labelcref{eq:epsilon_shape_shape}
        & \Cref{sec:LSS} \\

        \hline

        Central    &    Halo    Shape   &    $\theta_{\rm    CH}$    &
        \ref{fig:sketch_2halo},\ref{fig:Romano}
        &  \labelcref{eq:epsilon_shape_shape} &
        \Cref{sec:SHinternal,sec:LHshape,sec:SemiAnalytic}\\

        \hline

        Central   &    Central   Shape    &   $\Theta_{\rm    CC}$   &
        \ref{fig:sketch_2halo} &  \labelcref{eq:epsilon_shape_shape} &
        \citet{paper3} \\

        \hline

        Central   &   Central   Position   &   $\Theta_{\rm   Cc}$   &
        \ref{fig:sketch_2halo} & \labelcref{eq:epsilon_shape_shape}
        & \Cref{sec:LargeHydro} \\

        \hline

        Central  & Satellite  Distribution Shape  & $\Theta_{\rm CB}$  &
        \ref{fig:sketch_2halo}                                       &
        \labelcref{eq:epsilon_shape_shape} &
        \Cref{sec:LHshape} \\

         \hline

         Satellite &  Halo Position  of Centre  & $\theta_{\rm  Sh}$ &
         \ref{fig:sketch_satellites},    \ref{fig:satellitealign}    &
         \labelcref{eq:epsilon_shape_shape}                          &
         \Cref{sec:Satellite,sec:SHsatellite} \\

        \hline

        Satellite    &   Halo    Shape   &    $\theta_{\rm   SH}$    &
        \ref{fig:sketch_satellites}$^{*}$                            &
        \labelcref{eq:epsilon_shape_shape}                           &
        \Cref{sec:HaloHalo,sec:LSS} \\

        \hline

        Satellite & Central Position & $\theta_{\rm Sc}$ &
        \ref{fig:sketch_satellites}$^*$ &
        \labelcref{eq:epsilon_shape_shape} & \citet{paper3} \\

        \hline

        Satellite    &    Central    Shape    &    $\theta_{\rm CS}$    &
        \ref{fig:sketch_satellites}                                  &
        \labelcref{eq:epsilon_shape_shape} & \citet{paper3} \\

        \hline

        Satellite    &   Satellite    Shape    &   $\theta_{\rm SS}$    &
        \ref{fig:sketch_satellites}                                  &
        \labelcref{eq:epsilon_shape_shape} & \citet{paper3} \\

        \hline

        Satellite Distribution & Halo Shape & $\theta_{\rm BH}$ &
        \ref{fig:sketch_satellites}                                  &
        \labelcref{eq:epsilon_shape_shape}   &   \Cref{sec:Satellite,sec:SHsatellite} \\

        \hline

        Satellite  Distribution &  Central Shape  & $\theta_{\rm BC}$  &
        \ref{fig:sketch_satellites}&
        \labelcref{eq:epsilon_shape_shape}
        &   \Cref{sec:Satellite,sec:SHsatellite}
        \\

        \hline

         Satellite  Distribution  &  Satellite  Distribution  Shape  &
         $\Theta_{\rm      BB}$       &      \ref{fig:sketch_2halo}$^*$ &
         \labelcref{eq:epsilon_shape_shape}
         & \citet{paper3}\\

         \hline

        Satellite  Distribution  &  Satellite  Distribution  Position  &
        $\Theta_{\rm       Bb}$       &       \ref{fig:sketch_2halo}$^*$ &
        \labelcref{eq:epsilon_shape_shape} & \citet{paper3} \\

       \hline

        Galaxy    &    Halo    Shape    &    $\theta_{\rm    GH}$    &
        \ref{fig:sketch_satellites},         \ref{F:mgiifig}         &
        \labelcref{eq:epsilon_shape_shape} & \Cref{sec:LHshape}\\

        \hline

        Galaxy    &   Wall    Shape    &    $\theta_{\rm   GW}$    &
        \ref{fig:sketch_void}$^*$  & \labelcref{eq:epsilon_shape_shape}  &
        \Cref{sec:SHLSS} \\

        \hline

        Galaxy  &   Void  Position   &  $\theta_{\rm Gv}$  &
        \ref{fig:sketch_void}$^*$  & \labelcref{eq:epsilon_shape_shape}
        & \citet{paper3} \\

        \hline

        Galaxy   &   Filament   Shape    &   $\theta_{\rm   GF}$   &
        \ref{fig:sketch_filament}$^*$ & \labelcref{eq:epsilon_shape_shape}
        & \Cref{sec:SHLSS} \\

         \hline

      \end{tabular}
}
  \caption{Intrinsic alignment  observable misalignment  angles within
    one halo,  $\theta$, and  across two  haloes, $\Theta$,  for shape
    (upper  case letter)  and  position (lower  case  letter) of  dark
    matter haloes (H), central galaxies (C) satellites (or dark matter
    subhaloes; S)  the shape of  the satellite distribution  (which is
    defined by the  satellite positions; B), a  complete galaxy sample
    (includes centrals  and satellites;  G), walls (W),  filaments (F)
    and  the position  of the  centre of  a void  (v).  The  first and
    second of these indices denote the  object in the first and second
    columns  of the  table respectively.   Note  that not  all of  the
    alignments       listed       above       are       shown       in
    \Cref{fig:sketch_intra,fig:sketch_halo,fig:sketch_LSS}.   However,
    if the  alignment listed  is not explicitly  shown, the  figure is
    denoted with a $^*$ and it  should be straightforward to infer the
    alignment from the named figure.}

  \label{t:IAObservables}

\end{table*}

There  are a  number of  things that  must be  determined in  order to
measure the alignments of dark  matter haloes in $N$-body simulations.
The  first  step is  to  identify  the  haloes within  the  simulation
(\Cref{sec:haloes}). It is then possible  to determine the halo shapes
(\Cref{sec:shapes})  and the  angular  momentum or  spin  of the  halo
(\Cref{sec:momentum}).   Some  studies   investigate  alignments  with
respect to  the environment that  the halo resides in,  which requires
the   simulation   to  be   classified   into   cosmic  web   elements
(\Cref{sec:web}).  With  this information in  hand, it is  possible to
measure    the    alignments    of     the    dark    matter    haloes
(\Cref{sec:alignments}).   \Cref{t:IAObservables} provides  a list  of
possible  alignments,  indicating   where  individual  alignments  are
discussed in more detail either in this  review or if they can only be
found in  \citep{paper3}.  While this  section is focused  on $N$-body
simulations, these  measurements outlined  below are largely  the same
for hydrodynamic simulations (see \Cref{sec:Hydro}), with the shape of
the stellar and/or gas component also being measured.

\subsubsection{Dark matter haloes}
\label{sec:haloes}
In order  to study the shapes  and alignments of dark  matter halos in
$N$-body simulations, it is useful to provide a formal definition of a
halo  and  some  typical  methods  used  to  identify  them.   From  a
theoretical  perspective,  a  halo  is  defined  as  an  object  whose
constituent particles are gravitationally bound.  However, this can be
difficult to define in practice, given that simulations are discretely
sampled by particles, which trace  the local density and gravitational
potential.

A simple model  for the formation of haloes is  the spherical collapse
model  \citep[e.g.][]{PS74},  which  describes   the  formation  of  a
collapsed object by  considering the evolution of a  sphere of uniform
overdensity,  $\delta$,  in  a  smooth  background.   The  overdensity
initially expands with  the expansion of the Universe;  however, as it
is overdense, the  expansion slows, and eventually  begins to collapse
under its  own self-gravity.  The  overdensity required for a  halo to
collapse is related to the  underlying matter density.  After a period
of relaxation, the halo ultimately  reaches virial equilibrium.  For a
spatially flat universe  today, the overdensity of  a fully collapsed,
virialised  halo   (relative  to   the  critical  density)   is  given
by \footnote{Note that some studies consider overdensities relative to
  the  underlying matter  density, rather  than the  critical density.
  Such  overdensities can  be  easily  scaled, due  to  the fact  that
  $\Omega_M$  is the  ratio between  the mean  matter density  and the
  critical density.} $\Delta_c \simeq 178$,  which is often rounded up
to  $\Delta_c =  200$ for  simplicity in  simulation analyses.   It is
important to  note that the value  for $\Delta_c$ is dependent  on the
cosmology and  redshift of the  universe (at earlier  times $\Delta_c$
will be  a smaller  number).  In $N$-body  simulations, and  indeed in
weak lensing mass measurements, the virial  mass of a halo, $M_{vir} =
M_{\Delta_c}$, is typically computed within the virial radius, $R_{\rm
  vir} =  R_{\Delta_c}$, which is the  radius at which the  density of
the halo is $\Delta_c$ times the critical density of the universe.

The simplest method  to identify particles that have  coalesced into a
collapsed halo  is to simply  group particles together based  on their
proximity  to  neighbouring particles.   Such  a  scheme is  known  as
``Friends-of-Friends"  (FOF)  \citep[e.g.][]{DEF+85}.  A  typical  FOF
halo finder  will identify  all particles that  are separated  by less
than a  user-specified \textit{linking  length}, and will  group these
together  as   a  halo.    This  type   of  algorithm   is  relatively
straightforward to implement,  and can be run during  the evolution of
an  $N$-body  simulation, allowing  for  a  simple  way to  trace  the
evolution  of  structures  in  such simulations.   However,  the  mass
associated with a  FOF halo, computed by simply summing  the masses of
the  particles  identified  as  belonging to  the  halo,  is  strongly
dependent on  the choice of linking  length, and it is  not trivial to
relate this mass  to something more physically motivated,  such as the
virial mass.

An   alternative  approach   to  halo   finding  is   the  ``Spherical
Overdensity" (SO)  method \citep{LC94}.  The basic  principle involves
identifying local  density maxima, and identifying  a spherical region
around each maximum,  within which the mean overdensity  is above some
pre-defined level. The  link between this, and  the spherical collapse
model, should be immediately evident.  Many different variants of this
general  scheme have  been  adopted  in the  literature,  such as  the
``Bound  Density Maximum"  (BDM)  technique \citep[e.g.][]{KH97}.   In
contrast to FOF halos, those identified  with SO or BDM techniques can
overlap, and it is therefore straightforward with these latter methods
to identify subhaloes  within larger structures (i.e.   halos that lie
entirely within the virial radius of a larger structure).

Most commonly used halo finders  use modifications to, or combinations
of, the above methods; reviews of contemporary methods can be found in
\cite{BEF+07} and \cite{KKS+11}.

\subsubsection{Dark matter halo shapes}
\label{sec:shapes}

Once the dark  matter haloes have been identified,  their shapes (halo
axes) must  be identified  in order to  measure any  shape alignments.
The  shapes of  these  haloes  can be  determined  by considering  the
distribution of  their constituent  particles. Dark matter  haloes are
typically  modelled  as being  ellipsoidal,  and  although this  is  a
simplification of the  more complex $N$-body halo  shape, this appears
to work very  well.  One common approach to measuring  shapes uses the
so-called  inertia  tensor\footnote{This   equation  is  actually  the
  quadrupole tensor of  the mass distribution, but as  it is regularly
  referred to as the moment of  inertia tensor in the literature, this
  is the  convention followed throughout  this review.}  for  the dark
matter particles that have been identified as being a part of the halo
(see \Cref{sec:haloes}),
\begin{equation}
 I_{ij} = \frac{\sum_{n} w_n m_n x_{ni}x_{nj}}{\sum_{n} w_{n} m_n},
\label{eq:Inertia}
\end{equation}
where $m_n$  represents the mass  of the $n^{\rm th}$  particle, $w_n$
represents  a radial  weight  function to  be  discussed shortly,  and
$x_{ni}, x_{nj}$  represent the  position coordinates of  the $n^{th}$
particle with  respect to  the halo  centre (where  the centre  is the
location of  the gravitational potential  minimum), with $ 0  \leq i,j
\leq 2$  for 3D  and $0 \leq  i,j \leq 1$  for 2D.   The eigenvectors,
$\boldsymbol{e}_1, \boldsymbol{e}_2, \boldsymbol{e}_3$, of this tensor
represent the principal axes of the ellipsoids with the lengths of the
principal  axes  given   by  the  square  roots   of  the  eigenvalues
$\lambda_i$, such  that $\sqrt{\lambda_{1}}  \ge \sqrt{\lambda_{2}}\ge
\sqrt{\lambda_{3}}$.

The use of  $w_n=1$ results in the  so-called \emph{unweighted inertia
  tensor},  whereas  $w_n=1/d_{n}^2$  (where  $d_n$  is  the  weighted
distance  relative to  the centre  of the  halo) is  the \emph{reduced
  inertia tensor}, which  gives more weight to the  centres of haloes.
The  reduced  inertia  tensor  can be  computed  using  a  spherically
symmetric weight function proportional to  the inverse distance of the
particle   from   the   halo   centre,   i.e.    defining   $d_n^2   =
x_n^2+y_n^2+z_n^2$; or, alternatively,  using an ellipsoidal weighting
function that defines  $d_n^2 = x_n^2 + (y_n/q)^2  + (z_n/s)^2$, where
$q=b/a$  and  $s=c/a$,  and  $a  \ge b  \ge  c$  are  the  semi-major,
intermediate  and  semi-minor  axes  of  the  halo.   When  using  the
ellipsoidal weighting function, the inertia tensor is typically solved
for iteratively.

When measuring  the shapes of dark  matter haloes, it is  important to
consider  the  number  of  particles  being used  to  make  the  shape
measurement.  \citet{BEF+07}  performed resolution  tests on  their 3D
shape measurements  and showed  that shapes  were biased  toward being
less spherical (more prolate) when the number of particles in the halo
was  $\mathtt{\lesssim} 300$.   Similarly,  \citet{Jing02} found  that
their 2D projected shape measurements  converged with a minimum of 160
particles. It  was not  uncommon for some  early studies  mentioned in
this review to include haloes with as few as 10 or 20 particles, which
offers a reasonable  explanation for why some early results  may be in
contrast to more recent conclusions.   Given these known biases, it is
remarkable how  many early works  found results actually  in agreement
with the current consensus (see \Cref{sec:DMResults}).

An   alternative  shape   measurement   approach   was  suggested   by
\citet{JS02},  who noted  that while  the iterative  approach detailed
above worked well for low-resolution simulations, it does not converge
consistently in the case of high  resolution data due to the iterative
approach being  unstable in the presence  of significant substructures
(which   are   more   readily    resolved   in   the   high-resolution
simulations). The alternative  approach calculates isodensity surfaces
at  different overdensities  to determine  the axes  of the  halo.  To
start,  the  local density  is  calculated  at  the position  of  each
particle.   A  spherically  symmetric  spline kernel,  often  used  in
smoothed     particle    hydrodynamics     calculations    \citep[SPH;
  e.g.][]{HK89,ML85}, is adopted,

\begin{equation}
\label{eq:kernel}
\mathcal{W}(r_s,h_i)=\frac{1}{\pi h_i^3} = \left\{
\begin{array}{ll}
\displaystyle 1-\frac{3}{2}\left(\frac{r_s}{h_i}\right)^2
+ \frac{3}{4}\left(\frac{r_s}{h_i}\right)^3
& (r_s\le h_i) \\
\displaystyle \frac{1}{4} \left(2-\frac{r_s}{h_i}\right)^3
& (h_i<r_s< 2h_i) \\
0  & {\rm otherwise},
\end{array}
\right.
\end{equation}
where $h_i$ is the smoothing length  of the $i$th particle.  The local
density of a  particle, $\rho_i$ is computed by taking  the 32 nearest
neighbour particles, where $r_s$ is distance of the neighbour particle
from the $i$th particle, and  applying their contribution to the local
density using the weighting  in \Cref{eq:kernel}. The smoothing length
of  the  particle,  $h_i$,  is  often   set  to  be  half  the  radius
$r_s$. After the  local density of the particles  has been calculated,
isodensity  surfaces can  be  determined  at different  overdensities,
which correspond  to different halo radii,  and a triaxial fit  can be
used to determine the axes.

Another alternative approach  to measuring the shape of  a dark matter
halo  is  to  determine  the  shape  of  the  potential  of  the  halo
\citep[e.g.][]{KDM07}.   The   unweighted  potential   energy  tensor,
$W_{ij}  = \sum_n  x_{ni}  \mathrm{d} \Phi  /  \mathrm{d} x_{nj}$,  is
related to the (unweighted) kinetic energy tensor,
\begin{equation}
K_{ij} = \frac{1}{2}\sum_n m v_{ni} v_{nj}\ ,
\label{eq:kinetic}
\end{equation}
where $v$ is  the velocity of the particle, through  the tensor virial
theorem,
\begin{equation}
\frac{1}{2} \frac{d^2I_{ij}}{dt^2} = 2K_{ij} + W_{ij}\ ,
\end{equation}
where $t$ is  time.  For a relaxed  dark matter halo, the  term on the
left of  this equation should be  zero at $z  = 0$. In this  case, the
eigenvectors  of the  diagonalised  unweighted  kinetic energy  tensor
(\autoref{eq:kinetic})  will   provide  the  principal  axes   of  the
potential ellipsoid.

The axes derived using any of  the shape measurement methods above can
also be used to measure the triaxiality of the halo.
\begin{equation}
  \mathcal{T} = \frac{(a^2-b^2)}{(a^2-c^2)} = \frac{(1-q^2)}{(1-s^2)}
\end{equation}
The axis ratio  $s$ is typically referred to as  the sphericity of the
halo. In general,  a halo is oblate when  $\mathcal{T} < \frac{1}{3}$,
prolate when $\mathcal{T} > \frac{2}{3}$ and triaxial otherwise.

\subsubsection{Angular momentum and spin}
\label{sec:momentum}

Tidal torque theory  presents a framework for how  angular momentum is
built    up    in   galaxies    and    dark    matter   haloes    (see
\Cref{sec:tidal_alignments}).   Initial  angular momentum  predictions
can be  made analytically \citep[e.g.][]{Schaefer08},  while structure
formation is  in the linear regime.   At the time of  halo turnaround,
when a halo  begins to collapse and the processes  acting on it become
non-linear, tidal  torquing leaves  an imprint of  the tidal  field in
which the halo formed.  This is  the point where analytic models break
down and the continued build up of angular momentum, through processes
including mergers  and accretion,  is best traced  through simulations
(both  $N$-body and  hydrodynamic).  The  imprint of  the early  tidal
field is  expected to remain  present through these processes,  as has
been  shown   using  $N$-body  simulations  in   \citet{PDH02a}.   The
structures in the cosmic web  are physical manifestations of the tidal
field  and  these cosmic  web  elements  have reasonably  uniform  and
symmetric   morphologies   (filaments   in   particular),   which   on
large-scales present a uniform  tidal field \citep[i.e.  haloes within
  a filament would all experience similar tidal forces;][]{TLB13}.  As
a result, the  angular momentum vectors of  neighbouring haloes within
cosmic web elements should show  some alignments at small separations.
Moreover, the angular momentum of the  haloes should also align in the
direction  of the  overdensities  in  the large-scale-structure  (e.g.
along the semi-major axis  of a filament or in the  plane of a sheet).
However,  there  are a  number  of  influences  that can  cause  these
alignments  to change  over  time  and this  is  discussed further  in
\Cref{sec:Internal}.

Similar to the  shape measurements, the number of particles  in a halo
has an  affect on the  measurement of  the angular momentum  and spin.
\citet{BEF+07} performed  resolution tests on their  spin measurements
and found  that the  median spin  parameter was  biased high  when the
number of particles  in the halo was  $\mathtt{\lesssim} 300$. Caution
should also  be exercised  on results  from the  spin/angular momentum
literature that use fewer than 300 particles per halo.

The angular momentum of an $N$-body dark matter halo can be computed as:
\begin{equation}
\boldsymbol{J} = \sum_n m_n \boldsymbol{R}_n \times \boldsymbol{v}_n\
,
\label{eq:momentum}
\end{equation}
where the sum  is over the total number of  particles, $N$, identified
within a particular halo, $\boldsymbol{R}_n$ is the radial distance of
the  $n^{\rm   th}$  particle   relative  to   the  halo   centre  and
$\boldsymbol{v}_n$  is  the  velocity  of the  $n^{\rm  th}$  particle
relative to the  halo centre. Similarly, one can  compute the specific
angular momentum
\begin{equation}
\boldsymbol{j} = \frac{1}{N} \sum_n
\boldsymbol{R}_n \times \boldsymbol{v}_n \ ,
\end{equation}
and the cumulative  specific angular momentum profile  within the halo
radius $R$,
\begin{equation}
\boldsymbol{j}(\le  R)   =  \frac{1} {M(\le  R)}   \sum_{\le  R}   m_n
\boldsymbol{R}_n\times  \boldsymbol{v}_n.
\label{eq:specific}
\end{equation}

The  spin  parameter is  a  dimensionless  measure  of the  amount  of
rotation in  a dark matter  halo.  Under the standard  definition from
\citet{Peebles69}, the spin parameter is given by
\begin{equation}
\lambda = \frac{J |E|^{1/2}} {G M^{5/2}}\ ,
\label{spinparam}
\end{equation}
where $J$ is  the magnitude of the  angular momentum $\boldsymbol{J}$,
$E$ is the total energy of the halo, and $M$ is the halo mass.  Due to
the  difficulty in  measuring  the total  halo energy,  \citet{BDK+01}
introduced a modified spin parameter
\begin{equation}
\lambda^\prime = \frac{J} {\sqrt{2}MVR}\ ,
\label{modspinparam}
\end{equation}
where $V$ is the circular velocity at halo radius $R$ such that $V^2 =
GM / R$.

\subsubsection{Cosmic web elements}
\label{sec:web}

Numerical simulations and large-scale redshift surveys have provided a
picture  of   the  large-scale   structure  of  the   Universe.   This
\emph{cosmic web}  is comprised  of clusters,  filaments, sheets/walls
and voids.   Topologically, filaments are expected  join with clusters
and border sheets, which are expected to reside at the boundary of the
void surfaces.  The method to  recover the topology of the large-scale
density  field must  account for  the density  field being  discretely
sampled in  simulations (and  indeed in  observations as  well), while
still providing a consistent classification of web elements.

A number  of algorithms  exist to  define the  cosmic web  elements in
simulations,  and  three  of  these   are  based  on  the  same  basic
principle. For  each, a  locally symmetric  tensor is  diagonalised to
give three real eigenvalues, $\lambda_1  > \lambda_2 > \lambda_3$ with
corresponding eigenvectors  $\hat{e}_1$, $\hat{e}_2$  and $\hat{e}_3$.
Classification into the four different  cosmic web types, void, sheet,
filament and cluster, is related to  the number of eigenvalues above a
given threshold $\lambda_{th}$, where a common choice is $\lambda_{th}
=   0$  \citep[e.g.][]{HPC+07,HCP+07,CPD+12,TLB13}.    The  value   of
$\lambda_{th}$  is not  always zero  and  some studies  set its  value
through  visual inspection  of the  simulated cosmic  web, as  this is
thought  to  produce  a  cosmic   web  that  more  accurately  matches
observational evidence \citep[e.g.][]{LHK+12,LHF+13}.

This basic procedure to define web  elements is performed on the tidal
shear    tensor   (ignoring    the   factor    of   $4\pi    G$,   see
\Cref{sec:tidal_alignments};
e.g. \citealp{HPC+07,HCP+07,CPD+12,FCP14}),
\begin{equation}
T_{ij} = \frac{\partial^2 \Phi}{\partial x_i \partial x_j}\ ,
\label{eq:sheartensor}
\end{equation}
the velocity  shear tensor,  where the Hubble  constant $H_0  = 100h\:
{\rm      km}\:       \:{\rm      s}^{-1}\:       {\rm      Mpc}^{-1}$
\citep[e.g.][]{LHF+13,FCP14},
\begin{equation}
\Sigma_{ij} = \frac{1}{2H_0}\left(\frac{\partial v_i}{\partial x_j} +
  \frac{\partial v_j}{\partial x_i}\right)\ ,
\label{eq:velocitytensor}
\end{equation}
and the  field defined by the  Hessian of the smoothed  density field,
$\mathcal{S}(\rho)$ \citep[e.g.][]{AWJ+07,ZYF+09,TLB13},
\begin{equation}
  \mathcal{H}_{ij} =  \frac{\partial^2 \mathcal{S}(\rho)}{\partial x_i
    \partial x_j}.
\label{eq:densitytensor}
\end {equation}

For both $T_{ij}$  and $\Sigma_{ij}$, the classification  is such that
zero, one,  two or three eigenvalues  above $\lambda_{th}$ corresponds
to  void,  sheet,  filament  and cluster  respectively.   However  for
$\mathcal{H}_{ij}$, the sign of the  eigenvalues is opposite.  This is
due to  the potential  calculated in $T_{ij}$  being derived  from the
matter  density  distribution  through the  Poisson  equation,  making
$\Phi$ a Fourier transform of the density field.  Thus, zero, one, two
or  three eigenvalues  above  $\lambda_{th}$  corresponds to  cluster,
filament, sheet and void respectively.

While  the  basic  premise  of these  classification  schemes  remains
consistent within  the literature, the details  of the implementations
differ and many  works include extensions to the methods  to form more
robust  classifications.   The  reader   is  advised  to  consult  the
references   directly   for  the   fine   details   of  each   scheme.
Additionally,  many alternative  classification schemes  exist --  too
many to mention them all here. For a comprehensive list of alternative
schemes  and  the development  of  the  field, see  \citet{SCP09}  and
references therein.

\subsubsection{Alignments}
\label{sec:alignments}

The  simplest  method   to  measure  alignments  is   to  determine  a
misalignment angle, $\theta$ (or $\Theta$ between two haloes), between
the  axes   (semi-major,  intermediate,  semi-minor   or  spin/angular
momentum) of  the halo (or haloes)  such that $\cos(\theta) =  1$ is a
parallel alignment and $\cos(\theta) = 0$ is perpendicular alignment.

When   considering    the   alignments    between   two    haloes   in
three-dimensions, it is common to give each halo a 3D orientation unit
vector  $\boldsymbol{\hat{l}}$.    Again,  this  orientation   can  be
associated with the  direction of any of  the axes of the  halo or the
direction of the angular momentum/spin  vector.  The 3D unit vector in
the   direction   connecting   the    two   haloes   is   defined   as
$\boldsymbol{\hat{r}}$.   With these  definitions, it  is possible  to
determine whether  neighbouring halo  axes tend to  point in  the same
direction,
\begin{equation}
  \mathcal{A}_{ll}(r)   =  \langle   |  \boldsymbol{\hat{l}}_i   \cdot
  \boldsymbol{\hat{l}}_j | \rangle,
  \label{eq:axisalign}
\end{equation}
and if haloes tend to point in the direction of their neighbours,
\begin{equation}
  \mathcal{D}_{lr}(r)   =   \langle   |   \boldsymbol{\hat{l}}   \cdot
  \boldsymbol{\hat{r}}|        \rangle       \equiv        \frac{1}{N}
  \sum_{i,j}|\boldsymbol{\hat{l}}_i \cdot \boldsymbol{\hat{r}}_{ij}|.
\label{eq:haloalign}
\end{equation}
Both       of       these       correlations       correspond       to
$\langle|\cos(\Theta)|\rangle$,  where  $\Theta$ is  the  misalignment
angle between the two halo  axes in \Cref{eq:axisalign} or between the
halo   axis   and   the   vector  connecting   the   two   haloes   in
\Cref{eq:haloalign} \citep[e.g.][]{FGK+02,BS05,KE05}.   An alternative
is   to   investigate   the   square    of   the   dot   products   in
\Cref{eq:axisalign,eq:haloalign},   since  the   halo  axes   have  no
preferred   direction  (i.e.    $\pm   \cos(\Theta)$  are   physically
identical).       This      is       equivalent      to      measuring
$\langle|\cos^2(\Theta)|\rangle$ \citep[e.g.][]{Hopkins05}.

It  is  also common  to  measure  the shape  of  haloes  as viewed  in
projection  on  the plane  of  the  sky  to match  observations.   The
projected  shape of  a  halo is  typically defined  by  the moment  of
inertia  tensor  (\autoref{eq:Inertia}  and surrounding  text),  in  a
method that is analogous to  calculations of galaxy surface brightness
distributions \citep[e.g.][]{M91}.  An ellipticity can then be defined
as  $\epsilon =  \epsilon_1 +  \mathrm{i}\epsilon_2$, with  components
given by
\begin{equation}
  \begin{aligned}
    \epsilon_1 = \frac{I_{11}  - I_{22}}{I_{11}+I_{22}} \\ \epsilon_2
    = \frac{2I_{12}}{I_{11}+I_{22}},
  \end{aligned}
\end{equation}
such that
\begin{equation}
\epsilon_i = \frac{1-q^2}{1+q^2}\{\cos(2 \varphi),\sin(2 \varphi)\},
\end{equation}
where $q$ is the axis ratio and $\varphi$ is the position angle of the
halo.  From these ellipticity measurements, an ellipticity correlation
function can  be constructed that is  analogous to \Cref{eq:axisalign}
(see also \autoref{eq:epsilon_shape_shape}),
\begin{equation}
  \xi_{\epsilon\epsilon}(r)          =         \left\langle          |
  \epsilon_i(\boldsymbol{x})    \cdot   \epsilon_j(\boldsymbol{x}    +
  \boldsymbol{r})   |   \right\rangle    \equiv   \langle   \epsilon^S
  \epsilon^{\prime S}\rangle\ ,
\label{eq:2Dalign}
\end{equation}
computed  over  all  pairs  of  haloes with  a  3D  separation  vector
connecting    the    two    haloes   defined    as    $\boldsymbol{r}$
\citep[e.g.][]{CM00,HRH2000,Jing02}.

Comparing with the correlations  in \Cref{eq:iadef}, the alignments in
\Cref{eq:axisalign,eq:2Dalign} are  equivalent to the II  term and the
shape-density alignments in \Cref{eq:haloalign}  are equivalent to the
$\delta {\rm  I}$ term, which gives  rise to the GI  correlations when
also accounting for gravitational shear of the luminous galaxies.

\subsection{Results}
\label{sec:DMResults}

In  this  section,  the  results  from  the  $N$-body  literature  are
presented by alignment type.  \Cref{sec:spins} starts with the general
shapes  and spins  of dark  matter  haloes.  The  section then  treats
alignments as  a function  of their scale,  starting with  dark matter
halo  internal   alignments  in  \Cref{sec:Internal}  and   moving  to
halo-satellite  alignments  in  \Cref{sec:Satellite},  then  halo-halo
alignments  in  \Cref{sec:HaloHalo}  before  finally  presenting  dark
matter   halo   alignments   with   the   large-scale   structure   in
\Cref{sec:LSS}.

\subsubsection{Halo shapes and spins}
\label{sec:spins}

There is  consensus among many  $N$-body simulation studies  that dark
matter haloes  tend to be  strongly aspherical, with a  preference for
prolate                over               oblate                haloes
\citep[e.g.][]{DEF+85,FWD+88,WQS+92,KE05,SWO+06,AFP+06,SFC12}.    When
considering  a  full sample  of  haloes,  independent of  environment,
haloes   tend   to   become   more  prolate   with   increasing   mass
\citep[e.g.][]{KE05,PLP+06,HCP+07,SFC12}.   Dark  matter  haloes  also
tend to  collapse to  a shape  that is more  triaxial and  become more
spherical  over  time  \citep{Hopkins05}.   There  appears  to  be  no
significant dependence of shape  parameters on environment for massive
haloes  ($M_{vir}>2\times10^{12}\:h^{-1}\msun$), although  environment
does  play a  significant role  for less  massive haloes  in clusters,
which tend  to be  less spherical  and more  prolate, while  haloes in
filaments tend to  be more oblate \citep{HPC+07}.   In addition, there
is some evidence to suggest that  haloes that reside close to clusters
and haloes  with little substructure are  more spherical \citep{RP07}.
Virialised (relaxed) haloes  are more spherical than  haloes that have
experienced a  lot of mergers  and contain substructure  (which causes
them to  be more  elongated) and haloes  in high  density environments
like clusters  will have formed  earlier and  had more time  to relax,
resulting in their more spherical nature \citep{RP07}.

The distribution  of halo spin parameters  tends to be described  as a
lognormal                                                 distribution
\citep[e.g.][]{DEF+85,WQS+92,CL96,BS05,SWO+06,DN09},
\begin{equation}
p(\log\lambda) = \frac{1}{\lambda\sqrt{2\pi}\sigma}\exp
\left[-\frac{\log^2(\lambda/\lambda_0)}{2\sigma^2}\right]\ ,
\end{equation}
where $\lambda_0$  is the median  spin parameter (the location  of the
peak),    and   $\sigma$    is    the    scatter   in    $\log\lambda$
\citep{DEF+85,BDK+01,MDS02,BS05,ACG+05}.   This distribution  tends to
be rather  insensitive to the  choice of cosmological  parameters, and
remains essentially unchanged whether  the spin parameter includes all
the mass out to the virial radius, or only the mass within a truncated
radius  e.g.    $0.12R_{\rm  vir}$  \citep{BKG+05}.    The  log-normal
distribution  of  spins  can  be   modelled  equally  well  by  linear
tidal-torque theory applied to shells of collapsing material or with a
model based on the transfer of the orbital angular momentum of merging
satellites   to  the   internal  spin   of  the   halo  \citep{MDS02}.
Furthermore, \citet{VKK+02} found that a random walk model to build up
angular momentum  naturally produced  a lognormal distribution  of the
spin parameter, though they only  tracked the major progenitor of each
halo  during its  evolution, rather  than all  the mass  in the  halo.
\citet{HPC+07}  showed  that this  lognormal  model  did not  fit  the
distribution  for  spin  parameters  of   $\lambda  >  0.1$  in  their
simulations, while \citet{BEF+07} found  the lognormal distribution to
be   generally  a   poor  fit   to  the   spin  distribution   of  the
$\mathtt{>}10^6$ haloes in their  catalogue. They suggested that while
this fit  was sufficient for  low numbers  of objects, it  avoided the
very low spin values that were present in the catalogue.

\subsubsection{Internal alignments}
\label{sec:Internal}

Central galaxies form in the inner region of a dark matter halo. It is
therefore instructive  to consider  the shape  and orientation  of the
halo at  the radius of interest  to the galaxy  as well as a  range of
larger radii (and not simply at the virial radius).

The shape  of a  dark matter halo  changes with  radius \citep[e.g.][]
{FWD+88,DC91,WQS+92,JS02,BS05,AFP+06,KDM07,FLW+09,SFC12}.   While some
early works  found that the dark  matter haloes are more  spherical in
the centre \citep[e.g.][]{FWD+88}, more recent studies agree that dark
matter haloes become more spherical with increasing radius.  There is,
however,  some disagreement  on  how the  \emph{alignment}  of a  halo
changes   with   radius;    i.e.    $\theta_{\rm   H}(R_1,R_2)$,   see
\Cref{fig:sketch_intra} and \Cref{t:IAObservables}.

\begin{figure*}[t!]
  \centering
    \includegraphics[width=0.99\textwidth]{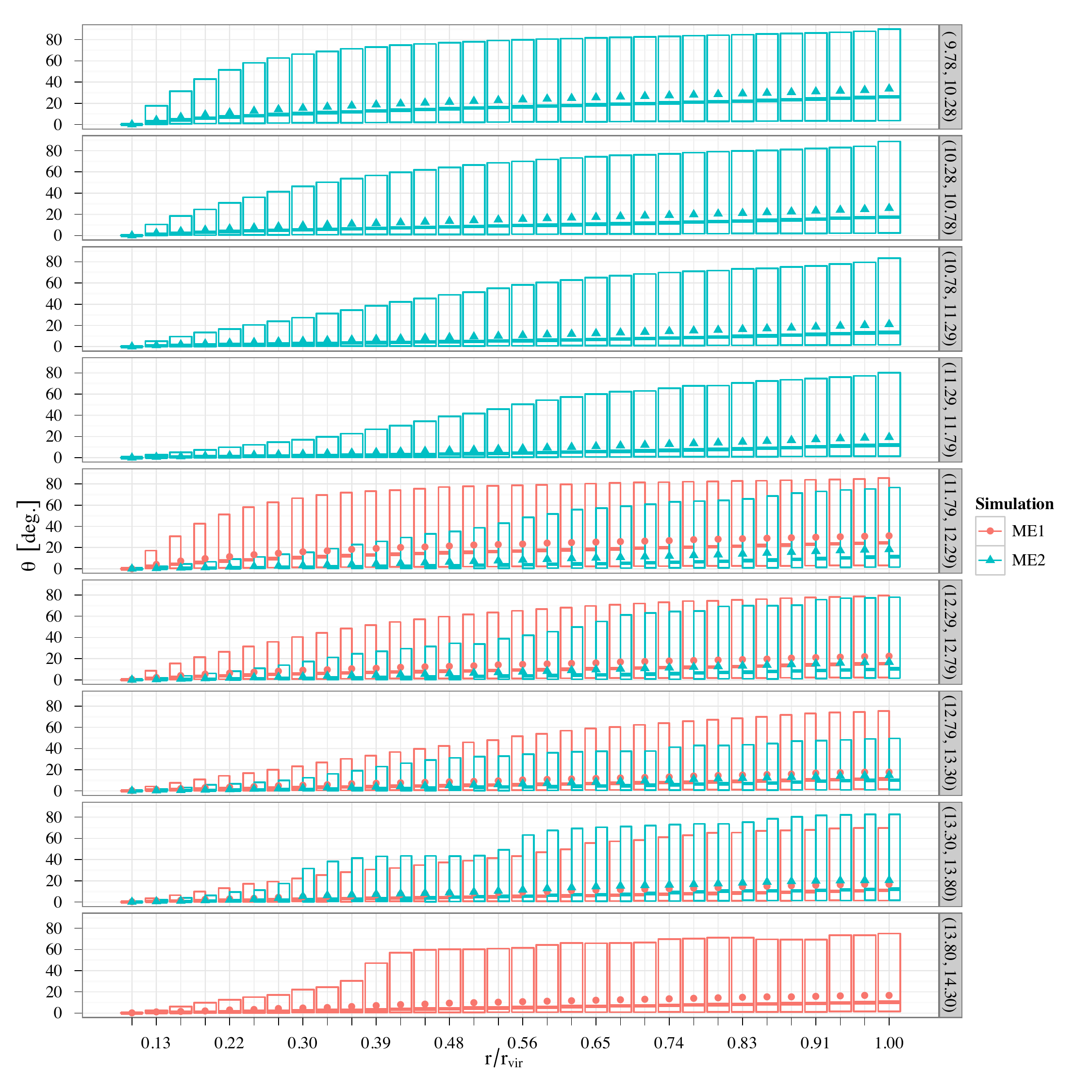}
    \caption{Angle between  the major  axes at different  radii within
      individual     haloes,     $\theta_{\rm    H}(R_1,R_2)$     (see
      \Cref{fig:sketch_intra}  and  \Cref{t:IAObservables}), at  $z  =
      0.5$.  The  lower resolution simulation, Millennium  1, is shown
      in red, while the higher resolution simulation, Millennium 2, is
      shown in blue.   Points indicate mean angles at  a given radius,
      while boxes show  the $25\% - 75\%$ central  quartile range with
      the  median of  the distribution  denoted by  a horizontal  line
      within each box.  Each panel  shows a different dark matter halo
      mass range;  this is  shown in  the grey box  on the  right with
      units of  $\log_{10}(M_{200}/h^{-1}\msun)$.  The halo  radius on
      the x-axis is normalized by $R_{\rm vir}$, which is $R_{131}$ at
      this redshift (see \Cref{sec:haloes}). \permjcap{SFC12}}
    \label{fig:haloalignments}
\end{figure*}

For example, \citet{SFC12} investigated  dark matter haloes using both
the   Millennium    1   \citep[ME1,][]{SWJ+05}   and    Millennium   2
\citep[ME2,][]{BSW+09}   simulations.   ME1   has   a  particle   mass
resolution of $8.6  \times 10^8 \:h^{-1}\msun$ and ME2  has a particle
mass resolution a little over two  orders of magnitude higher at $6.89
\times 10^6  \: h^{-1}\msun$.  This  study covered dark  matter haloes
over a range of masses from $10^{10} - 2 \times 10^{14}\: h^{-1}\msun$
and a wide range of halo  radii ($0.1R_{\rm vir} - R_{\rm vir}$). They
used the reduced inertia tensor  with an elliptical weighting function
to measure the  shapes of the haloes at different  radii.  In addition
to  showing  that  dark  matter  haloes  became  more  spherical  with
increasing  radius,  they also  showed  that  the haloes  became  more
spherical with decreasing halo mass at  a fixed fraction of the virial
radius.   \Cref{fig:haloalignments} shows  the internal  alignments of
individual dark matter haloes as a function of radius and halo mass at
a redshift  of $z  = 0.5$,  which corresponds to  an $R_{\rm  vir}$ of
$R_{131}$  (see \Cref{sec:haloes}).   They found  that the  semi-major
axes  of the  dark  matter haloes  had a  mean  misalignment angle  of
$\theta_{H}(R_1,R_2) \simeq  20^{\circ}$ between  the inner  and outer
radii.   There   was  a  large   scatter  on  the   overall  alignment
distribution  and  while  the  distribution was  skewed  toward  small
misalignment angles, around 25\% of the haloes had the semi-major axis
of the  outer halo perpendicular to  the semi-major axis of  the inner
halo.  More  massive haloes tended  to have less  misalignment between
the inner and  outer radii.  The authors noted that  while the highest
halo mass  bin that  could be investigated  with both  simulations had
good agreement  in halo alignments  between the two  simulations, this
agreement decreased with  decreasing halo mass bins.   This was likely
due to substructure  contamination in ME1, since  its lower resolution
made  it difficult  to accurately  remove all  substructure particles,
likely resulting  in the shape measurements  spuriously detecting more
twisted inner and outer haloes.

\citet{KDM07} investigated a single very high particle mass resolution
(particle   mass   $\mathtt{\simeq}  20,000\:h^{-1}\msun$),   isolated
Milky-Way scale halo and also used  the reduced inertia tensor with an
elliptical weighting function  to measure the shape of  their halo, in
addition to  measuring the  shape with  the unweighted  kinetic energy
tensor  (see  \autoref{eq:kinetic}).   Both  measurements  produced  a
prolate  halo, however  the velocity  shape (from  the kinetic  energy
tensor)   was  far   more  spherical   than  the   mass  distribution.
Additionally, the  halo became  slightly more  spherical in  the outer
radii in both  cases.  The semi-major axis of the  halo was aligned to
$1^{\circ}$ at all radii.  However, as there was only one halo in this
study,  it  is  possible  that   this  is  a  statistical  fluctuation
corresponding to  one of the  (not entirely rare)  well-aligned haloes
found  in  \citet{SFC12}.   Additionally,   since  the  halo  did  not
experience any  major mergers from  $z = 1.7$,  is was not  subject to
tidal interactions, making it much more likely to be well aligned.

It  is   still  worth   remembering  that   the  mass   resolution  of
\citet{KDM07} was significantly higher  than in \citet{SFC12}, so some
of the twisting in the ME2 haloes may be a resolution effect. However,
this is unlikely  to account for all of the  twisting measured and the
general consensus in  the field follows the  results of \citet{SFC12},
that dark matter haloes may experience misalignments between the inner
and  outer halo.   This  is  further confirmed  with  the addition  of
baryons (see \Cref{sec:SHinternal}).

\begin{figure}[t]
  \begin{center}
    \includegraphics[width=0.5\hsize]{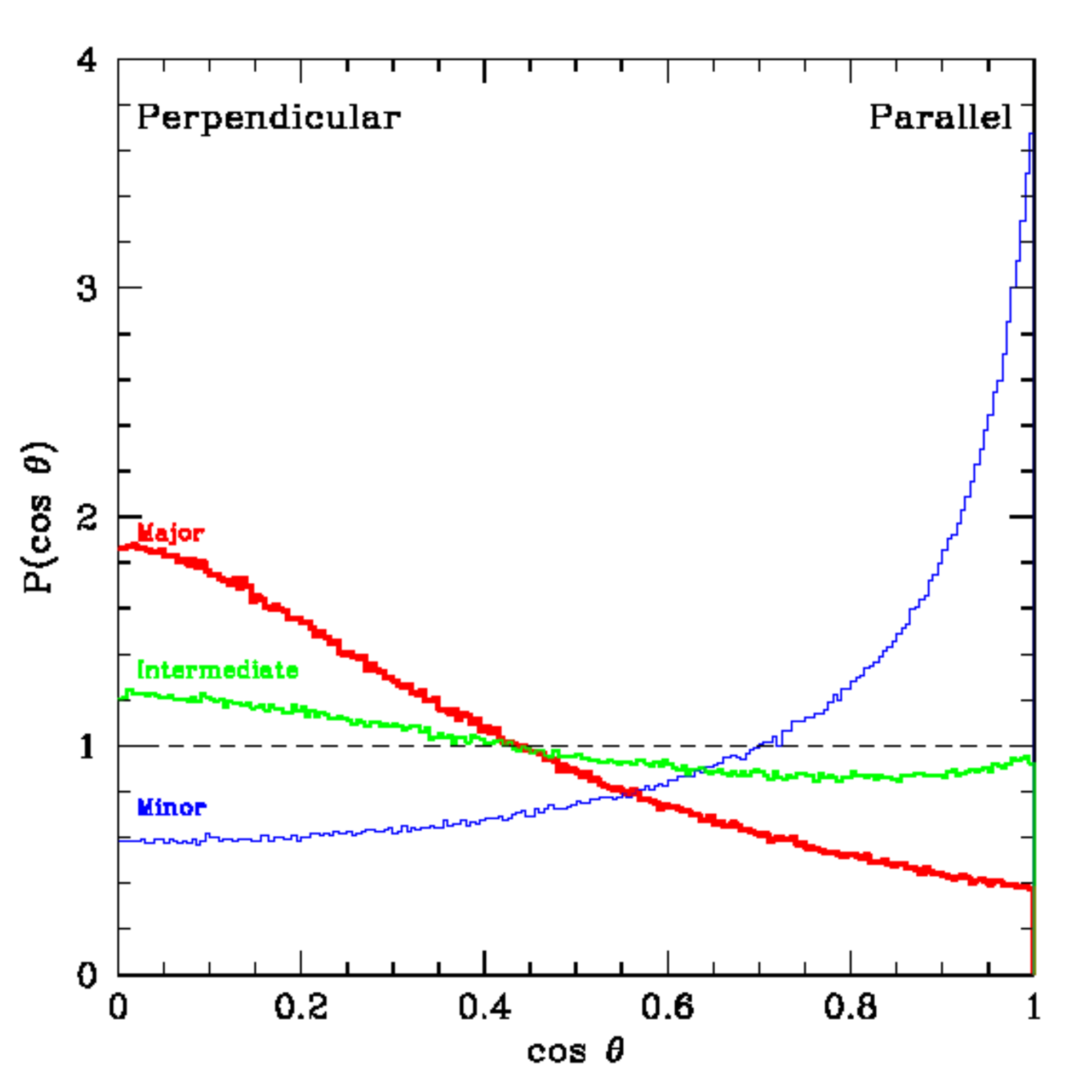}
    \caption{Normalized histograms of the cosines of the angle between
      the  angular momentum  vector and  the major  (thick red  line),
      intermediate (medium green line) and minor (thin blue line) axes
      of the dark matter haloes. A random distribution would be a flat
      line at $p\left( \cos \theta \right) = 1$. \permmn{BEF+07}}
    \label{fig:spinalign}
  \end{center}
\end{figure}

It is  common to measure  the alignment  of the halo  angular momentum
with the  halo axes,  $\theta_{\lambda_{\rm H}{\rm  H}}$.  There  is a
strong  consensus that  the  angular  momentum of  a  halo is  aligned
parallel with the semi-minor axis  and perpendicular to the semi-major
axis   of   the   halo    mass   distribution,   as   illustrated   in
\Cref{fig:spinalign}
\citep[e.g.][]{BE87,WQS+92,BDK+01,BS05,ACG+05,AFP+06,SWO+06,RP07,BEF+07,PSP08}.
\citet{BS05} found  a median misalignment of  $25^{\circ}$ between the
semi-minor axis  and the  angular momentum,  and a  stronger alignment
trend  in the  central $0.25  R_{\rm vir}$  with the  strength of  the
alignment  increasing as  a function  of  halo mass.   There was  also
evidence  to   suggest  that  haloes   containing  a  high   level  of
substructure  showed  a  stronger   alignment  between  their  angular
momentum and  semi-minor axes compared to  haloes without substructure
\citep{RP07}.  Dark  matter haloes  can gain angular  momentum through
merger  activity \citep[e.g.][]{VKK+02,MDS02},  and  haloes with  more
substructure  have  undergone  more  recent  mergers.   These  mergers
occurred   preferentially   in  the   plane   defined   by  the   halo
semi-major/intermediate  axes  (see  \Cref{sec:LSS}),  adding  to  the
angular momentum in the direction of the semi-minor axis.

\citet{BF12} investigated significant, rapid  changes to the direction
of the  angular momentum vector,  which they termed `spin  flips'. The
study focused  on Milky Way  sized dark matter  haloes ($\mathtt{\sim}
10^{12-12.5}\:h^{-1}\msun$ at  $z=0$) that  were relaxed.   They found
that these  spin flips  occurred in dark  matter haloes  regardless of
whether  they   had  experienced  a   major  merger  event   in  their
lifetime. Instead, the  spin flips were caused by  strong tidal forces
from  events  like  minor  mergers  or  even  flybys  of  neighbouring
haloes. They  showed that $10.5\%$  of their  haloes had at  least one
spin flip  of $\mathtt{>}45^{\circ}$  within a timescale  of $0.5$Gyrs
and  that $10.1\%$  of these  occurred without  a corresponding  major
merger  event.  They  also investigated  the spin  flips in  the inner
$0.25R_{\rm vir}$ and found that  $58.5\%$ of the haloes experienced a
spin  flip of  $\mathtt{>}45^{\circ}$ and  that all  but one  of these
occurred without  a major merger  event.  This result  is particularly
interesting as spin flips of the inner dark matter halo may disrupt or
morphologically  transform galactic  discs  that form  in this  region
\citep[e.g.][]{OEF+05,RSH+09,SWS+09}.

\citet{BDK+01} considered  the alignment between the  angular momentum
vectors  in  the  inner  and   outer  half-haloes  within  the  virial
radius. Between  $70\%$ and  $90\%$ of  the haloes  were aligned  to a
greater degree  than $\cos [\theta_{\lambda_{\rm H}}(R_1,R_2)]  = 0.7$
($\mathtt{\sim}45^{\circ}$),  with  most showing  significantly  lower
misalignments.  Similar  alignment results were seen  by \citet{BS05},
who considered the alignments at  several reference radii.  They found
that the alignment became progressively  worse as $R_1$ and $R_2$ were
further separated, and that the  median misalignment angle between the
angular momentum  vectors in the  innermost and outermost  regions was
$\mathtt{\sim} 50^{\circ}$.

\citet{BS05}  also  found  that  the  internal  alignment  of  angular
momentum was worse for higher mass  haloes.  This is likely related to
hierarchical structure formation; the  most massive haloes formed most
recently  and are  likely  to  have experienced  a  major merger  more
recently than smaller mass haloes, which perturbs the angular momentum
of the halo.

\subsubsection{Halo-subhalo alignments}
\label{sec:Satellite}

In  a   Universe  where  structures  form   hierarchically  \citep[see
  e.g.][]{PS74,EPS91},  it is  expected that  dark matter  haloes will
contain substructures  or satellites.   In the  hierarchical formation
scenario,  at  high  redshifts  low-mass objects  begin  to  collapse.
Through continuous accretion  of dark matter over  time, they increase
their mass. They  also undergo mergers with other haloes,  and in this
process  the smaller  halo will  lose its  loosely bound  outer layers
while the denser core survives as  a subhalo within its new host halo.
The alignment (shape,  spin and angular momentum)  and distribution of
dark matter subhaloes  within their host halo has been  the subject of
many                          $N$-body                         studies
\citep[e.g.][]{T97,KGG+04,KDM07,FJL+08,PBG08,KDP+08,KYK+08,ALB+09,WLK+14}.

One of  the alignment signals  investigated is the orientation  of the
subhalo distribution  within the shape  of the host dark  matter halo,
$\theta_{\rm     BH}$,     see    \Cref{fig:sketch_satellites}     and
\Cref{t:IAObservables}     \citep[e.g.][]{KGG+04,FJL+08,KDM07,WLK+14}.
Most  studies  used different  versions  of  the inertia  tensor  (see
\autoref{eq:Inertia} and  the surrounding text) to  determine the axes
of their host haloes.  The exception was \citet{WLK+14} who determined
the  axes  of  the  host  haloes at  given  local  mass  overdensities
following  the method  of  \citet{JS02}  (see \autoref{eq:kernel}  and
surrounding text).  Most  studies determined the shape  of the subhalo
distribution through the lines connecting  the centre of the host halo
with each  subhalo and  calculated the  alignment as  a shape-position
alignment, $\langle|\boldsymbol{l} \cdot \boldsymbol{\hat{r}}|\rangle$
\citep{KGG+04,FJL+08,WLK+14}. While \citet{KDM07}  determined the axes
of  the  subhalo  distribution  ellipsoid  through  diagonalising  the
weighted moment  of inertia tensor  constructed from the  positions of
each  subhalo,  finally  treating   the  alignment  as  a  shape-shape
alignment.  The  resulting alignments  determined whether  the subhalo
distribution ellipsoid  (or ellipse if  projected on to a  2D observer
plane; e.g.   \citealp{WLK+14}) was  random within the  host or  if it
exhibited some anisotropic signal.

Instead of calculating the alignment with the orientation angle of the
host  halo,  \citet{WLK+14} assumed  an  orientation  for an  imagined
central  galaxy, $\theta_{\rm  BC}$ (see  \Cref{fig:sketch_satellites}
and  \Cref{t:IAObservables}).  This  orientation  either followed  the
inner, intermediate  or outer  host halo orientations.   Their results
showed  that  while  the   satellite  distribution  aligned  with  the
``galaxy''  semi-major  axis  at   all  orientations,  the  alignments
increased in  strength for  orientations with the  inner, intermediate
and  outer haloes  respectively (implying  an alignment,  $\theta_{\rm
  BH}$, with  the outer  host halo  semi-major axis).   When comparing
with observations, the alignment signal was too strong when the galaxy
followed  the  outer   halo  orientation,  and  much   closer  to  the
observations when  the galaxy followed  the inner halo  orientation or
had a misalignment  drawn from a Gaussian distribution with  a mean of
$0^{\circ}$ and standard deviation of $25^{\circ}$.

There are differences  within all of these studies on  how the subhalo
population  was selected.   Additionally, a  number of  different halo
finding algorithms  were employed and  each had different  particle or
mass cutoffs.  Regardless of these  differences, the consensus is that
the orbits  of subhaloes within  a host halo are  strongly anisotropic
(not  random) and  the  semi-major axes  of  the subhalo  distribution
ellipsoids preferentially align with the  semi-major axis of the outer
host halo.  The  question now is why subhaloes  exhibit this alignment
and whether it is the result of anisotropic infall or if the subhaloes
have an isotropic infall and then experience dynamic effects that lock
in the alignment.  Given that the outer host halo may be significantly
misaligned with  the inner host halo  (see \Cref{sec:Internal}), which
would have an impact on the  alignment of the subhalo population, this
might  suggest  that  dynamical  effects  are  the  cause  of  subhalo
distribution alignments.   However, haloes merge  preferentially along
the direction of their host filament (see \Cref{sec:LSS}), which would
add an element of anisotropic infall.

The  alignment of  the subhalo  distribution with  the host  halo spin
axis,   $\theta_{\lambda_{\rm  H}{\rm   B}}$,   was  investigated   by
\citet{ALB+09}.  They found that  the subhalo distribution was aligned
perpendicular to the spin  axis.  As mentioned in \Cref{sec:Internal},
the  angular momentum  vector is  preferentially perpendicular  to the
major axis of  the host halo, indicating that this  work is consistent
with the results above.

Another quantity  that is studied is  the alignment of a  subhalo with
the  centre  of  mass  of   the  host  halo,  $\theta_{\rm  Sh}$,  see
\Cref{fig:sketch_satellites}         and        \Cref{t:IAObservables}
\citep[e.g.][]{KDM07,FJL+08,KDP+08,KYK+08,PBG08}.  The majority looked
at the 3D shapes of  the subhaloes but \citet{KYK+08} investigated the
2D projection on to the observers  plane.  The studies all agreed that
there is a strong preferential  radial alignment of subhaloes with the
centre of  mass of the  host.  \citet{FJL+08} investigated  this trend
over  a wide  range of  scales  and found,  unsurprisingly, that  this
radial  alignment was  strongest  on smaller  scales  and dropped  off
rapidly with  increasing distance  from the centre  of the  host halo.
The evolution  of this alignment showed  that while there is  a strong
preference for  a nearly radial  alignment, there are  some deviations
throughout the orbit of the subhalo around its host.

\begin{figure*}[t]
\begin{center}
\includegraphics[width=0.7\hsize]{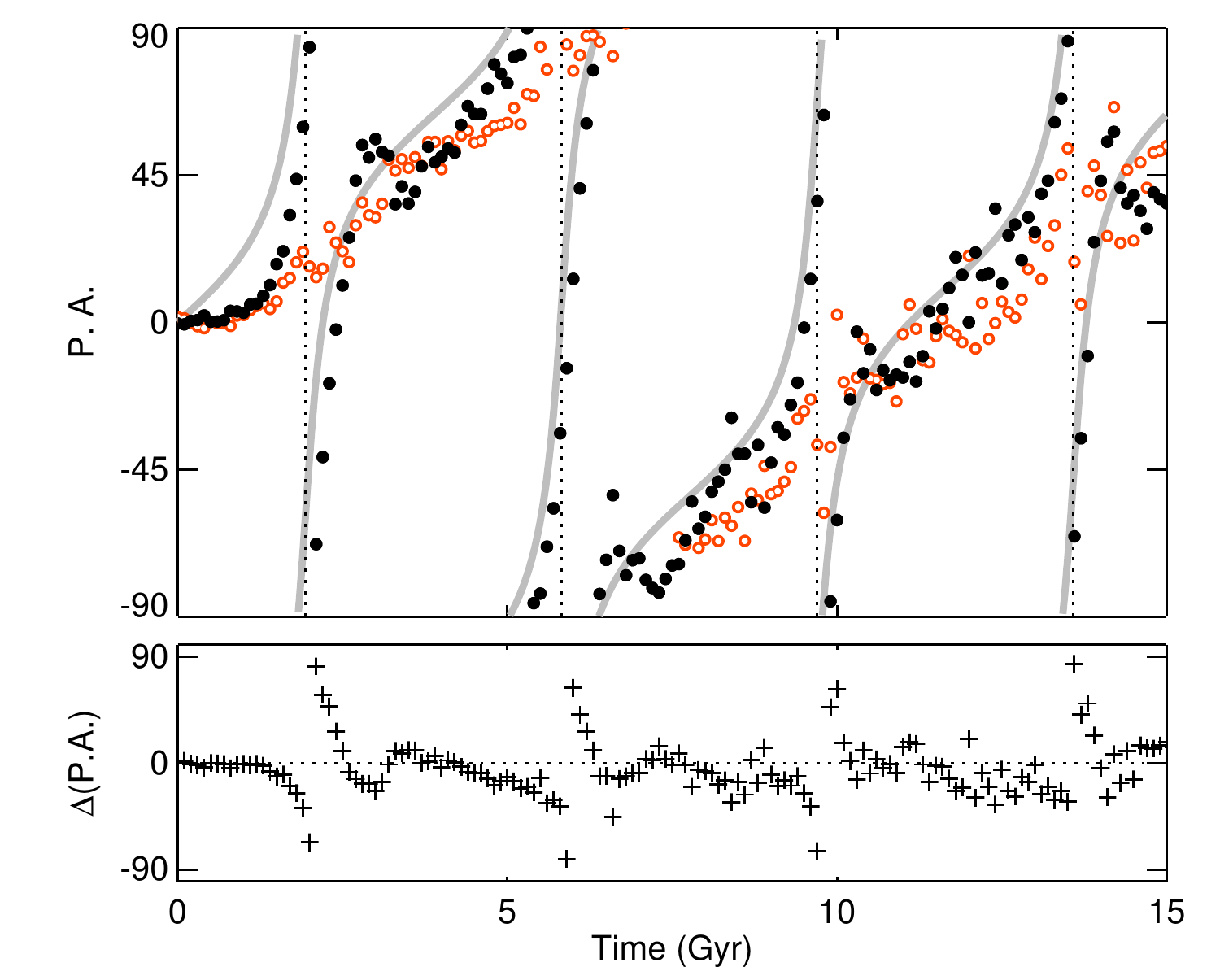}
\caption{\textit{Top:} Stellar and dark  matter position angles (P.A.)
  vs. time for  an eccentric orbit. Grey lines  indicate the direction
  to the cluster centre and dotted vertical lines represent pericentre
  passages.  While  the dark matter  particles (black dots)  appear to
  follow the orbital  motion closely, varying their  rotation speed to
  match  the  orbital velocity  changes,  the  stellar particles  (red
  circles)    appear    to    be     locked    in    uniform    figure
  rotation. \textit{Bottom:}  Misalignment between the  orientation of
  the  dark matter  and stars  $|\Delta{\rm P.A.}|$  as a  function of
  orbital  time.   The  misalignment   peaks  immediately  after  each
  pericentre  passage,  and   remains  small  for  the   rest  of  the
  orbit. \permapj{PB10}}
\label{fig:Pereira2010}
\end{center}
\end{figure*}

\cite{PB10}  took a  different  approach to  study  the alignments  of
satellite  galaxies  within  dark  matter haloes.   They  simulated  a
multi-component, $N$-body  (stars+dark matter) satellite,  orbiting an
external, analytical potential (which simulated the host cluster sized
halo), and showed that the  stellar component of the satellite reacted
significantly slower to  the tidal torque from the host  halo than the
dark  matter,  producing  a  radial twisting  of  the  satellite  that
depended on the eccentricity of the  orbit.  If the satellite was in a
circular  orbit,  both  dark  matter and  stars  were  tidally  locked
(semi-major axis  pointing toward  the centre of  mass) but  the stars
took approximately  twice as long  as the dark  matter to lock  to the
external potential.   In such  orbit, the stars  and dark  matter were
always  aligned with  each  other \textit{when  measured  at the  same
  radius}.  By contrast, in very eccentric orbits the change in torque
was too rapid  to tidally lock the stars,  which figure-rotated around
the potential,  while the  dark matter  maintained a  radial alignment
except   at   pericentre   where  a   sudden   misalignment   occurred
(\Cref{fig:Pereira2010},  top  panel).   Additionally, after  a  brief
period of  influence from  the host  halo, the  stars and  dark matter
aligned  with each  other, except  for short  periods surrounding  the
pericentre passage  (\Cref{fig:Pereira2010}, bottom  panel).  Overall,
\citet{PB10} also found  that there is a strong  preference for nearly
radial alignments  but they  also showed that  the magnitude  of these
deviations depends on the eccentricity of the orbit.

\begin{figure}[t]
\begin{center}
\includegraphics[width=0.7\hsize]{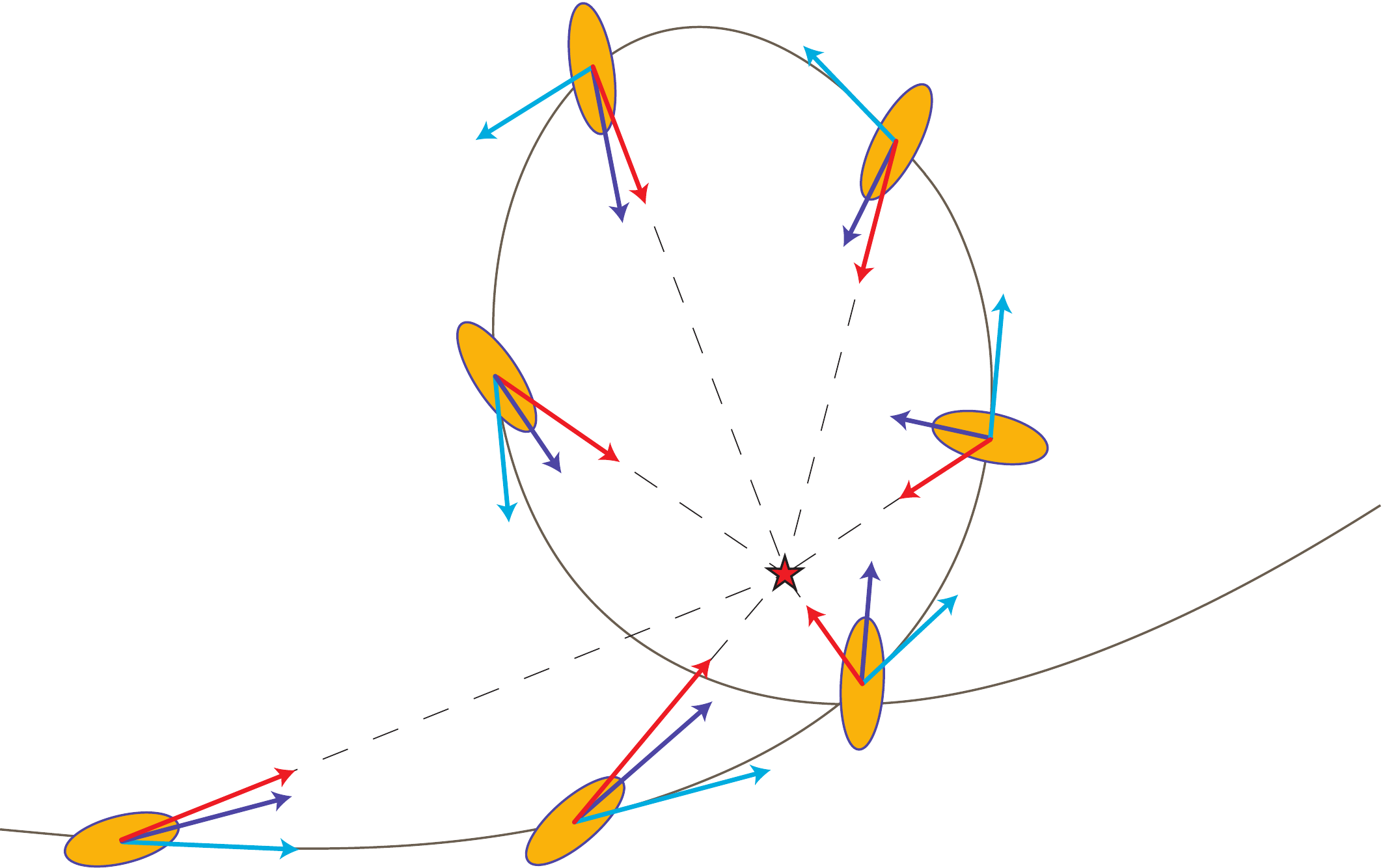}
\caption{Sketch of a subhalo in orbit  around its host.  The centre of
  mass of the host is represented  by a star.  The radial direction is
  drawn as a  red vector, the light blue vector  indicates the orbital
  direction, and  the dark blue vector  is the direction of  the major
  axis of  the subhalo. The  red and  dark blue vectors  are generally
  very close, apart from a short  mismatch at pericentre caused by the
  high orbital  velocities. The dark  blue and light blue  vectors are
  close before pericentre, but almost  orthogonal to each other in the
  second part of the orbit. \permapj{PBG08}}
\label{fig:satellitealign}
\end{center}
\end{figure}

\citet{KDM07,PBG08}  and  \citet{PB10}  used tidal  torque  theory  to
explain the motion  of a subhalo around the host;  this is illustrated
in \Cref{fig:satellitealign}.  A subhalo  is typically in an eccentric
orbit.   As it  approaches  pericentre, the  subhalo generally  points
toward the  centre of the host  halo mass because it  is being tidally
torqued  in the  direction of  the potential  gradient, which  is also
close to the direction of  motion.  As the subhalo reaches pericentre,
the torquing is less effective due to its high velocity at this point,
resulting in a lower radial  alignment signal.  The torquing continues
throughout the orbit, which keeps  the subhalo semi-major axis largely
aligned with the centre of mass  of the host.  The misalignment of the
subhalo semi-major axis with the orbital direction initially increases
before coming back into alignment after passing the apocentre.  As the
subhalo approaches pericentre again, a new cycle begins.

The  signals found  in the  simulations were  typically stronger  than
those in  observations (that  measure the  alignments of  the luminous
satellites) and \citet{KYK+08} also  attempted to explain this through
tidal torque theory.  Since the  subhaloes are subject to strong tidal
torquing  as they  pass  through apocentre,  the  loosely bound  outer
particles  will be  more highly  distorted than  the inner  particles.
Consequently,  the shape  of the  outer subhalo  could be  more easily
pulled into  alignment with the  center of mass  of the host  than the
inner  region  at  the  centre  of the  subhalo  (where  the  luminous
satellite would reside).  This would  result in the luminous satellite
and the inner region of the dark matter halo having a weaker alignment
with the centre of mass of the host than the outer subhaloes.

\subsubsection{Halo-halo alignments}
\label{sec:HaloHalo}

Determining whether dark  matter haloes have a tendency  to align with
nearby   haloes  has   been  the   focus  of   a  number   of  studies
\citep[e.g.][]{BE87,CM00,HRH2000,OT00,Jing02,BS05,Hopkins05,ACC06,SFC12}.
There are two different kinds of  alignments measured - the first is a
direct   alignment,   $\Theta_{\rm   HH}$,   that   measures   whether
neighbouring  halo axes  tend  to  point in  the  same direction;  see
\autoref{eq:axisalign} for 3D  alignments and \autoref{eq:2Dalign} for
2D alignments.   The second  alignment, $\Theta_{\rm  Hh}$, determines
whether haloes tend to point in the direction of their neighbours; see
\autoref{eq:haloalign}).     Both    alignments     are    shown    in
\Cref{fig:sketch_2halo} and  \Cref{t:IAObservables} and  are discussed
below.

Many  studies showed  that the  semi-major axes  of group  and cluster
sized haloes ($M_{\rm h} \gtrsim 10^{13}\:h^{-1}\msun$) preferentially
point    in     the    same    direction    in     a    3D    analysis
\citep[e.g.][]{BE87,FGK+02,KE05,Hopkins05}.    There    was   evidence
showing this signal to be significant  out to $30 \: h^{-1}$Mpc in the
simulations    \citep[e.g.][]{FGK+02,Hopkins05,KE05}.     \citet{BS05}
investigated  the alignments,  $\Theta_{\rm HH}$,  using predominantly
galaxy mass haloes and measured an  alignment signal that was lower in
amplitude than  the works  above, which they  attributed to  the lower
mass range of their haloes.  This  assumption was confirmed in a later
investigation by \citet{SFC12} that measured the alignment over a wide
range of halo masses from subhaloes  to clusters ($10^{10} < M_{\rm h}
< 2 \times 10^{14}\: h^{-1} \msun$).  This study showed that there was
a tendency for neighbouring halo semi-major  axes to point in the same
direction in all mass ranges, but  the strength of the alignment was a
strongly increasing function of halo mass.

In   similar   studies,    \citet{CM00,HRH2000}   and   \citet{Jing02}
investigated projected 2D axis alignments of cluster sized dark matter
haloes.  Projections  provide a  lower limit  on the  alignment signal
(since  projections necessarily  lose information  about the  system).
However, \citet{Jing02}  showed that  the results in  \citet{CM00} and
\citet{HRH2000} further underestimated the correlations as their shape
measurements included haloes with as  few as 20 particles, which could
lead  to a  factor of  two underestimate  of ellipticity  correlation.
Using a minimum  of 160 particles for their  final shape measurements,
\citet{Jing02}  showed that  the ellipticity  correlations had  a high
enough  amplitude  to contaminate  both  deep  and wide  weak  lensing
surveys.

A number of works considered  the alignment, $\Theta_{\rm Hh}$, of the
orientation of a  halo with the direction  to the centre of  mass of a
neighbouring                                                     halo,
\citep[e.g.][]{OT00,FGK+02,Hopkins05,KE05,ACC06,SFC12}.           This
alignment signal tended  to be stronger than the  halo axis alignments
\citep[e.g.][]{FGK+02,Hopkins05,KE05, SFC12}, and  was also a strongly
increasing  function  of  halo mass  \citep[e.g.][]{BS05,SFC12}.   The
signal was  detected out to  $\mathtt{\ge}100 \: h^{-1}$Mpc  for group
and   cluster    mass   haloes   ($M_{\rm   h}    \gtrsim   2   \times
10^{13}\:h^{-1}\msun$)          \citep[e.g.][]{FGK+02,Hopkins05,KE05}.
\citet{Hopkins05} also  studied cluster halo alignments  as a function
of redshift and showed that the alignments were greater at early times
and that  aligned halo  pairs were  more likely to  be connected  by a
filament  than  unaligned halo  pairs.   They  found this  result  was
consistent  with the  anisotropic merging  and infall  scenario, where
haloes form  through mergers  and accretion  of dark  matter traveling
coherently along the direction of the large-scale filaments, such that
the alignments should be present from the time of halo formation.

\citet{LKJ05} attempted  to provide an alternative  explanation to the
anisotropic infall model, arguing that the model is merely qualitative
and  that  primordial  alignments  should   be  damped  over  time  by
non-linear processes. Anisotropic infall has  been used to explain the
observational result  that brightest cluster galaxies  and dark matter
haloes have  their semi-major  axes preferentially aligned.   To mimic
this scenario in  the absence of baryons, they  identified the largest
central  subhalo  within  the  cluster-sized  dark  matter  haloes  to
represent  the   central  galaxy   and  investigated   the  alignment,
$\theta_{\rm SH}$.   In their  alternative model,  interaction between
the  host halo  tidal field  and the  subhalo should  account for  the
observed  alignments, assuming  that the  subhalo angular  momentum is
aligned parallel  to the  subhalo semi-minor axis.   Their simulations
showed that the  semi-minor axes of the  subhaloes were preferentially
perpendicular to the semi-major axis  of the host haloes, in agreement
with  observations and  the  predictions of  their alternative  model.
However,  in order  to rule  out  the anisotropic  infall model,  they
conceded that they  needed to measure the  angular momentum alignments
directly but  their simulations lacked  the resolution required  to do
this.

When  investigating  spin-spin  alignments,  $\Theta_{\lambda_{\rm  H}
  \lambda_{\rm H}}$,  tidal torque  theory suggests  that neighbouring
haloes  should experience  some  alignment of  their  spin vectors,  .
\citet{TLB13} showed that neighbouring halo  spins had a weak parallel
alignment  to  each  other,  but   only  for  halo  separations  under
$0.3\:h^{-1}\mathrm{Mpc}$.  On these scales, only subhaloes in massive
clusters would be able to exhibit  alignments.  The scale of this halo
separation is so small that this  is likely why other lower resolution
studies    did    not    detect    any    clear    alignment    signal
\citep{BE87,PDH02a,FGK+02,BS05}.   By   contrast,  \citet{HN01}  found
significant but  weak alignments on scales  from $1-30\:h^{-1}$Mpc and
\citet{HPC+07} also  found a  correlation in  alignments of  spins for
massive haloes  in clusters.   However, in an  extension of  this work
using a different selection  criteria for their haloes, \citet{HCP+07}
found  no halo-halo  spin  correlation.  They  used  a more  stringent
selection criteria to  remove haloes that had not  yet relaxed.  These
systems  were mostly  two close  neighbours  that were  either in  the
process of merging  or had been spuriously linked into  a single halo.
They also  removed the  haloes where  the distance  of the  most bound
particle from  the centre of  mass exceeded  a quarter of  the largest
distance between the outermost particle in  the halo and the centre of
mass (since the  most bound particle in a relaxed  halo would normally
reside close to the centre  of mass).  This cleaned catalogue modified
the  spin parameter  distribution,  and hence  reversed their  earlier
detection of spin alignments.  Since  halo selection criteria can have
a significant effect on results, this, and the fact that they measured
the spins of haloes with as  few as 20 particles, may offer reasonable
explanations   for  why   \citet{HN01}  found   alignments  in   their
simulations.

\subsubsection{Halo-LSS alignments}
\label{sec:LSS}

$N$-body simulations on a cosmological scale have shown that structure
formation   (i.e.     gravitational   collapse)   occurs    first   in
sheets. Filaments  then form and  matter in these  filaments collapses
into  dark matter  haloes.  These  dark matter  haloes move  along the
filaments  toward intersections  between multiple  filaments that  are
knots  of  potential minima  where  cluster  haloes are  formed.   See
\Cref{sec:web} for  more information  and details  on some  cosmic web
classification schemes.

There are  a number of studies  that measure the shape  alignment of a
halo  with  the   surrounding  large-scale  structure;  $\theta_{HW}$,
$\theta_{HF}$, $\theta_{SH}$\footnote{Note  that the subscript  `H' in
  this  alignment  is the  cluster  cosmic  web  element and  the  `S'
  represents the galaxy-sized and smaller halo substructure within the
  cluster.},                     $\theta_{\rm                     Hv}$
\citep[e.g.][]{PCP+06,HCP+07,BTP+07,CBG+08,ZYF+09,PSM+11,LHF+13,FCP14}.
The consensus is  that shape alignments with the LSS  are stronger and
more robust to measure than angular momentum/spin correlations.  There
is a consistently strong parallel  alignment of the semi-major axis of
a halo  with its  host sheet  plane, the semi-major  axis of  its host
filament or  cluster or perpendicular  to the radial direction  of the
centre  of voids  (the  semi-minor  axis of  the  haloes  tends to  be
parallel to the radial direction of  voids) in all but the lowest mass
haloes.   This alignment  increases in  strength with  increasing halo
mass.  \citet{FCP14} traced the shape  of the structure using both the
tidal     shear    field     (which    they     term    the     T-web;
\autoref{eq:sheartensor})      and     velocity      field     (V-web;
\autoref{eq:velocitytensor}).   However,  they  did not  divide  their
simulation   into   distinct   web    elements   using   a   threshold
$\lambda_{th}$, instead  choosing to classify the  alignments directly
with  respect to  the eigenvectors  $\hat{e}_1$ and  $\hat{e}_3$.  For
simplicity of reporting results, they  defined a strong alignment with
$\hat{e}_3$  to   be  a  strong  alignment,   $\theta_{\rm  HF}$  (see
\Cref{fig:sketch_filament}), with a filament, while a strong alignment
with  $\hat{e}_1$   is  an  anti-alignment,  $\theta_{\rm   HW}$  (see
\Cref{fig:sketch_void}  and  \Cref{t:IAObservables}),  with  a  sheet.
This choice has the benefit of being independent of $\lambda_{th}$ for
each classification scheme, however the signal may become diluted when
mixing  of  the  signal  occurs between  environments.   Despite  this
unusual classification scheme, their results for the tidal shear field
matched previous works.  However  in the momentum-based velocity field
measurement, they  found an anti-alignment  of haloes with  sheets for
haloes $M_{\rm h} > 10^{12}\:h^{-1}\msun$.

\begin{table*}[t]
 \resizebox{\textwidth}{!}{
      \begin{tabular}{cccccc}
        \hline\hline
Author & Web Method & Spatial Scale& Along &
Alignment & Mass dependence\\
 & & $h^{-1}$~Mpc & & & \\\hline

\citet{FCP14} & T-Web & $0.5-1$ &
$\hat{e}_3$  (filament) & $-$ & $>10^{12}\:h^{-1}\msun$\\

 &   & &
$\hat{e}_3$  (filament) & none & $<10^{12}\:h^{-1}\msun$\\

&   & &
$\hat{e}_1$ (wall) & none & $>10^{12}\:h^{-1}\msun$\\

&   & &
$\hat{e}_1$ (wall) & none & $<10^{12}\:h^{-1}\msun$\\\hline

\citet{FCP14} & V-Web & $0.5-1$ &
$\hat{e}_3$  (filament) & none & $>10^{12}\:h^{-1}\msun$\\

&   & &
$\hat{e}_3$ (filament) & none & $<10^{12}\:h^{-1}\msun$\\

&   & &
$\hat{e}_1$ (wall) & + & $>10^{12}\:h^{-1}\msun$\\

&   & &
$\hat{e}_1$ (wall) & none & $<10^{12}\:h^{-1}\msun$\\\hline

\citet{LHF+13} & V-Web & $1$ &
filament &$-$ & $>10^{12}\:h^{-1}\msun$\\

&   & &
filament &$+$ & $<10^{12}\:h^{-1}\msun$\\

&   & &
wall & $++$ & all masses\\\hline

\citet{TLB13} & Hessian density & $2-5$ &
filament & $-$ & $> 5\times 10^{12}\:h^{-1}\msun$\\
&   & &
filament & $+$ & $< 5\times 10^{12}\:h^{-1}\msun$\\\hline

\citet{CPD+12} & Morse Theory \& T-Web & $1-5$ &
filament & $--$ & $>10^{12.5}\:h^{-1}\msun$ \\

&   & &
filament & $++$ & $<10^{12.5}\:h^{-1}\msun$ \\

& & &
wall & $++$ & all masses\\\hline

\citet{ZYF+09}  & Hessian density &  $2.1$ &
filament & $++$ & if anticorrelated with shape\\

& &  &
filament & $--$ & if correlated with shape\\\hline

\citet{AWJ+07} & Hessian density & - &
wall & $++$ & $>10^{12}\:h^{-1}\msun$\\

& & - &
wall & $+$ & $<10^{12}\:h^{-1}\msun$\\

& & - &
filament& $-$ & $>10^{12}\:h^{-1}\msun$\\

& & - &
filament& $+$ & $<10^{12}\:h^{-1}\msun$\\\hline

\citet{HCP+07} & Tidal Web & $2.1$ & filament & $-$& none\\

& & &
wall & $++$ & $>10^{12}\:h^{-1}\msun$\\
& &    &
wall& $+$ & $<10^{12}\:h^{-1}\msun$\\\hline \hline

      \end{tabular}}

  \caption{Angular  momentum  alignment  with the  cosmic  web,  where
    $\hat{e}_1$ is the major and  $\hat{e}_3$ is the minor eigenvector
    of  the  corresponding tensor  (velocity  or  shear).  Summary  of
    theoretical     results    provided     by    similar     analysis
    methods. (-~-)$++$ indicates a  strong (anti-)alignment and (-)$+$
    indicates a weak (anti-)alignment. \permmn{FCP14}}
  \label{tab:momentumcomp}

\end{table*}

Many  works investigated  the  spins  of a  haloes  as  a function  of
environment
\citep[e.g.][]{AWJ+07,HPC+07,HCP+07,BTP+07,SPC+08,ZYF+09,CPD+12,TLB13,LHF+13,AY14,FCP14}. There
is  consensus  that this  measurement  is  far  less robust  than  the
alignments  with  halo  shape  and   results  are  more  dependent  on
measurement algorithm,  simulation and environment  definition.  There
is  little   $N$-body  simulation  literature  for   angular  momentum
alignments  around   voids,  $\theta_{\lambda_{\rm   H}v}$,  directly,
although comparisons may be made with alignments of sheets since these
reside at  the boundaries of the  void surfaces and in  the case where
the  void is  modelled as  spherical,  the alignment  would simply  be
perpendicular  to  the  sheet  alignment.   \citet{HWH+06,PCP+06}  and
\citet{BTP+07} investigated the void alignments directly and concluded
that there  were no angular  momentum alignments around voids  and the
orientation  of the  angular momentum  of the  haloes was  random.  By
contrast   \citet{CBG+08}\footnote{While   this  simulation   includes
  hydrodynamics, the  results are  focused on  dark matter  haloes and
  appear to be independent of the baryons, hence its inclusion in this
  section.}  did find alignments, where the angular momentum vector of
haloes  located   in  a  shell   at  the  void  surface   was  aligned
preferentially perpendicular  to the  direction of  the centre  of the
void.  These alignments decreased rapidly  with void radius, so taking
a wide shell would mask the signal (leaving them actually in agreement
with \citealp{HWH+06} who only used  wide shells).  They also showed a
slight trend of increasing alignments with increasing halo mass. It is
worth  noting  that this  result  is  in  contrast with  the  findings
presented in \Cref{sec:Internal}, that the  angular momentum of a halo
is preferentially parallel to the  \emph{semi-minor} axis of the halo;
\citet{CBG+08}  found   the  semi-minor  axis  of   their  haloes  was
preferentially parallel to  the radial direction to the  centre of the
voids  so  the angular  momentum  of  their haloes  is  preferentially
perpendicular  to  the semi-minor  axis  in  this work.   Given  these
discrepancies, further investigations would be required to clarify the
alignments of the halo angular momentum vector around voids.

When considering the alignment,  $\theta_{\lambda_{\rm H}{\rm W}}$, of
the angular momentum  of haloes with sheets, there is  a trend for the
angular   momentum    to   be   aligned   parallel    to   the   sheet
\citep[e.g.][]{AWJ+07,HCP+07,ZYF+09,CPD+12,TLB13,LHF+13}.    There  is
also a trend for the  alignment, $\theta_{\lambda_{\rm H}{\rm F}}$, of
the angular  momentum of  high mass  haloes to  be perpendicular  to a
filament,  while lower  mass  haloes align  parallel  to the  filament
\citep[e.g.][]{AWJ+07,HPC+07,SPC+08,CPD+12,TLB13,AY14}  and  the  mass
where  this  transition  occurs  is  around  $M_0^S  \simeq  5  \times
10^{12}\:h^{-1}\msun$      \citep[e.g.][]{HPC+07,AWJ+07,CPD+12,TLB13}.
\citet{CPD+12} investigated the redshift  dependence of the transition
mass  and found  that the  halo transition  mass $M_{\mathrm{crit}}^S$
decreased with increasing redshift such that
\begin{equation}
M^{S}_{\mathrm{crit}} \approx M_0^S(1+z)^{-\gamma_s},\,\,\,\, \gamma_s
= 2.5 \pm 0.2.
\end{equation}

Looking at  the alignments in  terms of the halo  model, \citet{PSP08}
found similar  results.  They  calculated the dark  matter correlation
function parallel and perpendicular to  the angular momentum vector of
the haloes.  They found that the halo angular momentum vectors aligned
preferentially  perpendicular   to  the   mass  distribution   of  the
large-scale structure  (the two-halo  term) at  high masses  while low
mass halo  angular momenta  may preferentially  point parallel  to the
mass  distribution on  large scales.   This inversion  occurred around
masses of $5.5 \times 10^{12.5}\:h^{-1} \msun$, which is qualitatively
consistent with the findings above.

In  contrast  to  these  findings  \citet{LHF+13}  disputed  that  the
transition  of  the  halo   angular  momentum  vector  alignment  from
perpendicular to the  filament in high mass haloes to  parallel in low
mass haloes  occurs only in  filaments. They showed evidence  for this
transition occurring in  all environments, with the mass  at which the
transition occurs decreasing with  environment type from clusters down
to voids (e.g.   in voids, haloes of intermediate  mass transition the
direction of  their angular  momentum from  perpendicular to  the void
centre  to parallel  to the  void  centre with  decreasing mass).   In
further  contrast,  \citet{FCP14}  found  results  that  only  broadly
matched many of these earlier works.   For web elements defined in the
V-web,  they found  that the  angular momentum  of haloes  with masses
above $10^{12}\:h^{-1}\msun$ tended to  align along walls, without any
clear  trend with  respect to  filaments, while  the alignment  signal
disappeared  for halo  masses  below  $10^{11}\:h^{-1}\msun$.  In  the
T-web, they  similarly found no  evidence for any alignment  of haloes
with masses below  $10^{12}\:h^{-1}\msun$, and only a  weak signal for
alignment at higher mass, with a trend for the angular momentum to lie
perpendicular to filaments. The discrepancies  in this work may be the
result of  the chosen  web classification  scheme diluting  an already
weak signal  even further  by mixing  environments, or  the alignments
could have  a high sensitivity on  small scales to the  method used to
construct   the   cosmic   web  (including   numerical   choices   for
interpolating the relevant fields).   \citet{FCP14} provided a helpful
comparison  of  angular momentum  alignments  found  in these  similar
works, which is reproduced in \Cref{tab:momentumcomp}.

\subsection{$N$-body simulations Roundup}

There is  a wealth of  literature that investigates the  alignments of
both  the  shapes  and  spins   of  dark  matter  haloes  in  $N$-body
simulations over a large range of scales.  This section introduced the
common  techniques  used  to  measure alignments  in  simulations  and
reviewed  the existing  literature on  the alignments  of dark  matter
haloes in $N$-body simulations.

Overall, the shape  alignments were stronger and more  robust than the
spin alignments.  It was established  that the dark matter haloes have
a strong  tendency to be  prolate and that the  shape of the  halo can
change with radius.  The subhalo  distribution semi-major axes tend to
align  with the  semi-major  axis of  the host  dark  matter halo  and
individual subhalo semi-major axes tend  to point toward the centre of
mass of the host dark matter halo throughout their orbit. Neighbouring
dark  matter haloes  tend to  have  semi-major axes  that are  aligned
parallel  and  a  stronger  tendency  to point  in  the  direction  of
neighbouring haloes  (in the direction of  the large-scale structure).
This is consistent with the findings  that the semi-major axes of dark
matter haloes tend to align parallel  with their host sheet plane, the
semi-major axis of their host  filament or cluster or perpendicular to
the direction to the centre of the void.  In general, shape alignments
on all scales were stronger with increasing halo mass.

The  literature  on angular  momentum  and  spin alignments  had  less
consensus.   A number  of works  found the  distribution of  halo spin
parameters  to be  well fit  by a  lognormal distribution,  while some
works found that the lognormal  was a poor fit.  Further investigation
with $N$-body simulations containing  a large statistical sample (many
thousands) of well  resolved haloes would be required  to resolve this
issue.  There  was strong consensus  that the angular  momentum vector
was aligned parallel with the semi-minor axis and perpendicular to the
semi-major axis  of the dark matter  halo.  Yet, there was  also clear
evidence that the haloes experience  significant, rapid changes in the
alignment of the  angular momentum vector over time (spin  flips) as a
result of  major and  minor mergers  and even  flybys from  other dark
matter haloes.   In addition,  the direction  of the  angular momentum
vector can  change as a function  of halo radius and  these spin flips
can occur in  the inner radii as  well as the outer  radii.  As events
like mergers and flybys cause the angular momentum vector to flip, the
selection criteria for the dark matter haloes included the studies has
a strong influence on the  findings.  The frequent halo spin-flips are
likely responsible for the less robust spin alignment results.

Nearby neighbouring  haloes (with small  separations) may show  a weak
alignment  of  their  spin  vectors  in the  same  direction,  but  no
alignments at  larger separations.  In the  large-scale structure, the
spin vectors  of the dark matter  haloes tend to be  parallel with the
plane of  their host sheet, while  there is no clear  consensus on the
orientation  with  voids.   Although  if  sheets  are  considered  the
boundaries of (spherical) voids then it  could be argued that the dark
matter  halo  spin vectors  are  preferentially  perpendicular to  the
direction to the centre of a void.   A number of works found that dark
matter  halo  spin vectors  tended  to  be  parallel with  their  host
filament  at low  masses and  transitioned to  being perpendicular  at
masses  above around  $5 \times  10^{12}~h^{-1}\msun$, with  this mass
decreasing with  increasing redshift.  It  is also possible  that this
transition occurs in  all cosmic web elements with the  mass where the
transition occurs  decreasing with  environment type from  clusters to
filaments, sheets, and voids.

It is  difficult to  relate these  studies to  the real  world because
observations  deal  with luminous  galaxies  and  we cannot  currently
measure the  alignments of the  dark matter haloes directly.   In some
studies, it was assumed that the  orientation of the central region of
the dark matter halo could be used  as a proxy for the central galaxy.
However,  \Cref{sec:Hydro}  shows  that the  introduction  of  baryons
significantly changes the  relative shapes and alignments  of the dark
matter   haloes,   particularly   in   the  inner   regions   of   the
halo. Consequently, caution should be exercised when exclusively using
$N$-body simulations to gain  insight into galaxy alignments.  Despite
these limitations, $N$-body simulations  continue to play an important
role in  studies of galaxy  alignments.  This is discussed  further in
\Cref{sec:SemiAnalytic,sec:Roadmap}.

\section{Hydrodynamic simulations}
\label{sec:Hydro}

A  significant shortcoming  of $N$-body  simulations in  the study  of
galaxy dynamics or intrinsic alignments  is that these simulations are
concerned only  with the  evolution of  structure under  gravity.  The
particles used  in the simulations interact  only gravitationally, and
any effects relating to baryon  or gas physics are completely ignored.
Detailed theoretical understanding of the (expected) shape and angular
momentum/spin alignments between galaxies inside haloes and their host
haloes (and combinations thereof) is  fundamentally limited due to the
not-well-understood role  that gravitational collapse of  the halo and
baryonic  physics has  on the  shapes of  galaxies. In  particular the
feedback processes  between galaxy  and cluster-scale physics  and the
surrounding   dark    matter   haloes   is   not    well   understood.
Hydrodynamic\footnote{Note   that  the   terms  ``hydrodynamic''   and
  ``gasdynamic'' are used interchangeably  in the literature, although
  hydrodynamic is  adopted in this review.}   simulations address this
issue  by treating  the  evolution  of the  gaseous  component of  the
Universe  using the  methods  of computational  fluid dynamics.   This
approach enables  the complex  interactions of the  different baryonic
components (gas,  stars, etc.)   to be treated  self-consistently with
the  dark  matter,  and  on   a  much  smaller  scale.   This  enables
simulations of the formation of galaxies within dark matter halos, and
the ability to probe the complex  physical processes that give rise to
the visual properties of galaxies that we observe today.

Hydrodynamic  simulations are  a  relatively new  way  to explore  the
intrinsic alignments of galaxy shapes  and spins, in part because they
are quite  computationally intensive.  The obvious  advantage of using
hydrodynamic  simulations is  that $N$-body  simulations require  some
sort of semi-analytic model to identify the positions and orientations
of  galaxies (see  \Cref{sec:SemiAnalytic}),  whereas in  hydrodynamic
simulations, the  process of  galaxy formation is  naturally included.
The  disadvantage  of  using  hydrodynamic  simulations  is  that  the
relevant sub-grid physics that is needed to form realistic galaxies is
not yet known.  The term ``sub-grid  physics'' is meant to include all
physical  processes  that take  place  on  a  smaller scale  than  the
resolution of the simulation, and therefore must be included with some
model.   This  includes  some  important aspects  of  star  formation,
accretion  onto supermassive  black holes,  radiative heating/cooling,
and feedback from supernovae (important  at low masses) and some other
mechanisms at high masses (e.g., from AGN).  While there are claims in
the  literature of  increasingly  realistic  galaxy populations  using
hydrodynamic  simulations  \citep[e.g.][]{KDC+14,VGS+14}, the  results
still show some  discrepancies with reality that  require caution with
their  usage. For  example, the  results of  the sub-grid  physics are
often  tuned to  match  particular observables  (e.g.  the  luminosity
function)  at  a particular  redshift  (typically  redshift zero),  so
extrapolation to different redshifts  should therefore be treated with
caution, as  observations at  high redshift  are limited  and matching
these  results is  not  guaranteed.  Ideally,  a  full exploration  of
intrinsic  alignments  in  hydrodynamic simulations  would  involve  a
comparison  between  several  independent simulations  with  different
implementations  of sub-grid  physics, to  check how  the results  for
galaxy intrinsic alignments depend on  the details of galaxy formation
and feedback.   Unfortunately, such a  comparison is not  yet possible
due   to   the   tremendous   expense  of   large-volume   {\em   and}
high-resolution hydrodynamic simulations.

\subsection{Small-volume hydrodynamic simulations}
\label{sec:SmallHydro}

In this  section, small-volume  and zoom hydrodynamic  simulations are
discussed.

\subsubsection{Internal alignments}
\label{sec:SHinternal}
One very interesting study made possible with hydrodynamic simulations
is how  the baryonic  galaxy aligns  with its  host dark  matter halo,
$\theta_{CH}$, and  whether baryons have  an effect on halo  shape and
alignment.

Broadly,  studies agree  that  the addition  of baryons  significantly
affects the  shape of the  host dark matter  halo.  Most agree  that a
galaxy forming  in the centre  of the halo will  cause the halo  to be
more  spherical overall  and there  is  a consensus  that haloes  with
galaxies tend toward  oblateness, rather than the  prolateness seen in
$N$-body                        only                       simulations
\citep[e.g.][]{KKZ+04,BKG+05,BS06,GFS06,ANF+10,DMF+11}.       However,
since these  studies used different  sized haloes or  simulations that
had  different  feedback  effects,  it is  difficult  to  make  direct
comparisons.  For example \citet{KKZ+04}  ran Adaptive Refinement Tree
$N$-body + hydrodynamic simulations of cluster sized haloes and showed
that the presence of baryons significantly increased the sphericity of
the host dark matter haloes with  a magnitude that decreased at larger
radii such that the sphericity was largely indistinguishable from dark
matter only  haloes at the  virial radius.  \citet{BKG+05}  found very
similar results,  an increase in  halo sphericity that  decreases with
increasing   radius,   using   seven  different   implementations   of
hydrodynamics   in  the   simulations  and   employing  a   multi-mass
resimulation  technique to  generate  high  resolution galactic  sized
discs.   \citet{ANF+10} ran  $N$-body  +  hydrodynamic simulations  of
galactic  sized discs  with  radiative cooling  but neglected  stellar
feedback.  By calculating equipotential axial ratios they found haloes
that  were   significantly  rounder   than  their  dark   matter  only
counterparts at all  radii out to nearly $2R_{\rm vir}$,  which may be
due  to strong  overcooling due  to neglecting  stellar feedback.   By
contrast, \citet{DMF+11}  used simulations including  stellar feedback
and used the galaxy density to characterise the halo shapes (which are
typically flatter  than equipotential surfaces).  The  haloes in their
sample were  almost spherical  and slightly more  oblate in  the inner
regions,  while  their dark  matter  only  haloes tended  toward  more
prolate in the  inner regions.  Thus, they agreed  that baryon physics
significantly affects the  shape of the halo in the  inner regions but
they  did not  see  significant  changes in  sphericity  in the  outer
regions between the $N$-body and hydrodynamic simulations.

\citet{KLK+10} performed a similar analysis by comparing a dark matter
only simulation with  a hydrodynamic simulation with  the same initial
conditions. This study  included 3 Milky Way sized  galaxy haloes with
their corresponding substructures. They tested  the shapes of the dark
matter haloes and subhaloes in  their simulations and found the haloes
to be more spherical in the hydrodynamic simulation, in agreement with
the general consensus  above.  However, the subhaloes  showed no shape
differences between the two  simulations, indicating that the addition
of  gas physics  had no  influence  on the  subhalo population.   They
suggested that the  lack of shape changes  may be due to  the low mass
nature of the subhaloes and that there may be more obvious changes for
cluster sized host haloes with large galaxy sized subhaloes.

\citet{BKG+05} tested  whether the orientation  of the inner  halo was
being  driven  by the  galactic  disc,  by  generating rings  of  test
particles tilted  by $10^{\circ}$ with  respect to the disc  plane and
comparing the  gravitational torque on  the test particles due  to the
dark  matter and  the baryons.   For test  particles in  rings $<  0.1
R_{\rm  vir}$,  the  baryons  and   dark  matter  exerted  torques  of
comparable magnitude,  while the dark  matter dominated the  torque at
larger  radii.  Therefore,  it was  concluded that  the alignments  at
inner radii were  due to simultaneous evolution of the  halo and disc,
rather than the disc driving the  halo orientation, or vice versa, and
that the orientation  of the inner and outer halo  in the hydrodynamic
simulations  was  uncorrelated.   Tidal torque  theory  suggests  that
correlations between  a galaxy and  its host halo are  expected during
the   formation  epoch   \citep[e.g.][]{HP88}.   However,   while  the
alignment of  a galaxy and its  host halo is initially  very close, at
later times  the outer halo  will continue to accrete  material.  This
could  alter the  net angular  momentum or  shape of  the outer  halo,
causing  the  misalignments.   This   distinction  is  important  when
considering the alignment of satellite populations and their intrinsic
alignment signal.

Thinking about how galaxy angular  momentum aligns with its host halo,
\citet{BKG+05,BEF+10} and \citet{DMF+11} all showed that the alignment
of  the  angular  momentum  vector  of  the  galactic  disc  with  the
semi-minor axis of the  host halo was close at $R  < 0.1 R_{\rm vir}$.
\citet{BKG+05} and \citet{DMF+11}  also showed that the  halo could be
significantly misaligned at  $R > 0.1 R_{\rm  vir}$ with misalignments
of $\theta_{\lambda_{\rm C}{\rm H}} > 45^{\circ}$ at $R_{\rm vir}$.

\citet{BAC+02,Chen03,vdBAH03}  and \citet{Sharma05}  made measurements
of the misalignments, $\theta_{\lambda_{\rm C}\lambda_{H}}$ at $R_{\rm
  vir}$ and found an average  misalignment of between $20^{\circ}$ and
$36^{\circ}$  for their  standard baryonic  simulations. In  a similar
study, \citet{HTC10}  found that  while the  alignment of  the angular
momentum for both  the central disc galaxy gas and  stars was close to
the   angular   momentum   of   the  inner   dark   matter   halo   at
$\theta_{\lambda_{\rm c}\lambda_{\rm H}} \sim 18^{\circ}$, there could
be significant misalignments, $\theta_{\lambda_{\rm c}\lambda_{\rm H}}
\sim 50^{\circ}$, with  the angular momentum of the  whole dark matter
halo.   This result  appeared  to have  no  dependence on  environment
density or stellar or halo mass.

There is a known problem in hydrodynamic simulations where cooling gas
forming the central  galaxies has an angular momentum that  is too low
to  produce  extended  galaxies  \citep[e.g.][]{NB91}.   This  may  be
partially  due  to numerical  limitations,  but  uncertainties in  the
baryonic feedback  processes are  certain to  contribute significantly
\citep[e.g.][]{MD02}.    While  the   feedback  processes   are  being
investigated  separately,   \citet{vdBAH03}  tested  the   effects  of
artificially pre-heating  the baryonic component of  the intergalactic
medium, and found that this pre-heating decoupled the baryons from the
dark matter,  resulting in significantly higher  misalignments between
the  angular momentum  of both  the gas  and the  dark matter  halo of
$58^{\circ}$ on average.

\begin{figure*}[t]
  \begin{subfigure}{0.5\textwidth}
 \includegraphics[width=\hsize]{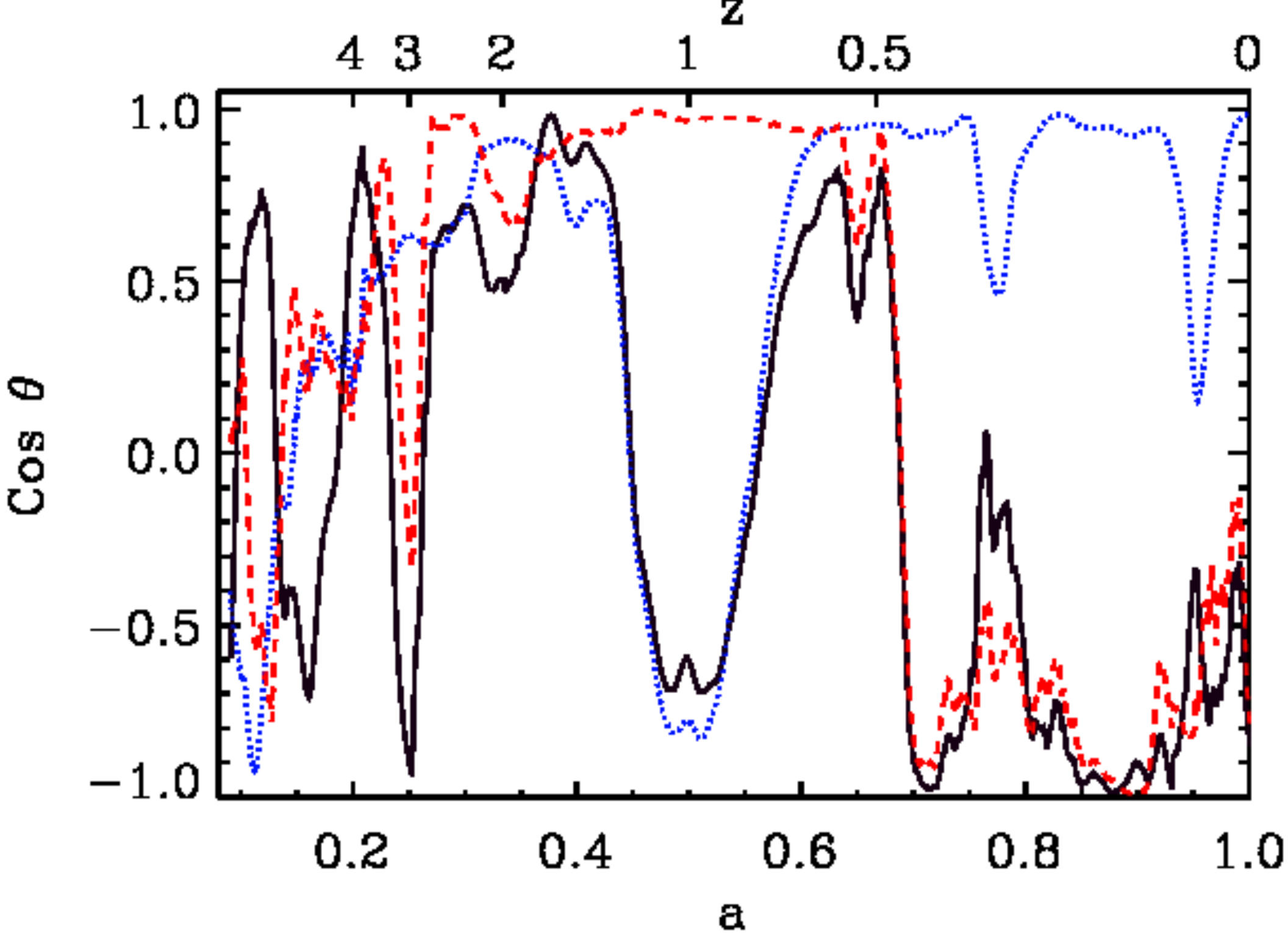}
 \caption{Dark matter halo: $R_{\rm vir}$, stars/gas: 8~kpc}
 \label{fig:Romano_Rvir}
\end{subfigure}
  \begin{subfigure}{0.5\textwidth}
 \includegraphics[width=\hsize]{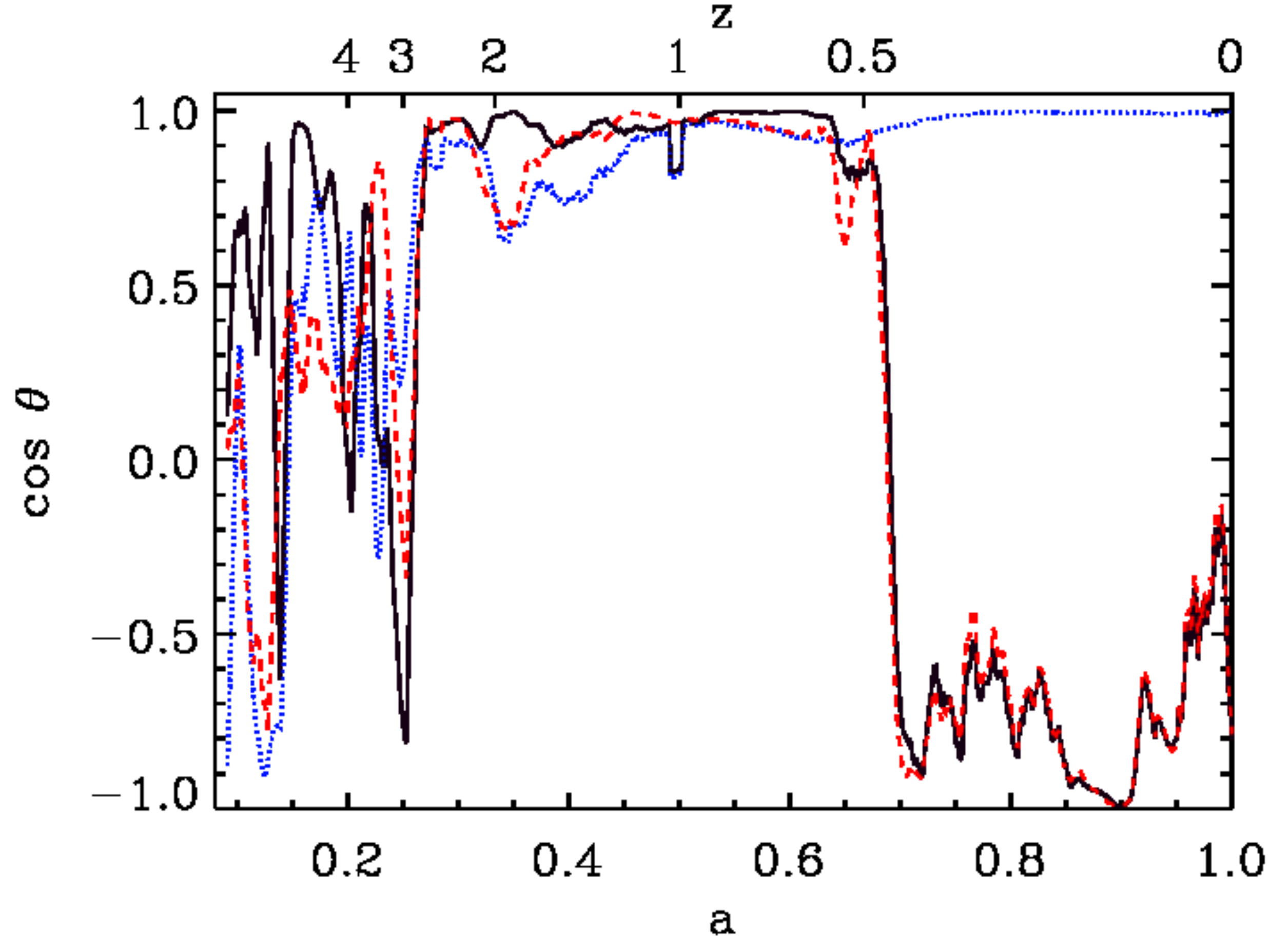}
 \caption{Dark matter halo/stars/gas: 8~kpc}
 \label{fig:Romano_8kpc}
\end{subfigure}
  \caption{Correlations  between  the  angular  momenta  of  the  dark
    matter,  gas, and  stars of  a $\mathtt{\sim}10^{12}h^{-1}\:\msun$
    galaxy system,  within the radii given  under each panel and  as a
    function of scale factor/redshift .  The angle $\theta$ is defined
    between     the     corresponding    angular     momenta,     e.g.
    $\theta_{\lambda_{\rm   gas}  \lambda_{\rm   H}}$.   The   colours
    represent the  dark matter-gas  (black, solid),  dark matter-stars
    (blue,      dotted)      and     stars-gas      (red,      dashed)
    correlations. \permapj{RSH+09}}
\label{fig:Romano}
\end{figure*}

\citet{RSH+09,SWS+09,DRV+13,Cen14}  and   \citet{DBR+15}  investigated
spin   orientation   changes   over  time.    When   investigating   a
$\mathtt{\sim}10^{12}\:h^{-1}\msun$,   somewhat  isolated   (no  major
mergers after  $z = 1.5$)  central galaxy, \citet{RSH+09}  showed that
the spin alignments of the dark  matter halo (within $R_{\rm vir} \sim
400$ kpc), stars  and gas (within the inner 8  kpc) changed frequently
over time  and even experienced $\mathtt{\sim}180^{\circ}$  flips over
short   timescales    and   at    times   with   no    major   mergers
(\Cref{fig:Romano_Rvir}).  When  also limiting the dark  matter to the
inner 8~kpc,  they found that at  $z \gtrsim 3$ major  merger activity
caused all alignments to change frequently.  From $3 \gtrsim z \gtrsim
0.5$, the  spins of the dark  matter, stars and gas  were well aligned
but at $z  \lesssim 0.5$, the alignments flipped when  cold gas became
concentrated in  the central regions  and had  a spin in  the opposite
direction  to  the  stars and  dark  matter  (\Cref{fig:Romano_8kpc}).
\citet{SWS+09} found  that misalignments between the  stellar disc and
newly accreted cold  gas could cause the stellar disc  to lose part of
its mass or become completely destroyed (but later regrow).

\citet{Cen14} simulated and investigated  spin alignments in more than
300 galaxies between $M_* =  10^{10} - 10^{12} \msun$, They calculated
the     orientations    of     the    specific     angular    momentum
(\autoref{eq:specific})  for  both gas  and  stars  within the  galaxy
radius over time and showed that orientations were dependent on galaxy
type  (elliptical or  spiral), environment,  luminosity and  redshift.
They  found  that  the   spin  alignments  decreased  with  increasing
redshift, increased with  increasing stellar mass and  were larger for
elliptical than spiral galaxies.

It is clear from \citet{Cen14} that frequent changes in spin direction
over time have a large effect on  the final shape and orientation of a
galactic disc.  Mergers  and cold gas accretion  cause more stochastic
and  rapid  orientation changes,  while  the  large-scale tidal  field
slowly acts  to bring galaxy  spins back into alignment.   However, as
noted   by  \citet{Cen14},   the  question   remains  whether   steady
anisotropic infall of  gas and stars along filaments  and sheets could
also  act as  a mechanism  to align  the spins  of haloes  along their
structures.

\subsubsection{Satellite alignments}
\label{sec:SHsatellite}
The alignment of  the distribution of satellite galaxies  in host dark
matter  haloes,   $\theta_{\rm  BH}$,  was  extensively   examined  in
\citet{LCF+07} and \citet{DMF+11}. The  method of analysis was similar
in both studies although \citet{LCF+07} had 9 parent haloes (with only
three of these having 11 or more satellites) and galaxies of all types
were included  in the  analysis, while  \citet{DMF+11} had  431 parent
haloes (of which 80 had 10 or more satellites) and only late-type disc
galaxies    (the    galaxies    were   classified    dynamically    as
dispersion-supported spheroids  or rotationally supported  discs) were
included in  the analysis.  Both  computed the shape of  the satellite
distribution  by  diagonalising  the unweighted  inertia  tensor  (see
\autoref{eq:Inertia}  and subsequent  explanations).  They  both found
that  the  distribution of  ten  or  eleven brightest  (most  massive)
satellites was significantly flatter (more planar) than the underlying
dark matter distribution.  They also found that satellite distribution
preferentially aligned with the  plane perpendicular to the semi-minor
axis  of  the  host  dark matter  halo.   \citet{DMF+11}  showed  that
accretion of  satellites occurred preferentially along  the semi-major
axis  of the  dark  matter  halo and  suggested  that the  anisotropic
satellite   distribution  was   due  to   anisotropic  infall.    This
distribution was preserved because present  day satellites had been in
orbit for  far less time than  the dark matter halo,  which would have
undergone relaxation and  phase mixing at an earlier  time.  Thus, the
satellites had only  ever experienced a static  dark matter potential,
which is  reflected in the preserved  anisotropic spatial distribution
with the  outer host halo.   \citet{DMF+11} went  on to show  that the
shape of  the satellite distribution showed  no preferential alignment
relative to the galaxy, which is  consistent with the alignment of the
inner host halo  and central galaxy being uncorrelated  with the outer
host   halo  where   the  satellite   distribution  is   located  (see
\Cref{sec:SHinternal}).   In fact,  they  found that  $20\%$ of  their
satellite  distributions semi-major  axes lay  within $10^{\circ}$  of
perpendicular to the plane of the galaxy, which is consistent with the
observed  distribution  of  the   Milky  Way  satellites.   Similarly,
\citet{LCF+07} found two  of their three galaxies with  eleven or more
satellites  had  the semi-major  axis  of  the satellite  distribution
aligned  within $20^{\circ}$  of  perpendicular to  the  plane of  the
galaxy.  Although  with this study  having only three galaxies,  it is
not possible  to infer whether  the satellite distribution  showed any
preferential alignments with the central galaxy.

In  a separate  analysis  that  accounted for  all  satellites in  the
systems, \citet{DMF+11}  showed that the  shape of the  full satellite
distribution continued to  be much flatter than the shape  of the dark
matter halo, which is in contrast to \citet{LCF+07} who found that the
full satellite distribution matched the  shape of the dark matter halo
well and  the distribution did  not exhibit the flattening  found when
only  considering the  eleven  brightest satellites.   However, it  is
important  to remember  that  there  were only  three  systems in  the
analysis  by \citet{LCF+07},  and  all galaxy  types were  considered,
making  a  direct  comparison   between  the  two  studies  difficult.
\citet{DMF+11}  also  showed that  the  orientation  of the  satellite
distribution shapes  showed only a weak  bias toward the plane  of the
galaxy   but  much   stronger   alignments  relative   to  the   plane
perpendicular to the semi-minor axis  of the host halo.  The alignment
of the satellite distribution was significantly more influenced by the
dark matter host halo, so in the presence of misalignments between the
inner and outer halo, the orientation  of the galactic disc is largely
decoupled from the satellites.

\citet{KLK+10} investigated  the alignment of subhalo  semi-major axes
with the centre  of mass of the host dark  matter halo, $\theta_{Sh}$,
in both a  hydrodynamic and dark matter only simulation  with the same
initial   conditions.    Both   simulations  were   a   $2\:h^{-1}$Mpc
resimulated  region in  a  $64\: h^{-1}$Mpc  parent simulation.   They
found that the  subhaloes pointed preferentially toward  the centre of
the host  halo in both simulations  and that the addition  of baryonic
physics had no  effect on the result.  This is  unsurprising given the
subhaloes did not  change shape with the addition  of baryonic physics
(as mentioned in \Cref{sec:SHinternal}).

\subsubsection{Large-scale structure Alignments}
\label{sec:SHLSS}
There is some question of what the alignment of the semi-major axis of
the baryonic components of galaxies is with respect to the large-scale
structure.   \citet{NAS04} used  $N$-body/hydrodynamic simulations  to
produce  four disc  galaxies  and investigated  their alignments  with
sheets, $\theta_{\rm  GW}$.  They  found that  the semi-major  axes of
some disc galaxies tended to be  highly inclined relative to the plane
of the large-scale (several Mpc) two-dimensional sheet that the galaxy
was  formed in.   This was  likely  due to  coherence in  anisotropies
across  a large  number of  scales  present during  the expansion  and
contraction of the protogalactic  material causing some galactic discs
to  be highly  inclined  relative to  the large-scale  two-dimensional
structure  where they  were embedded.   Investigating alignments  with
respect  to   filaments,  $\theta_{\rm  GF}$,   \citet{HTC10}  sampled
$\mathtt{\sim}100$ galactic discs spanning  two orders of magnitude in
stellar and  halo mass.  They  found a  strong alignment of  the major
axis of low  mass ($< 4 \times 10^{11}\:  h^{-1}\msun$) galactic discs
along  the  direction  of  the  filament  at  $z=1$.   This  alignment
decreased  with increasing  density, which  may be  due to  non-linear
effects like torques from ram pressure or accretion/mergers.  They did
not  see this  alignment at  $z=0$, however  this may  be a  numerical
effect  in  their  simulations.    While  admitting  some  statistical
limitations in  their studies, they  also found that the  most massive
disc galaxies  were aligned with  their major axis pointing  along the
direction of a filament.

\subsection{Cosmological-volume hydrodynamic simulations}
\label{sec:LargeHydro}

When studying  galaxy alignments, the use  of hydrodynamic simulations
over cosmological  volumes is obviously highly  desired.  Large volume
simulations enable a \emph{statistical}  approach to galaxy alignments
by providing  huge catalogues of  galaxies.  At  the same time,  if we
want to  use hydrodynamic simulations to  predict intrinsic alignments
on the scales that  are relevant for lensing, up to  tens of Mpc, then
the minimum simulation volume is $\mathtt{\sim}100\:h^{-1}$Mpc.

There  are  a number  of  recent  papers  on intrinsic  alignments  in
hydrodynamic   simulations   in    cosmological   volumes,   including
\citet{BKD+13,TMD+14,TMD+14b,TMD15,DPW+14,WDD+14}  and \citet{CGD+14}.
\citet{BKD+13}  used  the   OverWhelmingly  Large  Simulations  (OWLS)
\citep{schaye10} to study the effects of baryonic cooling on the spins
and  alignments  of  dark  matter haloes,  rather  than  studying  the
baryonic components of the galaxies themselves.

\subsubsection{Alignments with spin and angular momentum}
\label{sec:LHspin}

\citet{DPW+14,WDD+14} and  \citet{CGD+14} all focused on  the issue of
spin alignments  using the Horizon-AGN simulation  \citep{DPW+14} in a
$100 \:h^{-1}$~Mpc  box.  Using a catalogue  of $\mathtt{\sim}150,000$
galaxies, \citet{DPW+14}  investigated a  snapshot at  $z =  1.83$ and
focused on the  orientation of the spin of the  galaxies, with respect
to the direction of  the filaments, $\theta_{\lambda_{\rm G}{\rm F}}$.
They found that more massive galaxies have their spin perpendicular to
the  nearest  filament  while  lower mass  galaxies  have  their  spin
parallel to the nearest filament. The  mass of this transition for the
stellar  disc was  around  $M_*  = 3  \times  10^{10}\msun$, which  is
comparable to  a dark  matter halo  transition mass at  $z =  1.83$ of
$M_{\rm  h} =  5 \times  10^{11}\msun$.  This  is consistent  with the
redshift evolution of the transition mass found in \citet{CPD+12} (see
\Cref{sec:LSS}).   The study  went on  to determine  spin orientations
with respect  to the  nearest filament  for a  large number  of galaxy
properties.  The overall findings formed  a picture that old, massive,
red, metal-rich,  and dispersion-dominated (elliptical)  galaxies have
their  spins preferentially  perpendicular  to  the nearest  filament,
while young,  low-mass, blue, metal-poor, and  centrifugally supported
(disc) galaxies  have their  spins preferentially aligned  parallel to
the filament.

The Horizon-AGN  simulation was run  with an adaptive  mesh refinement
code that uses a Cartesian-based Poisson solver \citep{T02} to compute
the forces.  A common issue in such a solver is forces grid-locking on
the Cartesian  axes of the  box due to  a numerical anisotropy  in the
force calculation \citep[see e.g.][]{HE81}.  \citet{DPW+14} found that
gaseous   disc  spins   of  low-mass   galaxies  ($M_{\rm   gas}  \sim
10^9\:h^{-1}\msun$) in the simulation were preferentially aligned with
one of the Cartesian axes, while higher mass galaxies did not show any
grid-locking.   The cosmic  web  elements in  the  simulation did  not
experience any grid-locking  so the overall effect  of the grid-locked
galaxies  may make  the measured  alignments with  the filaments  more
noisy.  \citet{CGD+14} presented evidence that  if there is no spatial
correlation   between  the   grid-locking  experienced   by  different
galaxies, then  the grid-locking would not  significantly affect their
results.   However,  a comparison  of  all  of the  grid-based  solver
studies with a  solver that does not use a  grid-based technique (like
smoothed particle  hydrodynamics (SPH) for example)  would be required
to  test  these  assumptions  and   assess  the  full  impact  of  the
grid-locking on the results.

\citet{WDD+14}  used the  same simulations  as \citet{DPW+14}  between
$1.2 <  z < 3.8$ in  a study of the  effect of mergers on  galaxy spin
changes.  They showed that mergers  along filaments caused galaxies to
have a  spin perpendicular to the  filament and the more  mergers that
contributed  to  the  galaxy  mass,  the  stronger  the  perpendicular
alignment.  By contrast,  in the absence of  (successive) mergers, the
galaxy spins would (re)align with the direction of the host filament.

Addressing the concerns of  weak gravitational lensing, \citet{CGD+14}
used the spin  of the stellar component  of the galaxy as  a proxy for
ellipticity and investigated 160,000 simulated  galaxies at $z = 1.2$.
When determining the apparent axis ratio for their projected galaxies,
they  ignored the  disc  thickness (which  they set  to  zero) due  to
resolution  limitations in  the simulation,  which has  the effect  of
maximising the  alignment signal measured  and giving upper  limits to
the predictions.  To investigate the GI (\autoref{eq:iadef}) alignment
signal, where  correlations occur between the  gravitationally sheared
background galaxy  and the  intrinsically oriented  foreground galaxy,
they determined  the galaxy population  that was likely  to experience
coherent alignments as a result of  the local tidal field.  Similar to
\citet{DPW+14}, they found that  low-mass galaxies were preferentially
aligned  parallel  to filaments  and  higher  mass  ($M_* >  4  \times
10^{10}$) galaxies  were preferentially  aligned perpendicular  to the
filaments.  Additionally, red galaxies showed more random correlations
with the surrounding tidal field than blue galaxies.  Part of this may
be related  to the fact that  red galaxies are typically  more massive
than  blue  galaxies  and at  $z  =  1.2$,  the  red galaxies  in  the
simulation had  a mass  around the transition  mass (some  above, some
below).   The random  orientations  detected in  this  study might  be
expected  to become  more correlated  at later  times as  more of  the
population exceeds the transition mass.  However, it is also important
to   note   that   massive   red   galaxies   are   observed   to   be
dispersion-dominated  \citep[see][]{paper3} and  the  spin  of such  a
galaxy  is not  expected to  be  strongly correlated  with its  shape,
meaning that the physical mechanism causing the observed alignments of
massive red galaxies is unrelated  to the spin.  Further investigation
would require use  of the actual galaxy shapes rather  than using spin
as a proxy for the shape.

When  investigating  the  II   (\autoref{eq:iadef})  alignment  in  3D
orientations, \citet{CGD+14}  found a  strong correlation  between the
spins  of blue  galaxy pairs  out to  a distance  of $10\:h^{-1}$~Mpc,
while  red galaxy  pairs and  red-blue pairs  showed no  correlations.
There  was  also   a  correlation  of  the  spins   between  low-  and
intermediate-mass  galaxies,  while  massive  galaxies  had  a  signal
consistent  with  zero at  all  separations.   These conclusions  were
unchanged when projecting the spin onto the plane of the sky.

Observational studies  of spin alignments have  shown some correlation
between the  spins of disc-like  galaxies \citep[e.g.,][]{LE07,Lee11},
but all  were for $z <  0.2$, while this Horizon-AGN  simulation study
was performed at $z = 1.2$.  Observations of shape alignments across a
wide range  of redshifts show  red elliptical galaxies  dominating the
intrinsic  alignment signal  with blue  disc-dominated galaxies  being
consistent  with  zero   alignments  \citep[see][for  a  comprehensive
  overview of  observational studies]{paper3}.  At later  times in the
simulation, it will be interesting  to directly compare the amplitudes
of the signal  with observations.  For these to agree,  it is possible
that late-time  effects that reduce the  correlation of disc-dominated
galaxy  spins   will  be   necessary.   However,  the   alignments  of
dispersion-dominated red ellipticals would  be better captured through
study of the shapes of the galaxies directly.

\subsubsection{Alignments with shape and position}
\label{sec:LHshape}

\begin{figure*}[t!]
\begin{center}
$\begin{array}{c@{\hspace{0.1in}}c}
\includegraphics[width=0.99\textwidth,angle=0]{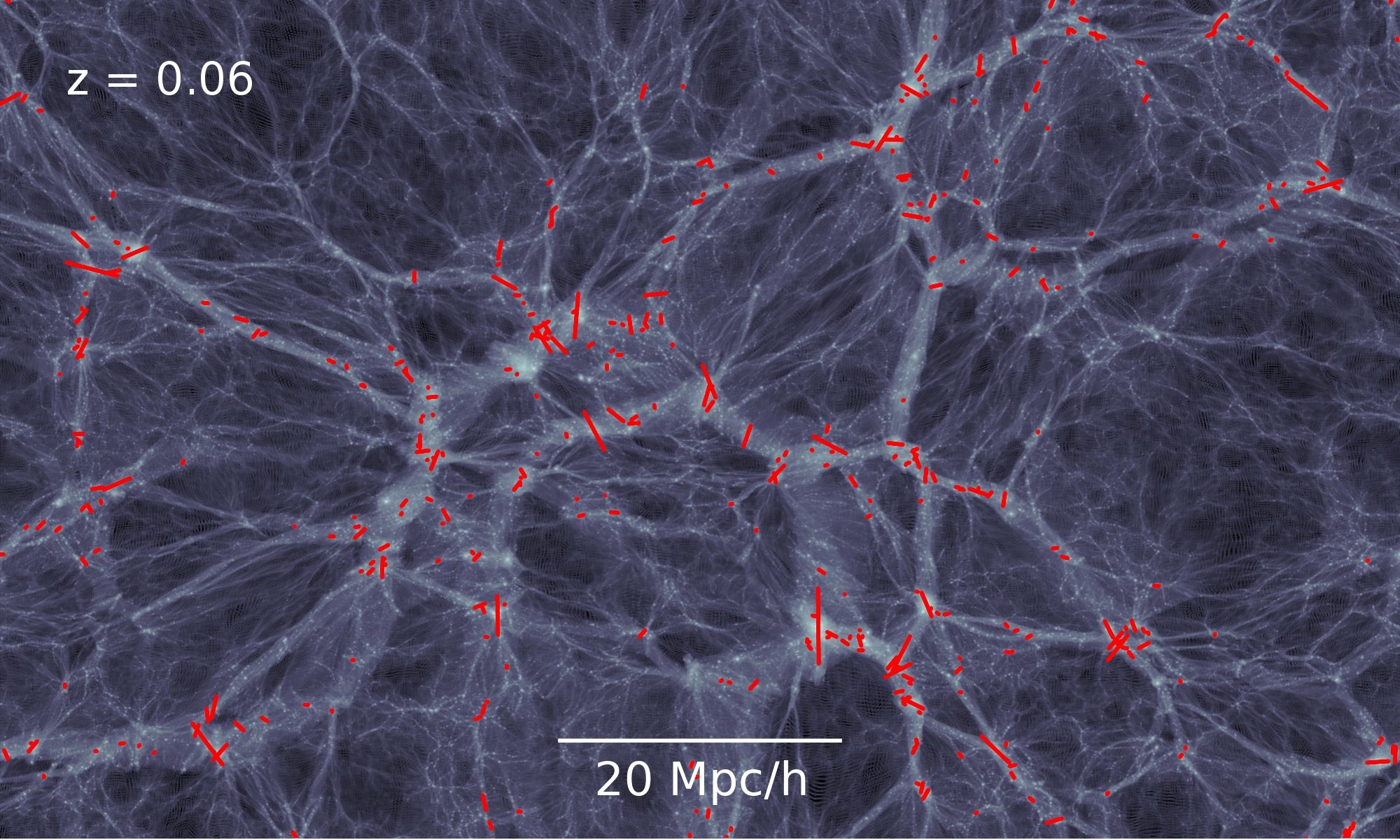} \\
\includegraphics[width=0.327\textwidth,angle=0]{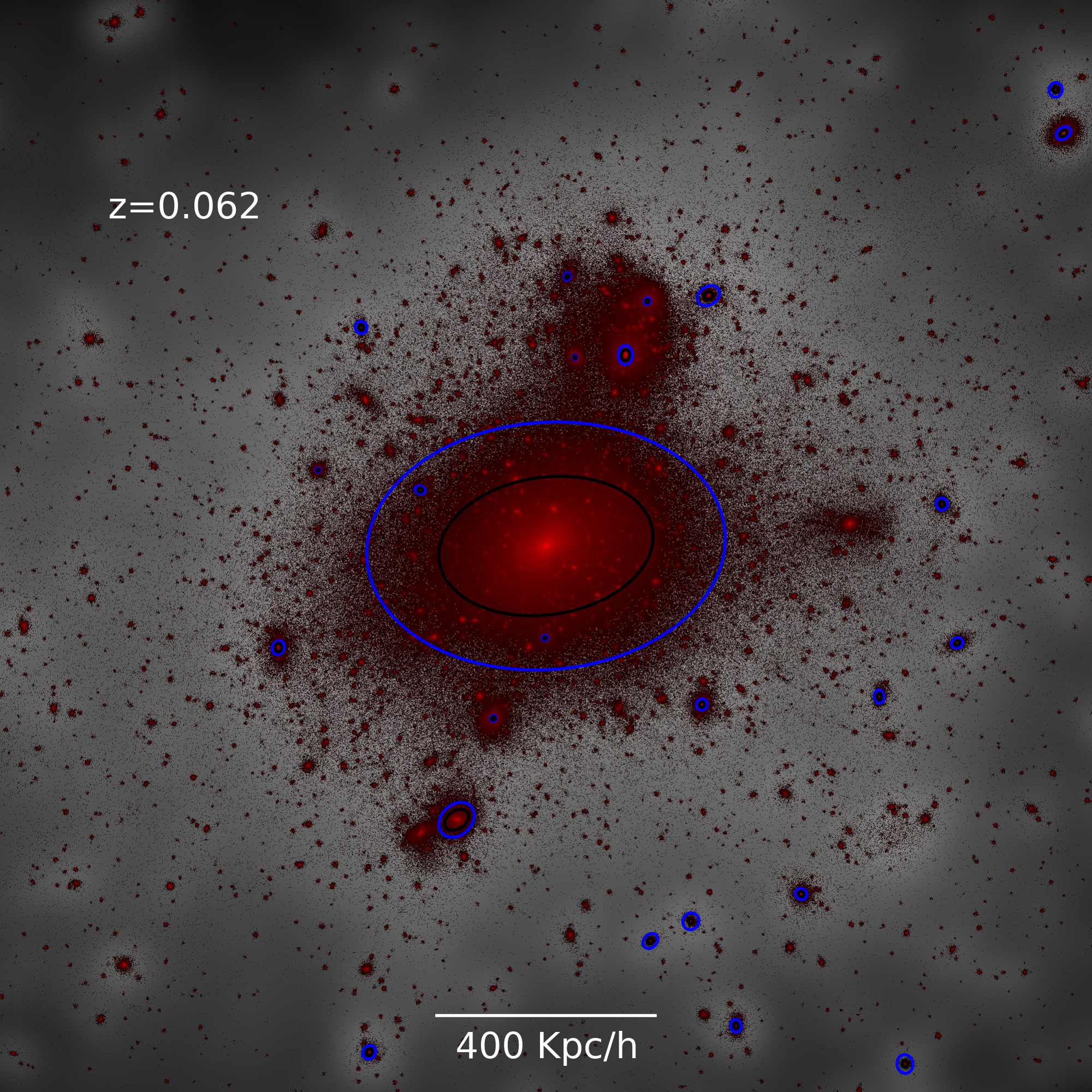}\
\includegraphics[width=0.327\textwidth,angle=0]{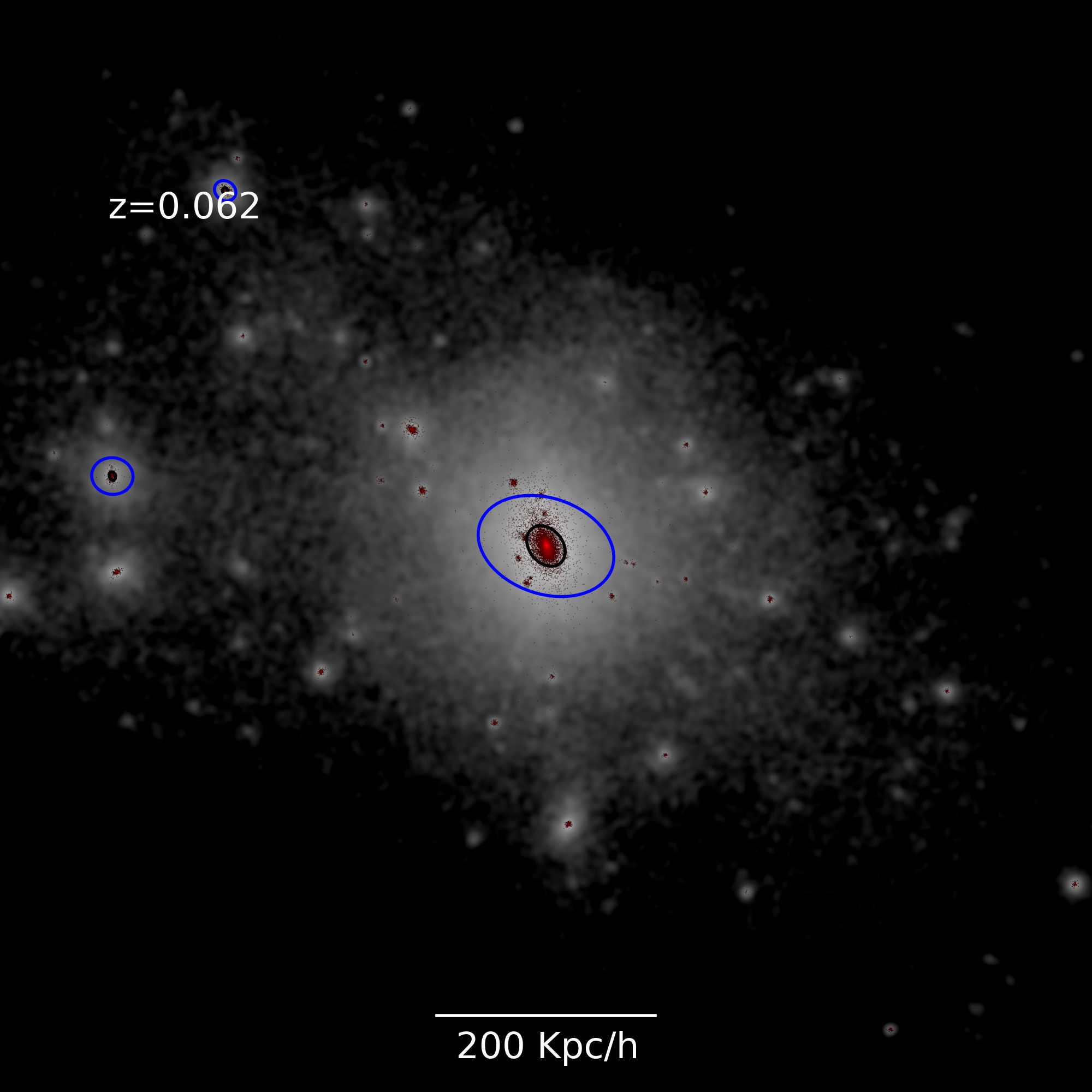}\
\includegraphics[width=0.327\textwidth,angle=0]{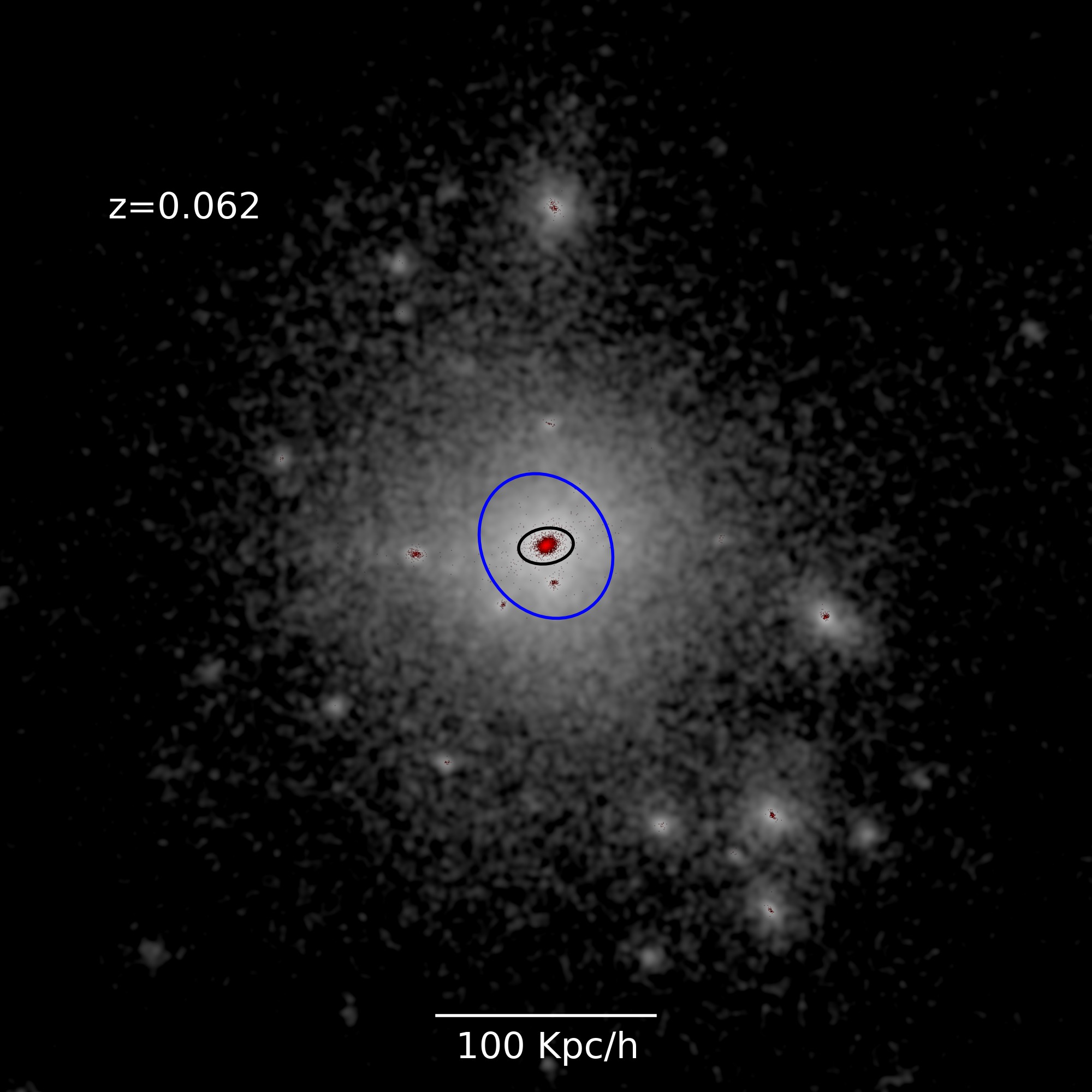}\\
\end{array}$
\caption{\label{F:mgiifig}{\em  Top:}   Snapshot  of  the   MB-II  SPH
  simulation in a  slice of thickness $2\:h^{-1}$Mpc at  redshift $z =
  0.06$. The  bluish-white colours represent  the density of  the dark
  matter  distribution and  the red  lines show  the direction  of the
  major axis of ellipse for the projected shape defined by the stellar
  component. The length  of the lines is linearly  proportional to the
  semi-major axis  length.  {\em Bottom  Left:} Dark matter  (shown in
  grey) and  stellar matter  (shown in red)  distribution in  the most
  massive   group    at   $z   =    0.06$   of   mass    $7.2   \times
  10^{14}\:h^{-1}M_\odot$.   The  blue  and black  ellipses  show  the
  projected  shapes of  dark matter  and stellar  matter of  subhaloes
  respectively. {\em  Bottom Middle:}  Dark matter and  stellar matter
  distribution in a group of mass $3.8 \times 10^{12}\:h^{-1}M_\odot$.
  {\em Bottom Right:} Dark matter and stellar matter distribution in a
  group of mass $1.1 \times 10^{12}\:h^{-1}M_\odot$.  \permmn{TMD+14}}
\end{center}
\end{figure*}

Unfortunately, any reliable measurement of the shape of a distribution
of  particles requires  the distribution  to be  sampled with  a large
number  of  particles.   Convergence tests  in,  e.g.,  \citet{TMD+14}
suggested that  hundreds of (but  preferably 1000) star  particles are
necessary  to accurately  infer  galaxy shapes  and therefore  measure
intrinsic alignments\footnote{There are measurements in the literature
  that use as few as 20  particles.  The results in this paper suggest
  that  those results  should  be approached  with  caution.} (with  a
similar requirement applying  to the number of  dark matter particles,
when studying dark  matter halo alignments).  This  leads to stringent
requirements  on  resolution;  for   example,  a  study  of  intrinsic
alignments  of galaxies  with  stellar mass  of $M_*\ge  10^{10}\msun$
requires a star particle mass of $10^7\:h^{-1}\msun$.  Relevant to the
understanding of intrinsic  alignments is the study of  the shapes and
alignments  of  galaxies with  respect  to  the (mostly  dark)  matter
distribution  in  their  host  haloes.  In  this  case,  a  stochastic
misalignment, $\theta_{\rm GH}$  (see \Cref{t:IAObservables}), between
galaxies  and  their  host  haloes  suppresses  the  galaxy  intrinsic
alignments with respect to those from $N$-body simulations.

\begin{figure*}[t]
\begin{subfigure}{0.45\textwidth}
  \includegraphics[width=\hsize,angle=0]{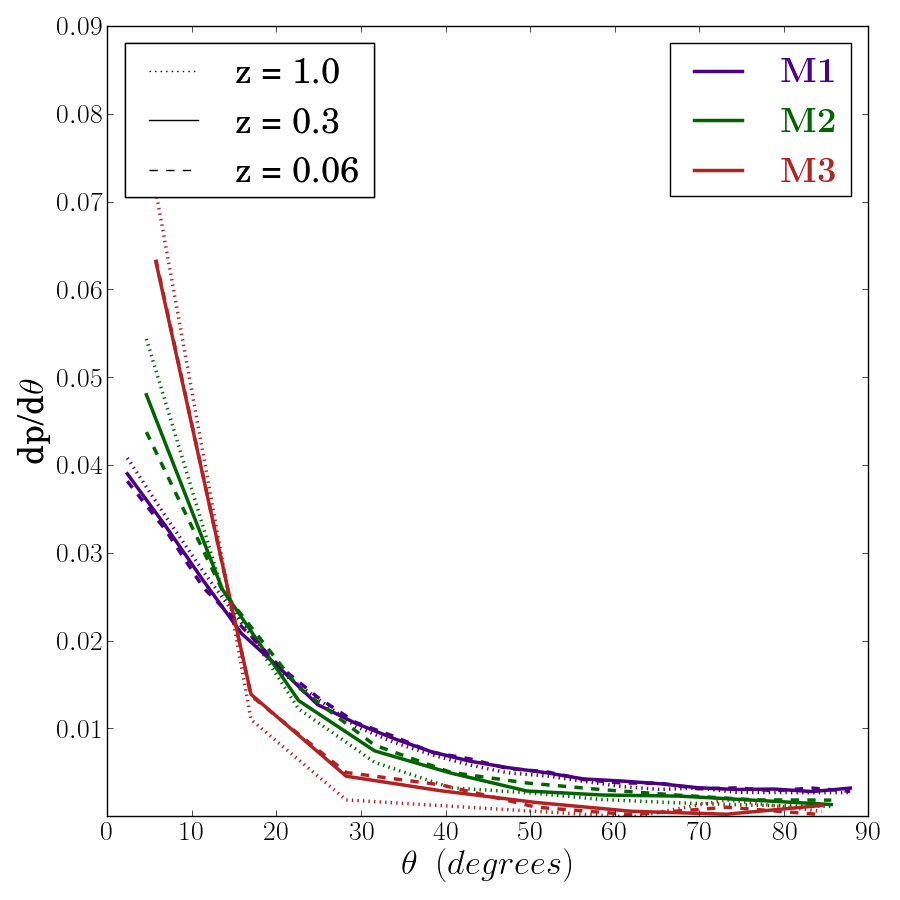}
  \caption{MassiveBlack II}
  \label{F:misalignment}
\end{subfigure}
\begin{subfigure}{0.515\textwidth}
  \includegraphics[width=\hsize]{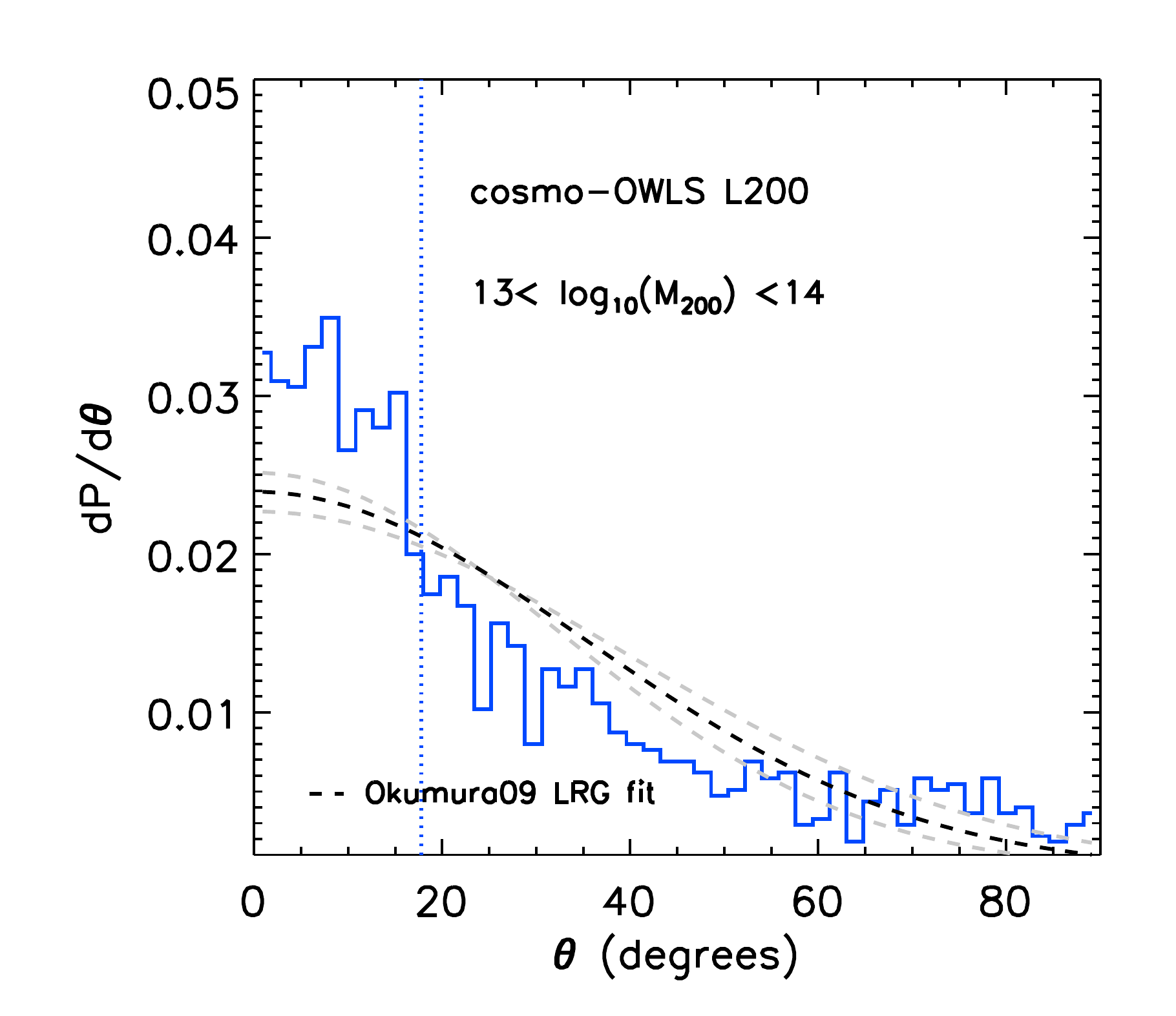}
  \caption{cosmo-OWLS}
  \label{F:misalignmentLRG}
  \end{subfigure}
\caption{\textit{Left}:   Histogram    of   2D    misalignments   from
  MassiveBlack  II  between  galaxy   and  dark  matter  halo  shapes,
  $\theta_{\rm GH}$, at  redshifts $z = 1.0, 0.3$, and  $0.06$ in mass
  bins  $M1, M2$  and $M3$  that  go from  lower to  higher mass  (see
  \citealp{TMD+14}  for  details).  The  $M3$  mass  bin is  the  most
  similar to  the mass  bin used  in the right  panel of  this figure.
  \permmn{TMD+14} \textit{Right:}  Histogram of 2D  misalignments from
  cosmo-OWLS between  stars (inside the stellar  half-mass radius) and
  the total  matter distribution for  haloes in the mass  bin $10^{13}
  \le M_{200}/ [h^{-1}\msun]  \le 10^{14}$ and at $z  = 0$, reproduced
  from \citet{VCS+15}.   The vertical line indicates  the median value
  of the  misalignment angle.   The black (mean)  and grey  (one sigma
  deviation) curves  are the  analytic Gaussian functional  forms that
  were employed  in the \citet{OJL09} observational  study of Luminous
  Red Galaxies (LRG).  \textit{Figure Credit: Marco Velliscig.}}
\label{fig:misalign}
\end{figure*}

\citet{TMD+14}    used    the     MassiveBlack-II    SPH    simulation
\citep[MB-II;][]{KDC+14},  with star  formation and  AGN feedback,  to
study  the shapes  and  alignments of  galaxies with  $\mathtt{>}1000$
particles (\Cref{F:mgiifig}). In a  similar study, \citet{VCS+15} used
the  Evolution  and  Assembly   of  GaLaxies  and  their  Environments
\citep[EAGLE;][]{schaye15,crain15}   and  cosmo-OverWhelmingly   Large
Simulations \citep[  cosmo-OWLS;][]{schaye10,LeBrun+14,McCarthy+14} to
study the  alignment and shape  of dark  matter, stellar, and  hot gas
distributions  with  $\mathtt{>}300$ particles.   Both  investigations
found  that   misalignments,  $\theta_{\rm   GH}$,  were   larger  for
lower-mass  haloes  than  for higher-mass  haloes  and  \citet{TMD+14}
showed that the  misalignments were only weakly  dependent on redshift
(see \Cref{F:misalignment}).  \citet{VCS+15} showed that, in the EAGLE
simulation,   the   median   misalignment   angle   was   reduced   to
$\mathtt{\sim}25-30^{\circ}$ in the most  massive haloes ($10^{13} \le
M_{200}/[h^{-1}\msun]  \le  10^{15}$)   and  both  \citet{VCS+15}  and
\citet{TMD+14} showed that the misalignment angles were systematically
reduced  at   all  masses  when   considering  2D,  rather   than  3D,
misalignment       angles.       Some       observational      studies
\citep[e.g.][]{HBH+04,OJL09} have  assumed a Gaussian  functional form
for the distribution  of the 2D stochastic  misalignment angle between
galaxies  and their  host  haloes.  However,  both \citet{TMD+14}  and
\citet{VCS+15}  showed  that  the  probability  distributions  of  the
misalignment   angle   were    not   Gaussian   (\Cref{fig:misalign}).
\Cref{F:misalignmentLRG}   shows   an   example  of   a   non-Gaussian
distribution of misalignment angles for  dark matter haloes, which are
potential hosts  of Luminous  Red Galaxies (LRG),  from \citet{VCS+15}
compared   with   the   assumed   Gaussian   functional   forms   from
\citet{OJL09}.   These  results  show  the  importance  of  exercising
caution when employing oversimplified assumptions about the stochastic
misalignment  between galaxies  and haloes  and when  interpreting the
measurements of galaxy intrinsic alignment correlation functions.

\citet{VCS+15} calculated the misalignment  of the baryonic components
(hot  gas and  stars) of  the  galaxies with  their host  haloes as  a
function of halo mass, radius, and galaxy type. Overall, galaxies were
found to align well with the  inner host haloes.  However, the stellar
distributions  exhibited  a   median  misalignment  of  $\mathtt{\sim}
45-50^{\circ}$ with respect to the outer host haloes.  One simulation,
calibrated to  reproduce the stellar  mass function at  redshift zero,
was used for comparison with the simulations that were generated using
different feedback models.  Importantly,  the simulations that did not
reproduce  the  basic  properties   of  galaxy  populations,  such  as
abundance  (number  density  per  comoving volume)  and  galaxy  size,
exhibited  differences  in  the   median  misalignment  angle  between
galaxies and their  host haloes of the order of  $10^{\circ}$ and this
could be as large as $20^{\circ}$  for the case where AGN feedback was
neglected in the simulation.

Both  \citet{TMD+14}  and  \citet{VCS+15}  used  only  the  unweighted
inertia tensor  defined in  \Cref{sec:shapes}.  However,  a subsequent
study by \citet{TMD+14b}  noted that this might have an  impact on the
conclusions, as  shape parameters  and alignment  angles significantly
depend on  the way they  are computed.  For  instance, \citet{TMD+14b}
included a study of how the  results depended on the use of unweighted
or  reduced  inertia tensors;  the  latter  weighting more  the  inner
regions of galaxies.  However,  weighting star particles by luminosity
rather than stellar mass did not lead to a noticeable change.

\citet{TMD+14b} took  the work  in \citet{TMD+14}  a step  further and
measured the intrinsic alignment  2-point correlation functions in the
MB-II  simulation.   These  included  measurements  of  position-angle
statistics   and  projected   correlation   functions  like   $w_{g+}$
(\autoref{eq:wg+}) for galaxies as a function of mass, luminosity, and
stellar mass,  with tabulated fits  to the non-linear  alignment model
(see \Cref{sec:IntermediateScales}) that demonstrate the scalings with
these parameters.  A preliminary  study was made of colour-dependence,
with blue  galaxies having  lower intrinsic alignment  amplitudes than
red  galaxies; however,  the colours  of the  galaxies were  not fully
realistic, so  a more detailed  study would be required  to understand
these  results.   Comparison of  $w_{g+}$  for  central and  satellite
galaxies within fixed subhalo mass ranges suggests that at lower mass,
central  and satellite  galaxies have  similar small-scale  alignments
($<1\:h^{-1}$Mpc)   but  satellites   have   smaller  (but   non-zero)
large-scale    alignments.     At   intermediate    subhalo    masses,
$10^{11.5}<M<10^{13}\:h^{-1}\msun$,  central  and satellite  subhaloes
were found to  have similar intrinsic alignments on  all scales, which
is somewhat  in tension with observations  (see \citealt{paper3}). The
$w_{g+}$  predictions  for a  massive  galaxy  sample were  reasonably
consistent with SDSS luminous  red galaxy measurements, which provided
some validation  that the  simulation included the  relevant processes
for galaxy intrinsic alignments of at least massive galaxies.

More recently, \citet{TMD15} investigated a critical ingredient of the
previous results: the impact of  galaxy formation physics on intrinsic
alignments.  This work relied on  direct comparison between MBII and a
dark  matter-only  simulation  with  the  same  box  size,  cosmology,
resolution, and  initial conditions,  so that  any differences  in the
results are  due only  to the additional  physics in  the hydrodynamic
simulations.  The  study matched  dark matter  haloes between  the two
simulations and showed that the shapes of the matched full dark matter
haloes  had  similar  orientations,  especially at  the  highest  halo
masses.  They also measured the  alignment of the stellar component in
MBII with that  of the inner part  of the dark matter halo  in the two
simulations.   A correlation  between  the orientations  was found  in
either case.  However, the correlation  between the orientation of the
stellar  component  in  MBII  and   the  matching  halo  in  the  dark
matter-only  simulation  was  weaker  than that  between  the  stellar
component in MBII  and the halo in MBII.  Further,  the authors showed
that misalignments between  the stellar component and  the dark matter
halo increased with  increasing radii in MBII,  while the misalignment
of the stellar component in MBII with the dark matter-only halo showed
no  significant change  with increasing  radius.  This  indicates that
there is  some coevolution of the  baryons and the dark  matter at the
inner halo  radii, which is  similar to the results  of \citet{BKG+05}
(see   \Cref{sec:SHinternal}).     However,   unlike   \citet{BKG+05},
\citet{TMD15} did find  a weak correlation between the  MBII inner and
outer haloes that may have been  missed in this earlier work (and also
in   the   satellite   alignment  study   of   \citealp{DMF+11},   see
\Cref{sec:SHsatellite})   due  to   the  smaller   number  statistics.
\citet{TMD15} concluded  that it  would not be  useful to  measure the
shape of the  dark matter component at  the inner radii as  a means to
trace the  shape and orientation of  the stellar component and  that a
mapping  between the  hydrodynamic  and  dark matter-only  simulations
should utilize  the shapes  of the  full haloes.   For more  detail on
those results, see \citet{TMD15}.

\citet{DLK+14}   investigated   the   alignments  of   the   satellite
distribution  using  the positions  of  satellites  within their  host
halos.   An  SPH simulation  was  used  with a  $100\:h^{-1}$Mpc  box,
including  the effects  of gas  cooling, star  formation, and  stellar
feedback  (although  no AGN  feedback,  which  is necessary  to  avoid
over-cooling).  They  investigated the  alignment of  central galaxies
with the positions of satellites, finding stronger alignments both for
red  centrals and  red satellites,  consistent with  the observational
picture (see \citealt{paper3} for  more details).  Interestingly, they
found  a physical  reason  for this  effect, which  was  that the  red
satellites stayed closer to the inner regions of the dark matter halo,
which correlated  strongly with the  central galaxy shape  (unlike the
outer regions of the dark matter halo, which were less correlated with
the central galaxy shape).  They found  trends with the host halo mass
and  with satellite  galaxy  metallicity,  providing predictions  that
could be tested with future datasets.

\subsection{Hydrodynamic simulations Roundup}

In this  section, small-volume and zoom  hydrodynamic simulations were
reviewed   as   well   as   large   cosmological-volume   hydrodynamic
simulations.   There   is  far  less   literature  in  this   area  as
hydrodynamic simulations are time consuming and the baryonic processes
that occur  on scales below the  resolution of the simulation  are not
well  understood.   Small-volume  simulations can  be  run  reasonably
quickly, but suffer from small number statistics.  Cosmological-volume
simulations contain a good statistical sample of galaxies but are more
difficult and time consuming to  run, meaning that there are currently
very few cosmological-volume simulations suitable for galaxy alignment
studies.

It is clear that the addition of  baryons cause the shapes of the dark
matter  haloes to  change.  Haloes  with baryons  tend toward  oblate,
rather  than the  prolate shapes  found in  the $N$-body  simulations.
There is  a discernible correlation  between the galaxy and  the inner
halo, which is the result of  coevolution of the baryons and the inner
dark matter halo, but the addition  of baryons causes the alignment of
the inner and outer haloes to become only weakly correlated.

Satellite distributions appear to be  well aligned with the outer host
halo when late-type disc galaxies are considered, while red early-type
galaxies  tend to  be better  aligned with  red central  galaxies.  It
appears that  red satellites  may be generally  located closer  to the
centre of  the halo, meaning  that they would  be more subject  to the
tidal forces of the inner halo.

The relative alignments of the dark  matter, gas and star spin vectors
within  a  galaxy  system  may  experience  numerous  flips  over  its
lifetime,  which will  have  a  significant effect  on  the shape  and
orientation of the galaxy over this time.  From a galaxy formation and
evolution  perspective,  this is  certainly  an  interesting field  to
pursue. However,  it is important  to note that while  spin alignments
are likely the dominant alignment mechanism for rotationally-supported
galaxies, the study of spin  alignments may omit some critical details
for intrinsic  alignment contamination  of weak  lensing measurements,
given  that  observed  intrinsic   alignments  are  dominated  by  red
galaxies, which  are pressure-supported and thus  subject to intrinsic
alignments due to different physics (e.g., tidal effects).

There is  some limited (to  date) but compelling evidence  that galaxy
alignments have  some dependence on  the sub-grid physics,  within our
current  abilities  to  determine  this  given  the  small  number  of
hydrodynamic  simulations with  sufficient  volume  and resolution  to
answer  this  question.   Further  studies  on  this  topic  would  be
useful.  It is,  however, clear  that the  future of  galaxy intrinsic
alignment studies will certainly benefit from extensive research using
hydrodynamic simulations for progress and this is discussed further in
\Cref{sec:Roadmap}.

\section{Semi-analytic modelling}
\label{sec:SemiAnalytic}

It  will  remain   impossible  for  the  foreseeable   future  to  run
hydrodynamic simulations with the  volumes, resolutions, and number of
realisations  required  to  generate  the  high  precision  covariance
matrices  required for  upcoming  telescope missions,  or for  direct,
robust  and   precise,  measurements  of  intrinsic   alignments.   An
established way  to link  the output of  pure $N$-body  simulations to
observables of galaxies is the use of analytic prescriptions that take
the properties of dark matter haloes as input. They can be informed by
simplified and parametrised physical  models or by effective relations
determined from higher resolution or hydrodynamic simulations.

Such  semi-analytic  models  of  galaxy formation  and  evolution  are
routinely  used  to predict  observables  like  galaxy clustering  and
luminosity  functions  (see  \citealp{BBM+06} for  a  review).   These
semi-analytic models  of galaxies  have been embedded  in a  number of
$N$-body simulations  in order to  investigate weak lensing II  and GI
signals \citep[e.g.][]{OJL09,OJ09} and  satellite alignments in haloes
\citep[e.g.][]{ZKG+05,KMG+05,AB06b,AB10,KBY+07,MKL08,BPN+08,FLW+09,LFC+09}\footnote{Some
  of  these studies  use  a halo  occupation distribution  \citep[HOD;
    e.g.][]{BW02}  approach to  reproduce observed  galaxy properties.
  This is  a more statistical  approach to matching  galaxy properties
  than  semi-analytic  modelling.}.   The  requirement  for  realistic
galaxies  lies  in the  fact  that  observations have  indicated  that
satellite  alignments appear  to be  different for  a range  of galaxy
types;  e.g. early-  and late-type  host galaxies,  red and  blue host
galaxies,  satellites   with  low   and  high  star   formation  rates
\citep[see][for   further   discussion  on   observational   satellite
  alignment  results]{paper3}.   If  the  central  galaxy  is  naively
assumed to follow the shape of  the host halo, the semi-analytic model
predicts   a   larger   alignment   signal   than   the   observations
\citep[e.g.][]{AB06b,KBY+07}.  It is also interesting to note that, in
order to correctly  reproduce magnitudes and colours,  the models have
to take  into account  a notion of  size such as  the scale  length of
discs.   Therefore  it seems  plausible  to  extend the  semi-analytic
approach to the shapes and orientations of galaxies.

The  simplest ansatz  assumes that  galaxies are  homologous with  the
underlying dark  matter halo (in  3D or  in projection), which  may be
applicable  to elliptical  galaxies.   In  their simulation  analysis,
\citet{HRH2000}  proposed additionally  a model  for spiral  galaxies,
which  are assumed  to  be  thin discs  perpendicular  to the  angular
momentum of  the host  dark matter  halo. \citet{HBH+04}  extended the
spiral  model in  two ways:  (a)  they now  assumed a  thick disc  via
rescaling  all  ellipticities by  a  constant  factor 0.73  which  was
estimated  from  observed  ellipticity  distributions,  and  (b)  they
allowed for a misalignment between  the angular momentum of the galaxy
and the halo, randomly drawn from a distribution fitted to the results
from the simulations of \citet{BAC+02}.

\citet{HWH+06}  made the  first attempt  to model  a realistic  galaxy
population  with   intrinsic  alignments  based  on   the  conditional
luminosity  function   by  \citet{CM05},   which  provides   the  halo
occupation with elliptical or spiral  galaxies of a certain luminosity
for a given  redshift and halo mass. Using only  haloes with occupancy
one,  i.e.   central   galaxies,  \citet{HWH+06}  assigned  elliptical
galaxies the shape of the parent  halo and modelled spiral galaxies as
in \citet{HBH+04}. The resulting `mix'  model of elliptical and spiral
galaxies produced an II signal that is in good agreement with the SDSS
main sample measurements of  \citet{MHI+06}. \citet{HWH+06} also found
a clear GI signal that showed no redshift evolution and increased with
host halo mass.

While  the halo  occupation formalism  captures the  average alignment
properties of galaxies with a  certain halo mass, a full semi-analytic
model is capable of tracking any dependencies on the merger history of
an individual halo. Thus, only the latter can account for correlations
between alignment  and galaxy luminosity, colour,  morphological type,
etc. induced  by the  evolution of that  galaxy. \citet{JSB+13,JSH+13}
complemented the  Millennium Simulation  and the  Durham semi-analytic
models \citep{BBM+06} with a range of prescriptions for galaxy shapes.

Elliptical galaxies  were still assumed  to follow the shape  of their
haloes; however,  while the aforementioned  works did not  specify how
the  halo ellipticity  was measured,  \citet{JSB+13,JSH+13} considered
explicitly  an ellipticity  based on  the simple  inertia tensor  (see
\autoref{eq:Inertia}  and  following  text)  and  a  rescaled  version
approximating  the  reduced  inertia   tensor,  the  latter  generally
producing more  circular galaxy shapes. Spiral  galaxies were modelled
again as a  thick disc, but the misalignment  distribution was updated
to one  from \citet{B12} and  the disc  was approximated as  an opaque
cylinder  with two  different  values of  thickness.  Satellites  were
included, but  as no subhalo  properties were available,  their shapes
were  assumed  to  be  determined  by the  same  prescription  as  the
respective central galaxy type, in some cases with rescalings based on
the observations by \citet{KDP+08}.   All satellites point towards the
centre  of  the  host,  with  misalignment  distributions  taken  from
\citet{KDP+08} and \citet{B12}.

\begin{figure}[t]
\centering
\includegraphics[width=0.7\textwidth]{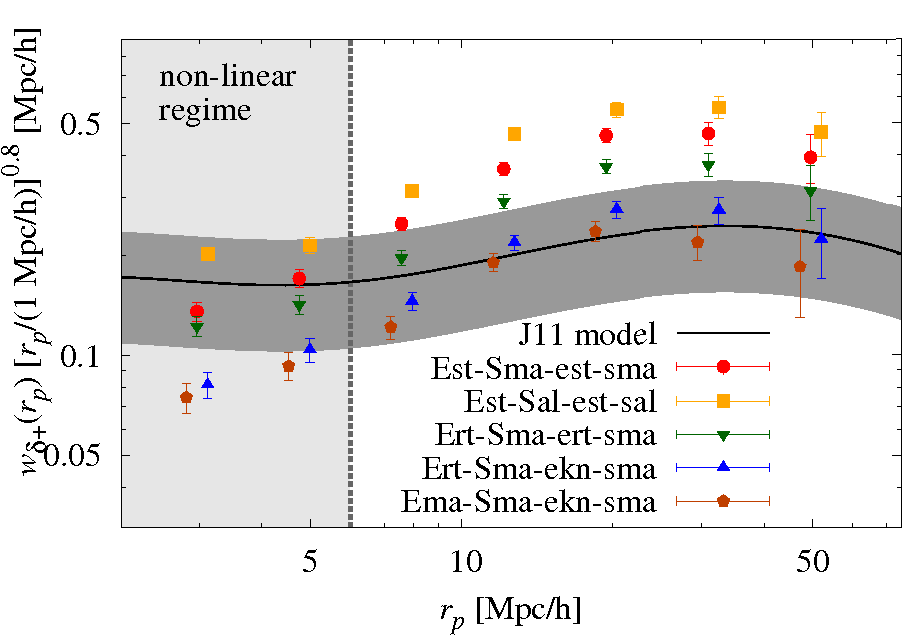}
\caption{Correlation  function between  matter density  and tangential
  galaxy  ellipticities,  $w_{\delta +}$,  as  a  function of  comoving
  transverse  separation $r_p$.   The signals  resulting from  various
  parametrisation choices in the  semi-analytic prescription are shown
  as  coloured symbols  (for a  detailed  explanation see  Table 1  of
  \citealp{JSH+13}).   Error bars  are  determined  from the  variance
  between 64 simulated fields. The black line is the prediction by the
  \citet{JMA+11} form  of the non-linear alignment  model, using their
  best-fit  parameters  for  the  amplitude  and  the  luminosity  and
  redshift  dependencies. The  dark  grey band  marks the  uncertainty
  propagated from  the $1\sigma$ errors on  these parameters.  Signals
  from the mocks have been divided  by linear galaxy bias, so that the
  measurements below $6\:h^{-1}$Mpc cannot directly be compared to the
  prediction. \permmn{JSH+13}}
\label{fig:joachimi_megaz_semianalytic}
\end{figure}

These models were confronted  with intrinsic ellipticity distributions
for various galaxy  samples extracted from the COSMOS  survey, as well
as   the   GI   correlation   measurements   of   \citet{JMA+11}   and
\citet{MBB+11}.  No  combination  of  models   was  able  to  fit  all
observables  simultaneously,  although the  semi-analytic  predictions
generally reproduced at least the correct trend and order of magnitude
(see \Cref{fig:joachimi_megaz_semianalytic}), despite the numerous
simplistic  assumptions. A  notable  exception is  the GI  correlation
function $w_{\delta  +}$ for early-type  galaxies with $M_r <  -19$ for
which  all  models  strongly  over-predict the  amplitude.  This  could
indicate a  failure of  the alignment or  galaxy evolution  models for
satellite galaxies, which dominate these  samples, or suggest that the
linear luminosity  scaling of  $w_{\delta +}$ found  by \citet{JMA+11}
cannot be extrapolated to fainter magnitudes.

There are clear  avenues for improving semi-analytic  models of galaxy
shapes  and  orientations,  such  as updating  scaling  and  alignment
relations from state-of-the-art simulations \citep[e.g.][]{TMD+14} and
iterating on  the alignment mechanisms, e.g.  by including information
on  the local  tidal  shear tensor  instead of  just  halo shape.  The
requirements on the underlying $N$-body simulations are quite demanding,
especially on  the mass  resolution, which should  at least  allow for
about 300 particles (ideally more  than $10^3$ particles) per halo for
which   shapes    and   angular    momenta   are   to    be   measured
\citep{BEF+07,JSB+13,TMD+14}. If these  halo properties are determined
only for  the central part of  haloes, then the same  requirement will
translate  into  even lower  particle  mass  for the  simulation.  For
sufficiently small scales and high  precision the impact of baryons on
the properties of the dark matter halo will become relevant, making an
analytic link  between luminous and  dark matter alignments  even more
challenging.

\section{Roadmap/wish list}
\label{sec:Roadmap}

When  planning to  connect theoretical  predictions and  observations,
what would an idealized roadmap or wish list look like?

In an  ideal scenario with  unlimited resources, simulations  would be
run in a  large ($\mathtt{\simeq}4$Gpc) box with a  very high particle
mass  resolution\footnote{The large  box  size would  account for  the
  large  scales  that  should  be  accounted for  in  studies  of  the
  large-scale structure and the high mass resolution would resolve the
  shapes  and details  of  individual galaxies  with high  precision.}
($\mathtt{\simeq}10^6\msun$).   Suites   of  hydrodynamic  simulations
would  be   generated  using  many  different   simulation  codes  for
comparison and  a wide  exploration of  baryonic physics  and feedback
effects would be included to gauge their impact on the simulations and
to  compare  with  observations   for  accuracy.   In  reality,  these
simulations  would be  impossible to  run, using  far more  CPU hours,
memory, disk  space and time  than are  available in the  world today.
Given a realistic landscape, the  following list represents tasks that
would be useful to pursue for future galaxy alignment analyses.

\subsection{Simulations  with  sufficient resolution  and  numbers  of
  haloes}
\label{sec:Wish1}

\begin{figure}[t]
\begin{center}
\includegraphics[height=0.782\textwidth,angle=270]{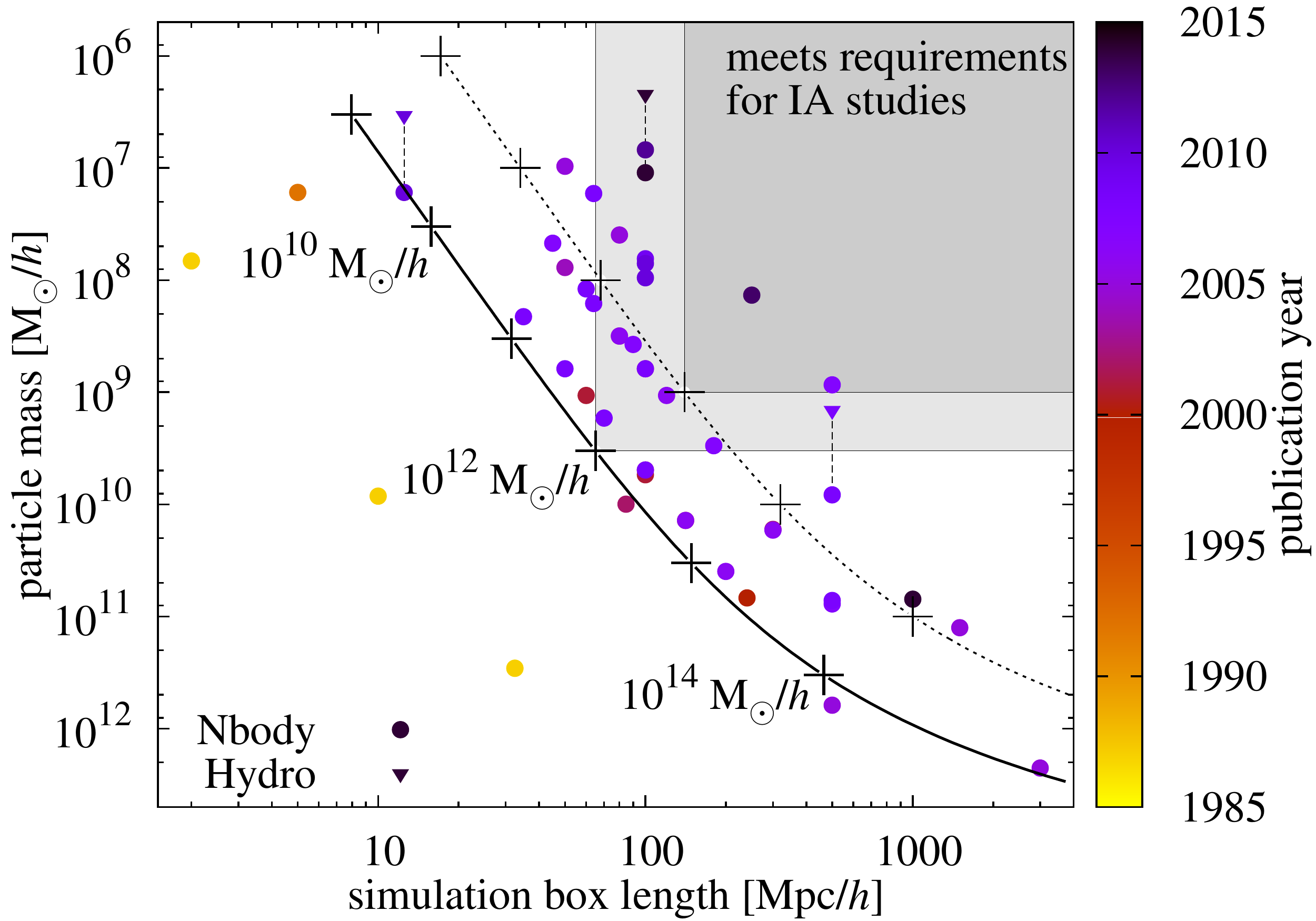}
\caption{Simulation specifications critical  for halo/galaxy alignment
  studies. Plotted are the particle mass and simulation box length for
  simulations  underlying a  selection  of alignment  works, with  the
  publication  year of  said works  given by  the colour  coding. Dots
  (triangles)  represent   $N$-body  (the  gas/stellar   component  of
  hydrodynamic) simulations. Dashed lines link the gas and dark matter
  parts of a given simulation.  The black solid (dotted) line connects
  points at  which one  expects to  find $10^3$  ($10^4$) haloes  of a
  given   mass   that   are   sampled  with   300   (1000)   particles
  each. Multiples  of ten for these  halo masses are indicated  by the
  crosses.  Areas to the  top and right of a given  point on the lines
  cover  simulations  that  fulfill  the  minimum  (solid)  or  strict
  (dotted) criteria  for accurate alignment measurements  of haloes of
  that mass,  as illustrated by the  grey areas for a  Milky Way sized
  halo ($10^{12}\:h^{-1}\msun$).}
\label{fig:simulation_specs}
\end{center}
\end{figure}

Simulations  are  currently  still  the best  way  to  test  intrinsic
alignment models  and mitigation techniques.   The first task  on this
wish list  is to  determine the size  and resolution  that simulations
should   have   for   future  intrinsic   alignment   studies.    When
investigating the properties of  $N$-body haloes, \citet{BEF+07} found
that  in order  to mitigate  numerical artifacts,  each halo  required
greater  than  300  particles.   However,   a  more  recent  study  on
hydrodynamic simulations found that  results converged when haloes had
$\mathtt{\sim}1000$  particles  \citep{TMD+14}.    A  simulation  with
$10^3$  haloes resolved  will  provide a  minimum statistical  sample,
while  $10^4$ haloes  will generate  a more  statistically significant
result. These numbers are roughly  informed by the galaxy sample sizes
of observational  intrinsic alignment  studies \citep[e.g.][]{JMA+11},
where a few thousand galaxies  are typically required for a detection,
while  a  few  ten-thousand  galaxies  allow  for  the  definition  of
subsamples  and  higher signal-to-noise.   \Cref{fig:simulation_specs}
shows  a census  of  published  simulations that  have  been used  for
halo/galaxy alignment studies to date.   Only simulations with a fixed
resolution, to  allow for a  direct and fair comparison,  are included
(zoom simulations and resimulations are excluded).  The area above and
to the  right of the solid  (dotted) line is where  a simulation would
need  to lie  to meet  the  minimum (strict)  criteria for  performing
galaxy alignment studies  of galaxies in haloes of a  given mass.  The
example shown in  this figure is for studies wishing  to resolve Milky
Way sized  haloes ($10^{12}\:h^{-1}\msun$)  and currently only  two of
the  simulations  published  meet  the strict  criteria  required  for
intrinsic alignment studies.  Therefore, when planning simulations for
future  intrinsic  alignment  studies,  this figure  can  be  used  to
determine  minimum box  size and  particle resolution  requirements to
ensure that  the simulations produce meaningful  results.  Needless to
say, the  simulations described  in the following  tasks in  this wish
list should all conform to the requirements outlined here.

\subsection{A mapping between hydrodynamic and dark matter-only simulations}
\label{sec:Wish2}
The addition  of baryons  to a  simulation has an  impact on  the dark
matter distribution,  as was  shown in  \Cref{sec:Hydro}. In  order to
quantify  these effects,  the  second task  in this  wish  list is  to
provide   a  mapping   between  hydrodynamic   and  dark   matter-only
simulations.   This  will  require  a detailed  comparison  between  a
hydrodynamic simulation and a dark matter-only simulation with exactly
the same initial conditions.  The resulting simulations should contain
the  same  dark  matter haloes  in  general\footnote{Some  differences
  between  the hydrodynamic  and  dark  matter-only simulation  haloes
  should  be expected  and  may  arise if  the  halo  finders split  a
  structure  into  different haloes  or  some  haloes fall  below  the
  minimum  particle threshold  between the  realisations etc.},  which
will  enable  a  quantative comparison.   \citet{TMD15}  performed  an
initial analysis of dark matter halo and subhalo shape and orientation
misalignments, as a  function of radius from the  halo centre, between
two such  simulations.  They showed  that the shapes  and orientations
measured  between the  simulations was  far  from random  and that  it
should  be possible  to  develop a  mapping  between the  simulations.
Further investigation is still required to quantify such a mapping and
then to utilize this to produce more precise semi-analytic models.

\subsection{Accurate galaxy properties in hydrodynamic simulations}
\label{sec:Wish3}

Beyond the technicalities involved in  the estimation of galaxy shapes
and misalignments in \Cref{sec:Nbody,sec:Hydro}, the open challenge in
studying galaxy  alignments with hydrodynamic simulations  remains the
fact that the  same processes that determine galaxy  shapes and mutual
alignments are connected (if not exactly  the same) to those that give
rise  to several  other  observed  scaling relations.   Unfortunately,
hydrodynamic simulations  have not yet converged  on an implementation
of  the  physics  behind  these  processes  and  often  they  fail  in
reproducing basic observables  such as the number  density of galaxies
as  a function  of  their luminosity  and/or  the mass-size  relation.
Although  these  simulations are  ideal  tools  to highlight  possible
hidden  connections between  different  processes, they  have not  yet
reached the maturity to have predictive power.  The third task in this
wish list  is the  most challenging  and time  consuming -  to produce
hydrodynamic simulations with more realistic galaxy properties.

For galaxy alignment  studies, there is good evidence  to suggest that
the alignment signal is a function of halo mass \citep[e.g.][]{SFC12},
but  observations provide  luminosity  or stellar  mass. However,  see
\citet{SMM14}  who  investigated  galaxy  intrinsic  alignments  as  a
function  of halo  mass as  determined from  simultaneous galaxy  weak
lensing measurements.  On the largest  scales, theory appears to match
the      observed      galaxy     alignments      reasonably      well
(\Cref{sec:LargeScaleTheory};   \Cref{f:ModelComparison})   and   they
appear to be  well modelled phenomenologically on  the smallest scales
(\Cref{sec:SmallScaleTheory};  \Cref{f:ModelComparison}).   Therefore,
if hydrodynamic simulations are able to match the observed stellar and
halo  masses, matching  the  galaxy alignment  signal to  observations
should be possible.  However, modelling  of the intermediate scales is
still      largely      unknown       and      poorly      constrained
(\Cref{sec:IntermediateScales})  and the  transition  from the  linear
regime to  the non-linear (high  density) regime is  where simulations
will  provide  significant  insight  to the  understanding  of  galaxy
alignments (see Figure 12 in \citealp{paper1}).  It is also clear that
the  environment (local  and  large-scale) should  have  an impact  on
intrinsic alignments.   As an  initial request, the  simulation should
match the abundances  of galaxies as a function of  stellar mass, type
(early- and late-type) and colour. As  an additional step, it would be
desirable to match the  size and ellipticity distributions\footnote{In
  order to  match ellipticity distributions,  hydrodynamic simulations
  would need to  resolve the thickness of the  discs accurately, which
  would require resolving interstellar medium processes accurately. It
  is  not  clear  that  this  final point  will  be  possible  in  the
  simulations, so this task may have  to remain on the ``wish'' list.}
of galaxies to observations.

\subsection{High precision semi-analytic simulations}

The  hydrodynamic simulations  outlined  in  \Cref{sec:Wish3} will  be
essential to providing  insight into how galaxies form  and evolve and
in modelling  the galaxy alignment signal.   However, generating these
simulations will  use enormous computing  resources -- both  CPU hours
and disk  space -- so  it will not be  possible to generate  the large
suites   of  simulations   that   are  required   for  upcoming   weak
gravitational lensing  surveys.  Therefore,  a faster way  to generate
simulations that  contain realistic intrinsic alignments  is required.
The final task in this wish list  is to take the galaxy alignments and
galaxy   properties   derived   from  these   increasingly   realistic
hydrodynamic  simulations  in  \Cref{sec:Wish3} and  use  the  mapping
identified in \Cref{sec:Wish2} to  populate $N$-body simulations.  The
semi-analytic simulations in  \Cref{sec:SemiAnalytic} used theoretical
models of alignments to populate  the $N$-body simulations.  This task
represents  the  next  generation  of  semi-analytic  simulation  that
contains the  increasingly realistic galaxy properties  and alignments
derived  from  hydrodynamic  simulations,  likely in  the  form  of  a
probability  distribution  to  account for  the  generally  stochastic
nature  of the  processes involved.   Another option  to populate  the
$N$-body simulations  may be to  use machine learning  algorithms that
use hydrodynamic simulations or, use current observations of intrinsic
galaxy alignments, as  a training set.  This task  should be completed
concurrently with the  previous tasks.  For each  advance in knowledge
coming from hydrodynamic simulations, many $N$-body simulations may be
populated for testing and development of mitigation techniques.

\subsection{Moving Forward}

The efforts  outlined here represent areas  where significant progress
can be made in  understanding galaxy alignments.  However, simulations
alone  will  not   be  able  to  provide   a  complete  understanding.
Additional  progress  will  be  made  by  combining  simulations  with
observations  and analytic  modelling, especially  in the  transitions
between  the   linear  and   non-linear  regimes.    Hydrodynamic  and
semi-analytic simulations must be compared with observations to ensure
accurate reproduction of galaxy properties.  Observations also provide
constraints on galaxy alignments directly,  which can then be compared
with  the  simulations  and  used   to  inform  the  analytic  models.
Similarly, insights coming from the  simulations can be used to inform
the analytic  models with the  hopes of  unifying the models  over all
scales (see \Cref{sec:roundup}).

These tasks are time consuming, challenging  and in some cases may not
be possible (particularly some of  the tasks in \Cref{sec:Wish3}), but
these are  areas where efforts  should be focused to  ensure continued
progress in understanding how intrinsic alignments form and evolve.

\section{Concluding Remarks}
\label{sec:Conclusions}

While the  title of this  review refers to  galaxies, as they  are the
primary observable  accessible to large-scale  structure observations,
this  review  has shown  that,  within  a theoretical  and  simulation
perspective, it is possible to  measure the shapes and orientations of
structures over  a wide range  of scales, from satellite  galaxies and
subhaloes through to the large-scale structures of the cosmic web.

Although  a reasonably  consistent  picture exists  for alignments  of
\emph{dark matter  haloes} in $N$-body simulations,  these are proving
to   be  of   limited   use  in   understanding   the  alignments   of
\emph{galaxies}, which  remain the focus  of interest for  both galaxy
formation and  evolution and  weak gravitational lensing  studies.  To
this  end,  future  efforts  would  be  best  spent  on  understanding
alignments  through  the  theory,   modelling  and  simulations  (both
hydrodynamic  and semi-analytic)  of galaxies  within their  host dark
matter haloes.

The theories and  models of the alignments of galaxies  span large and
small scales for  both early- and late-type  galaxies, with reasonable
success.  However, even for samples of bright elliptical galaxies, for
which  detailed  and   high-signal-to-noise  measurements  exist,  the
intermediate-scales  between   $1-10\:h^{-1}$Mpc  are   still  proving
difficult to model accurately. Work is  ongoing in this area and it is
expected  that a  model that  accounts for  clustering, halo  bias and
redshift-space   distortions  will   effectively   unify  the   small,
intermediate,  and  large  scales.  In  the  case  of  disc  galaxies,
observations generally yield marginal detections at best, so that only
future data  will be able  to tell if  the prevailing models  based on
tidal  torque  theory  provide  good  descriptions  of  a  variety  of
alignment processes.

Simulations  also provide  much needed  insight into  galaxy intrinsic
alignments  on  all   scales.   Only  with  suites   of  $N$-body  and
hydrodynamic  simulations will  it be  possible to  quantify alignment
correlations in  the non-linear  regime, investigate the  link between
the morphology of  bright and dark matter, and  establish the physical
mechanisms behind galaxy alignments -- whether these are driven by the
well-understood tidal stretching and  torquing paradigms, or by purely
non-linear  causes  such  as   vorticity  effects.   It  is,  however,
important to be  realistic about the limitations  of simulations.  For
example, the  sub-grid physics  in hydrodynamic simulations  are still
poorly understood  and represent  a large area  of uncertainty  in the
simulations.    Additionally,    reproducing   realistic   ellipticity
distributions in hydrodynamic simulations may never be possible due to
resolution effects.  This is not a reason to disregard simulations; it
is  simply a  reminder to  use caution  when interpreting  the results
presented,  particularly  when  the  studies  do  not  adhere  to  the
resolution requirements outlined in \Cref{sec:Wish1}.

The continued study of alignments  is important for both understanding
galaxy  formation  and  evolution  and for  mitigating  the  intrinsic
alignment effect in cosmological weak lensing surveys.  In particular,
upcoming   billion  dollar-class   telescope   missions  like   Euclid
\citep{LAA+11},     the     Large    Synoptic     Survey     Telescope
\citep[LSST;][]{LSST09} and  the Wide Field InfraRed  Survey Telescope
\citep[WFIRST;][]{SGB+15} can  only reach their full  potential if the
effects   of  intrinsic   alignments   can  be   mitigated  from   the
observations.  Continued efforts into  modelling the mildly non-linear
scales and pursuit of the roadmap outlined here will both increase our
understanding of  the formation and evolution  of intrinsic alignments
on all  scales and  also provide  data sets  with known  parameters to
develop and test mitigation techniques.

\section*{Acknowledgments}
\label{sec:acknowledgements}

We  acknowledge  the  support   of  the  International  Space  Science
Institute Bern for two workshops at which this work was conceived.  We
would like to thank Henk Hoekstra  for very useful discussions and for
his comments on drafts of this work.   We would also like to thank the
anonymous referee  for their careful reading  and detailed suggestions
that helped to improve the clarity of this review.  We are grateful to
J.  Blazek,  S.  Singh and  M.  Velliscig  for sharing their  data and
figures.  We would also like to  thank the participants of the Lorentz
workshop: Extracting  information from weak lensing:  Small scales=Big
problem, for their useful discussions and insights.

AK was supported in  part by JPL, run under a  contract by Caltech for
NASA.  AK was  also supported in part by NASA  ROSES 13-ATP13-0019 and
NASA  ROSES  12-EUCLID12-0004. MC  was  supported  by the  Netherlands
organisation for Scientific Research  (NWO) Vidi grant 639.042.814. BJ
acknowledges support  by an  STFC Ernest Rutherford  Fellowship, grant
reference ST/J004421/1.  TDK is supported  by a Royal Society  URF. AL
acknowledges  the  support of  the  European  Union Seventh  Framework
Programme  (FP7/2007-2013) under  grant  agreement  number 624151.  RM
acknowledges   the  support   of  NASA   ROSES  12-EUCLID12-0004.   CS
acknowledges  support from  the  European Research  Council under  FP7
grant number 279396. MLB is supported by the European Research Council
(EC  FP7  grant  number  280127)   and  by  a  STFC  Advanced/Halliday
fellowship (grant number ST/I005129/1).

\bibliographystyle{apj_title}

{\small
\bibliography{bibliography} 
}

\end{document}